\documentclass[fleqn,usenatbib, maxcitenames=3]{mnras}
\pdfoutput=1

\usepackage{newtxtext,newtxmath}
\usepackage{lscape}
\usepackage[T1]{fontenc}
\usepackage{ae,aecompl}

\usepackage{graphicx}	% Including figure files
\usepackage{amsmath}	% Advanced maths commands
\usepackage{amssymb}	% Extra maths symbols
\usepackage{enumitem}
\usepackage{float}
\title[V341 Ara: The Nova-Like Cataclysmic Variable That Has It All]
{Bow-shocks, nova shells, disc winds and tilted discs: The Nova-Like {\em V341~Ara} Has It All}
\author[N. Castro Segura et al.]{N. Castro Segura,$^{1}$\thanks{E-mail: N.Castro-Segura@Soton.ac.uk}
C. Knigge,$^{1}$
J. A. Acosta-Pulido$^{2,3}$,
D. Altamirano$^{1}$, 
S. del Palacio$^{4}$,
\newauthor
J.V. Hernandez Santisteban$^{5}$,
M. Pahari$^{1}$,
P. Rodriguez-Gil$^{2,3}$,
C. Belardi$^6$,
\newauthor
D.A.H. Buckley$^{7}$,
M.R. Burleigh$^6$,
M. Childress$^1$,
R.P. Fender$^{9,8}$,
D.M. Hewitt$^{7,8}$,
\newauthor
D.J. James$^{10}$,
R.B. Kuhn$^{7,11}$
N.P.M. Kuin$^{12}$,
J. Pepper$^{13}$,
A.A. Ponomareva$^{9,16}$,
\newauthor
M.L. Pretorius$^{7,8}$,
J.E. Rodr\'iguez$^{10}$,
K.G. Stassun$^{14}$,
D.R.A. Williams$^{15,9}$,
P.A. Woudt$^8$
\\
\newline
\emph{\normalsize Affiliations are listed at the end of the paper}
}
\date{Accepted XXX. Received YYY; in original form ZZZ}

\pubyear{2020}

\begin{document}
\label{firstpage}
\pagerange{\pageref{firstpage}--\pageref{lastpage}}
\maketitle

%Name       -- parallax -- gmag in Gaia
%TT Ari     -- 3.88     -- 10.92
%IX Vel     -- 11.04    -- 9.32
%V3885 Sgr  -- 7.54     -- 10.25
%RW Sex     -- 4.23     -- 10.63
%V341 Ara   -- 6.40     -- 10.71

\begin{abstract}

{\em V341~Ara} was recently recognised as one of the closest ($d \simeq 150$~pc) and brightest (V$\simeq 10$) nova-like cataclysmic variables. This unique system is surrounded by a bright emission nebula, likely to be the remnant of a recent nova eruption. Embedded within this nebula is a prominent bow-shock, where the system's accretion disc wind runs into its own nova shell. In order to establish its fundamental properties, we present the first comprehensive multi-wavelength study of the system. Long-term photometry reveals quasi-periodic, super-orbital variations with a characteristic time-scale of 10--16 days and typical amplitude of $\simeq$1~mag. High-cadence photometry from TESS reveals for the first time both the orbital period and a ``negative superhump'' period. The latter is usually interpreted as the signature of a tilted accretion disc. We propose a recently developed disc instability model as a plausible explanation for the photometric behaviour. In our spectroscopic data, we clearly detect anti-phased absorption and emission line components. Their radial velocities suggest a high mass ratio, which in turn implies an unusually low white dwarf mass. We also constrain the wind mass-loss rate of the system from the spatially resolved [O{\sc iii}] emission produced in the bow-shock;  this can be used to test and calibrate accretion disc wind models. We suggest a possible association between V341~Ara and a ``guest star'' mentioned in Chinese historical records in AD1240. If this marks the date of the system's nova eruption, {\em V341~Ara} would be the oldest recovered nova of its class and an excellent laboratory for testing nova theory.

\end{abstract}


%%%%%%%%%%%%%%%%%%%%%%%%%%%%%%%%%%%%%%%%%%%%%%%%%%

%%%%%%%%%%%%%%%%% BODY OF PAPER %%%%%%%%%%%%%%%%%%

\section{Introduction}

    Cataclysmic variable stars (CVs) are interacting binary systems in which a Roche lobe-filling secondary star loses mass to a white dwarf (WD) primary. In systems where the magnetic field of the WD is dynamically unimportant, accretion takes place via a disc. Nova-like variables (NLs) are a sub-class of CVs in which the mass-transfer rate from the companion is high enough to keep this disc in a fully ionized, optically thick, geometrically thin state \citep{Lasota2016}. In non-magnetic systems, the disc extends all the way down to the surface of the more slowly rotating WD. 
    %{\bf At these accretion rates the interface between these components is expected to be a narrow, optically thick boundary layer that radiates up to half of the total accretion luminosity. While at lower accretion rates is expected to be optically thin.
    %\underline{or} }
    The interface between these components is expected to be a narrow boundary layer (BL) that radiates up to half of the total accretion luminosity. This BL is usually expected to be optically thick at the high accretion rates found in NLs, although some recent models suggest a more nuanced picture \citep{Hertfelder2013AA...560A..56H}.
    NLs are perhaps the only astrophysical systems where standard, steady-state, non-relativistic accretion disc theory \citep{ShakuraSunyaev1973} should apply without  modification. This makes them extremely useful laboratories for testing the theory.
    
    The 10$^{th}$ magnitude variable star {\em V341~Ara} was first catalogued by \cite{Leavitt+1907}. It was then (mis-) classified by \citet{Hoffmeister1956} as a Type II Cepheid, based on the presence of a $P \simeq 12$~d period in their light curve of the system. In spite of {\em V341~Ara}'s extremely blue colour, this mis-classification stood for more than 60~years, with subsequent photometric studies reporting similar, though not identical periodic variability ($P \simeq 11$~d, \citealt{Hipparcos1997ESASP1200.....E}; $P \simeq 14$~d, \citealt{Berdnikov1998AcA....48..763B}).

    \begin{figure}
        \centering
        \includegraphics[width=.45\textwidth]{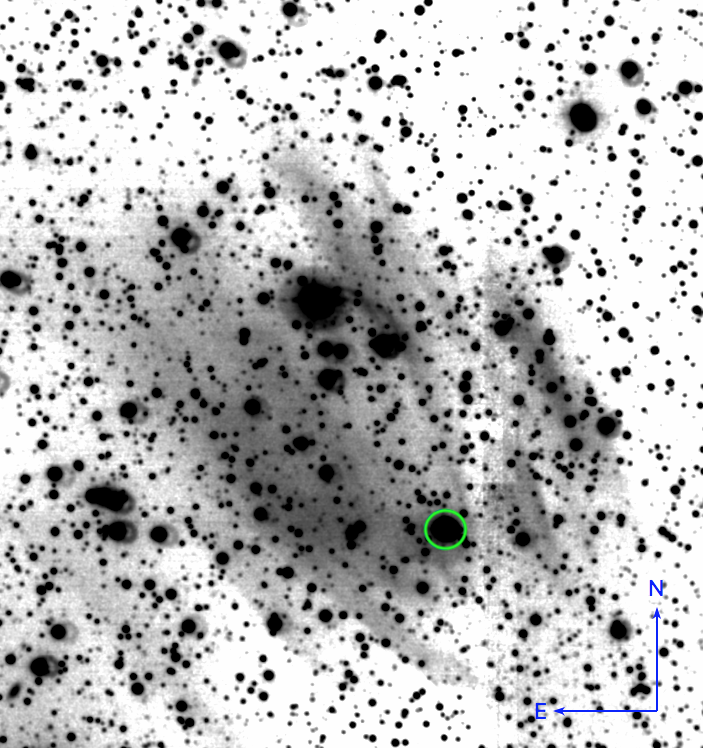}
        \hspace{0.05\textwidth}
        \includegraphics[width=.45\textwidth]{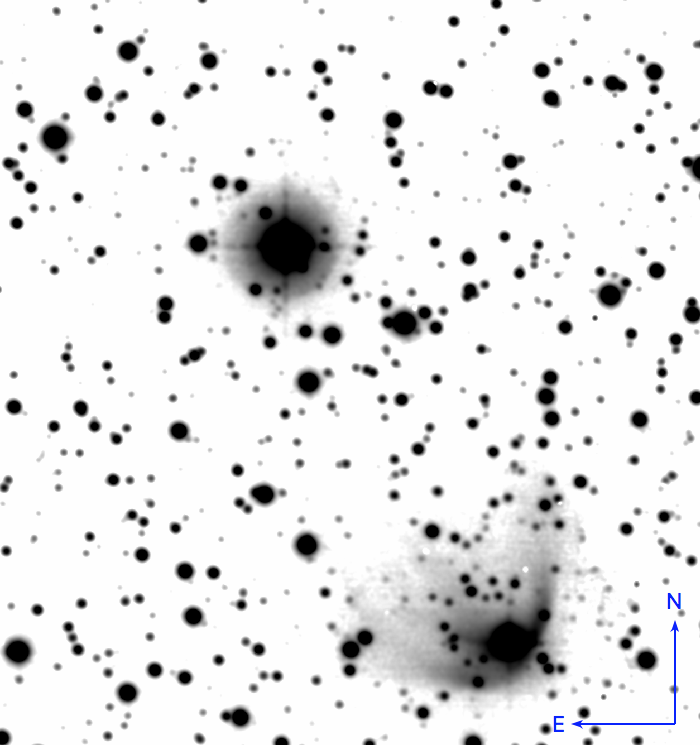}
        \caption{ESO/WFI narrow band images of the field around {\em V341~Ara}. Top: H$\alpha$ illustrating the emission  nebula in the vicinity of the nova-like variable, the star is marked with a green circle. Note the cosmetic defects North-West from the source due to gaps between different chips in the image. Bottom: Same field, but with 2x zoom, in [O{\sc iii}] illustrating the bowshock around {\em V341~Ara}. Images were processed using {\sc theli} software \citep{THELI2005AN....326..432E,THELI2013ApJS..209...21S}.}
        \label{fig:V341Ara image}
    \end{figure}
    
    In fact, {\em V341~Ara} is one of the brightest CVs known, though it was not identified as such until recently. This identification was made by \citet[][hereafter F08]{Frew2008PhDT.......109F}, based on a combination of spectroscopic signatures, flickering, blue colour and its identification with a ROSAT X-ray source
    \footnote{\citet{Samus+2007PZ.....27....6S}, as well as \citet{kiraga2012AcA....62...67K}, independently proposed the classification of {\em V341~Ara} as a CV, based on the X-ray match.}.
    F08's primary science objective was actually the identification of new planetary nebulae (PNe). Their study of {\em V341~Ara} was prompted by  its association with a previously uncatalogued large ($8'\times 6'$)  H$\alpha$ nebula. More recently, \citet[][hereafter B18]{BondMiszalski2018PASP..130i4201B} confirmed the NL classification and carried out the first radial velocity study for the system. They reported an orbital period of $P_{orb} = 3.65$~h, consistent with the bulk of the NL population.  
    
    NLs with periods in the range $3~{\mathrm h} \lesssim P_{orb} \lesssim 4~{\mathrm h}$ represent a critical evolutionary phase. The observed CV population exhibits a well-known ``period gap'' between $P_{orb}\simeq 2-{\mathrm h}$ and $P_{orb}\simeq 3-{\mathrm h}$. Individual CVs are thought to approach the upper edge of the gap from above (i.e. longer periods), but to evolve {\em through} the gap as detached systems. NLs like {\em V341~Ara} are therefore systems whose donors will soon lose contact with the Roche lobe, possibly as they make the transition from partially radiative to fully convective stars (see \citealt{Knigge+2011ApJS..194...28K} for a review). Many, if not all, of these systems also share at least some of the peculiar observational characteristics collectively known as the 
   ``SW Sex syndrom''  \citep{Dhillon_1990PhDT,Thorstensen1991AJ....102..272T,Rodriguez-Gil2007MNRAS.377.1747R,Schmidtobreick2012MmSAI..83..610S,Dhillon2013MNRAS.428.3559D}.
  
    Based on its H$\alpha$ luminosity, F08 estimated the mass of the nebula around {\em V341~Ara} as $M_{H\alpha}\simeq 5\times 10^{-3} ~\mathrm{M_\odot}$, roughly 100$\times$ less massive than a typical PN. Both F08 and and B18 suggested that the nebula may therefore be an old nova remnant associated with {\em V341~Ara} itself. Intriguingly, the star is located $\simeq$2$'$ from the center of the H$\alpha$ nebula, which might be due to its high proper motion if the nova eruption occurred $\simeq$1000~yrs ago. 
    
    However, perhaps the most important and unusual feature of {\em V341~Ara} is that it lies near the apex of an impressive parabolic bow-shock, thought to be due to the interaction of its accretion disc wind with the ISM (F08, B18). This bow-shock is seen in both the [O{\sc iii}] and [N{\sc ii}] nebular emission lines and is located within the larger scale H$\alpha$ nebula (see fig.\ref{fig:V341Ara image}). Only two other CVs surrounded by wind-driven bow-shocks are known: {\em BZ~Cam} \citep{Hollis+1992ApJ...393..217H} %, Te~11 \citep{Miszalski2016} 
    and {\em V1838~Aql} \citep{HernandezSantisteban2019MNRAS.486.2631H}
    \footnote{A bow-shock is also present around the NL IPHASX J210204.7+471015, but this is probably due to the interaction of this system's nova shell with the ISM \citep{Guerrero+2018ApJ...857...80G}.}.
    These objects are critical for testing disc wind models for CVs \citep[e.g.][]{Knigge+1997ApJ...476..291K,LongKnigge2002ApJ...579..725L,Matthews+2015MNRAS.450.3331M}, since 
    the observed shock properties can provide a model-independent mass-loss rate estimate (e.g. \citealt{Kobulnicky2+018ApJ...856...74K}). 
    
    Here, we present the first comprehensive multi-wavelength study of {\em V341~Ara} in order to establish the fundamental properties of the system. In Section~\ref{sec:data}, we introduce the data sets we use. Section~\ref{sec: Results} presents our analysis of the data, with Tables~\ref{tab:system param} and~\ref{tab:xrt fit} in this section collecting some of our main results. Finally, in Section~\ref{sec:discuss}, we discuss the unique nature of this system, given all of the observational evidence available that has been collected to date. In Section~\ref{sec:summary} the main results are summarized.

    \begin{figure}
        \centering
        \includegraphics[width=0.45\textwidth]{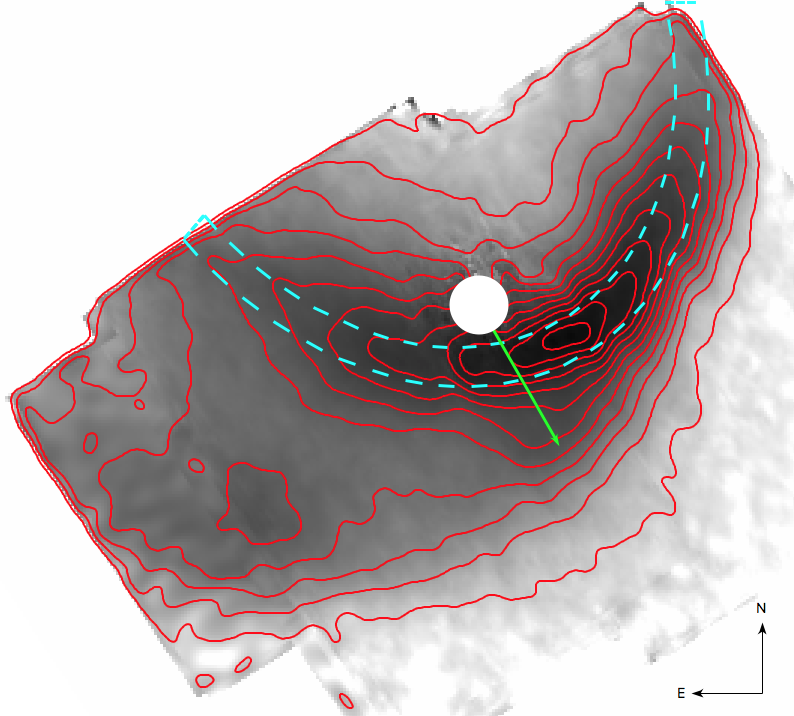}
        \caption{Combined image with background subtracted of the numerically integrated [O{\sc iii}] $\lambda 5007$\AA~ emission line from the integral field spectroscopy. The 4 arcsecond diameter circle is centered in the position of the source, while the green arrow represent the direction of the proper motion of the star derived from {\em Gaia DR2}. The dashed cyan paraboloid symmetry axis is aligned with the proper motion of the star and has a thickness of 3 arcseconds. Contours with equal flux density are plotted in red.}
        \label{fig:oiii WiFeS}
    \end{figure}

\section{Data Sets} \label{sec:data}
%{\em Table: Summary of the main data sets used in this study. (Should perhaps also include ROSAT, GALEX, SkyMapper, 2MASS, %VISTA, WISE --- Do we want to include an endless table?}
%\hline

    %In order to determine the orbital period and other the basic properties of the binary system, we carried out a %spectroscopic campaign with the {\em SAAO 1.9m telescope} in 2017. This data set proved to be insufficient, %therefore we performed time-resolved spectroscopy with the integral field unit {\em WiFeS} at {\em  SSO} in 2018. %This data is supplemented by recent and archival data obtained with the {\em CTIO 1.5m} and new observations %gathered with {\em SALT}. All together, the data set spans a maximum wavelength range of $\sim 0.4-1\mu m$, with a %total of 117 spectra, in three different years from 2006 to 2018. 
    
    %We also obtained the long-term photometric light curves of {\em V341~Ara} taken by the {\em TESS}, {\em ASAS-SN} and {\em %KELT} surveys. As well as archival data from the {\em Hipparcos satellite}, {\em ROSAT}, {\em GALEX}, {\em %SkyMapper}, {\em 2MASS}, {\em VISTA} and {\em WISE} which allow us to construct a full SED. 

    %Finally, we obtained $\sim$ 4 kilo second exposures in X-rays, along with (low resolution) UV grism spectroscopy, %with {\em Neil Gehrels Swift Observatory}. 

    For the present study, we have collected multi-wavelength data covering the radio, infrared, optical, near-ultraviolet and X-ray bands. This includes multiple sets of time-resolved optical photometry and time-resolved optical spectroscopy. It also includes 2-D optical spectroscopy covering the nebular emission around {\em V341~Ara}. In the remainder of this section, we briefly describe these data sets.
    A summary of the observations is provided in Tables~\ref{tab:sp log}~and~\ref{tab:phot log} for the spectroscopic and photometric time-series respectively.
    For reference, Fig.~\ref{fig:LCs} shows the epochs at which the various data sets were obtained, superposed on our long-term optical light curves.
    
\subsection{Time-resolved photometry}
%{\em Mention HIPARCOS in intro to this section.}
%\\
%\\\hline

    %This Nova-like system is known to exhibit (quasi-)periodicity of about 10 days (see fig. \ref{fig:hipparcos_folded}) %%which lead to the miss-classification as a Cepheid star clearly present in the {\em Hipparcos catalogue} %\citep{ESA1997} taken from 1990 to 1993 \citep[e.g.][]{Berdnikov1998StudyCepheids}. In this work we compile %extensive light curves in order to search for the orbital period and determine the presistency of these %super-orbital periodicities.
    
    %In figure \ref{fig:long term lc} the long-term optical light curve is presented. The datasets of our various other %observation epochs are marked. 
    
\subsubsection{ASAS-SN}

    The All-Sky Automated Survey for Supernovae ({\em ASAS-SN}) aims to provide near-continuous all-sky optical monitoring. It reaches a depth of $V \simeq 17$ and provides 3x90~s sequential exposures each night for the entire observable sky \citep{Kochanek+2017PASP..129j4502K}. The survey uses five stations, each of which consists of four 14 cm aperture Nikon telephoto lenses. The observations are made in the $V$-band (two stations) or $g$-band (three stations). {\em V341~Ara} has been monitored by {\em ASAS-SN} since March 2016. Here, we use all 893 $g$-band and 616 in $V$-band measurement that were gathered prior to October 2019.

\subsubsection{KELT}\label{sec:kelt}

    The Kilo-degree Extremely Little Telescope ({\em KELT}) project \citep{Pepper+2007PASP..119..923P,Pepper2018haex.bookE.128P} is a survey designed primarily for the detection of planetary transits around bright stars. Here, we use data from {\em KELT-South}, a 4-cm telescope that monitors the southern sky with a spatial resolution of $23''$ per pixel and a cadence of about 20-30 minutes (for airmass < 1.5). In order to maximize sensitivity, {\em KELT} does not employ filters, but the response function of the system peaks around $6000$\AA\ \citep{Pepper+2012PASP..124..230P}. {\em KELT} gathered data for {\em V341~Ara} from 2013 to 2018, providing a a total of 5840 photometric measurements across this 6-year period.  
    
\subsubsection{TESS}

    The {\em Transiting Exoplanet Survey Satellite} \citep[{\em TESS},][]{TESS2015JATIS...1a4003R} is a space-based telescope launched in 2018. Like {\em KELT}, {\em TESS}'s primary mission is to detect transiting exoplanets around nearby stars. It provides optical images with an effective spatial resolution of $21''$, covering a bandpass of $\simeq 0.6-1 \mathrm{\mu m}$. 
 
    {\em TESS} monitors the sky in sectors, with each sector being observed continuously for 27 days. Full-frame images are provided at a cadence of 30~minutes, but $10\times 10$ pixel postage stamps around a set of pre-selected targets are downloaded at the optimum 2~minute cadence. The data for these fortunate targets is reduced by the Science Processing Operations Center (SPOC) pipeline \citep{TESSpipeline2016SPIE.9913E..3EJ}, during this process a pixel-by-pixel correction is applied, the optimal photometric aperture along with an estimation of the contamination by nearby stars is calculated before extracting the light curves. Barycentered light curves (``LCF files'') are available from the {\em Mikulski Archive for Space Telescopes\footnote{\url{http://archive.stsci.edu/tess/}} (MAST)}.
    
    {\em V341~Ara} was one of the pre-selected targets in {\em TESS} Sector 12. As such, it was observed with 2-minute cadence from 2019 May 21 to 2019 June 19. %The 27-day {\em TESS} light curve is presented in Figure~\ref{fig:tess} and covers roughly 2 super-orbital periods. 
    
%\hline
%Figure: The long-term optical light curve of {\em V341~Ara}. The dates of our various other observation %epochs are marked.

\subsection{Time-resolved 1D and 2D spectroscopy}

%        Optical spectroscopy was reduced using the standard {\em IRAF} routines for {\em SAAO 1.9} and %{\em CTIO 1.5}, while {\em  PySALT} and {\em PyWiFeS} packages were used for {\em SALT} and {\em %WiFeS} respectively. In all cases, the LACOSMIC cosmic ray rejection routine was applied to the 2D %spectra before subtraction. %With the exception of {\em SALT}, 
%        All the observation provided arc spectra bracketing the science exposures, allowing a %time-dependent wavelength calibration. 

         %The data from this campaigns were reduced by using standard IRAF routines. 
%         In the data reduction process bias, flat-field, illumination and response corrections are %performed. Time-dependent wavelength calibration was carried out by interpolating between the two %closest calibration lamp exposures. 
%        %All together, we dispose of XX blue and XY red spectra.
%%In order to determine the fundamental binary parameters a series of spectra were taken using %different instruments. The log of all the observations and configuration can be seen in table %ref{table:sp log}

\subsubsection{CTIO: RC Spectrograph}

    B18 obtained medium resolution time-resolved spectroscopic observations of {\em V341~Ara} with the {\em RC Spectrograph} on the \emph{SMARTS} 1.5 m telescope at the \emph{Cerro Tololo Inter-american Observatory (CTIO)}. All of their data were collected between April and July of 2006, but most of their observations are missing from the CTIO archive. In our long-term radial velocity study, we therefore directly adopt the 59 measurements provided by B18 for the centroid of the H$\alpha$ emission line core in their red exposures.
    
\subsubsection{CTIO: Chiron}

    \emph{Chiron} \citep{Chiron2013PASP..125.1336T} is a high dispersion bench-mounted echelle spectrograph on the \emph{SMARTS 1.5m telescope} at \emph{CTIO}. We used {\em Chiron} in fiber mode, covering the 4083-8901~\AA\ region in 75 orders at a resolution of $R=\lambda/\Delta\lambda\simeq 27800$. We obtained a total of 26 exposures on two nights in 2018. Three exposures were taken on 25 March 2018, the remainder on 12 May 2018. All spectra were extracted and reduced using the {\sc Ch\_Reduce} code (Walter 2017\footnote{\url{http://www.astro.sunysb.edu/fwalter/SMARTS/NovaAtlas/ch_reduce/ch_reduce.html}}). 
    %Since the spectrograph is fiber-fed, sky subtraction is not possible. As part of the data reduction, the images were flattened, and the data %orders were extracted using a boxcar extraction algorithm with local instrumental background subtraction. Wavelength calibration was based on %a similarly-extracted ThAr comparison lamp. The instrumental signature was removed using spectra of the spectroscopic standard star $\mu$~Col %obtained with the identical instrumental setup. Finally, the spectra were resampled to a linear wavelength scale, and the orders were spliced %together to produce 1-dimensional spectra covering the full wavelength range.

\subsubsection{SAAO: SpUpNIC}

\emph{SpUpNIC} \citep{SpUpNIC2016SPIE.9908E..27C} is a low- to medium-resolution long-slit spectrograph mounted on the {\em 1.9-m Radcliffe telescope} at the {\em South African Astronomical Observatory (SAAO)}. We used the {\em gr4} grating for our observations, which provides a wavelength coverage of $\lambda\lambda3750-5100$~\AA\ with a dispersion of $\simeq\! 0.6$~\AA/pix. We obtained 12 exposures covering slightly less than 1~hr on each of the nights of 2017 March 12 and 16.  One additional exposure each was obtained on the nights of 2017 March 15 and 17. The spectra were reduced using standard {\sc IRAF} routines. Cosmic ray rejection was performed with {\sc lacosmic} \citep{LACOSMIC2001PASP..113.1420V}.

\subsubsection{SAOO: SALT/HRS}

The {\em High Resolution Spectrograph}  \citep[{\em HRS}; ][]{SALT-HRS2014SPIE.9147E..6TC} is a double-beam, fiber-feed echelle spectrograph mounted on the {\em Southern African Large Telescope (SALT)}. We used the low resolution mode of HRS to obtain spectra covering the range $\lambda\lambda3800-8900$\AA\ with a resolving power of $R\simeq15000$. Our observations were carried out on four nights between 2018 April 15 and  2018 April 21, with 5 exposures being taken on each of these nights. Data reduction was performed with the  {\sc  PySALT} pipeline.

\subsubsection{SSO: WiFeS}

The {\em Wide Field Spectrograph} \citep[{\em WiFeS},][]{WiFeS2007Ap&SS.310..255D} is a double-beam, image-slicing integral field unit (IFU) mounted on the Australian National University's 2.3~m telescope at Siding Spring Observatory. It provides spectral images across a contiguous $25'' \times 38''$ field of view, using $25\times 1''$ ``slitlets'' along the short axis and $0.5''$/pixel spatial sampling along the long axis. We used {\em WiFeS} in the B7000 (blue arm) and R7000 (red arm) configurations, which provides a resolution of $R\simeq7000$ across the full wavelength range $\lambda\lambda5300-7100$~\AA. Our observations were obtained on the night of 2018 April 26, during which the target field was monitored for 4.5~hours without interruption, using 300~s integrations. In order to fully cover the bow-shock surrounding {\em V341~Ara}, we used several different pointings to cover a total field of approximately $50''\times 40''$. The shifts and rotations between pointings were typically $20''$ and 90$^\circ$, respectively. 

Since {\em WiFeS} provides {\em spatially-resolved} spectroscopy, we used this data set for two distinct purposes. First, we extracted time-resolved 1-D spectroscopy for the central point source using a $3\times3$ spaxel aperture centered on the peak of the continuum count rate distribution. Second, we constructed narrow-band images by aligning all pointings to the central source and then combining them. We used a resampling factor of 5 for this mosaic, yielding a plate scale of $0.2''$ per spaxel for the final images. In order to create pure emission line images, we created narrow-band images centered on specific transitions and then subtracted a linear fit to the local continuum in each spaxel. Given that seeing was better than $0.9''$ throughout our observations, these are the highest-resolution, unsaturated images of {\em V341~Ara} and its bow-shock to date. Figure~\ref{fig:oiii WiFeS} shows the resulting emission line image for the [O {\sc iii}] $\lambda 5007$~\AA~transition.

\subsection{{\em Swift} X-ray and ultraviolet observations}
\label{sec:swift}

We observed {\em V341~Ara} twice with the {\em Neil Gehrels Swift Observatory} \citep{XRT2005SSRv..120..165B}, on 2018 July 04 and 2019 February 28, for a total exposure time of 4.6~ks.
We used standard {\em Swift/XRT} routines ({\sc xrtpipeline 0.13.4} in \emph{heasoft 6.25}) to reduce the XRT data, extract counts from a circular aperture of $5''$ centered on the source and construct a spectrum covering the 0.3-10 keV range for each of the two epochs. There was no significant variability in the X-ray data within or between the two epochs. 

We also used {\em Swift} to obtain near-ultraviolet (NUV) imaging observations with the {\em UV/Optical Telescope} \citep[{\em UVOT}][]{UVOT2005SSRv..120...95R}. In July 2018, we imaged the system for 326~s with the UVW1 filter ($\lambda_{eff} \simeq 2590$~\AA), while in February 2019, we used the UWM2 filter ($\lambda_{eff} \simeq 2230$~\AA) for a total of 680~s. Since {\em V341~Ara} was saturated in all of the {\em UVOT} images, we used readout-streak photometry \citep{Page2013MNRAS.436.1684P} to estimate source fluxes in the UVW1 and UVM2 bands. The  scatter in each band was consistent with being due to photometric errors, so we combined the exposures in each band to produce a single flux estimate for each filter.

Finally, in February 2019, we also obtained a total of 2.2~ks of NUV spectroscopy using the {\em Swift/UVOT} UV grism, which covers the wavelength range $\lambda\lambda 1800 - 5200$~\AA\ at a resolution of $R\simeq 150$. The spectroscopic data were reduced using the  {\em Swift} UVOT Grism {\sc uvotpy} package (version 2.4, \citealt{UVOTPY2014ascl.soft10004K}, an implementation of the calibration from \citealt{UVOTPY2015MNRAS.449.2514K}). The combined 1-D spectrum was corrected for a systematic wavelength shift by ensuring that the center of the Mg~{\sc ii} emission line lies close to 2800~\AA, its expected position. Since the detector sensitivity below 2000~\AA\ is decreasing more rapidly than at longer wavelengths, the flux calibration for $\lambda \lesssim 2000$~\AA\ is currently unreliable. We therefore removed this part of the spectrum.

\subsection{{\em MeerKAT} radio observations}
\label{sec:meerkat}

{\em MeerKAT} \citep{MeerKAT2009IEEEP..97.1522J} is a radio telescope consisting of 64$\times 13.5$~m antennas located in the Karoo region of South Africa. %It is a pre-cursor to the {\em Square Kilometer Array (SKA)} and will be incorporated into Phase~I of the SKA Mid-Frequency Array. 
The field containing {\em V341~Ara} was observed with {\em MeerKAT}  at a central frequency of 1.284\,GHz on 2019 March 29 as part of the {\em ThunderKAT} Large Survey Project \citep{ThunderKAT2017arXiv171104132F}. The full bandwidth of 856\,MHz was split into 4096 channels, visibilities were recorded every 8\,seconds, and 60 of the {\em MeerKAT} antennae were utilised. The Seyfert 2 galaxy {\em J1939-6342} was used as a bandpass and flux calibrator. The total integration time on {\em V341~Ara} was approximately $120$\,mins. The data were flagged using {\sc AOFlagger} version 2.9.0 \citep{AOFlagger2010ascl.soft10017O} to remove radio frequency interference, averaged by a factor of 8 in frequency and then calibrated using standard procedures in {\sc casa} version 5.1.1 \citep{CASA2007ASPC..376..127M}. The calibrated data was imaged using the multi-facet-based radio imaging package {\sc DDFacet} \citep{DDFacet2018A&A...611A..87T}, with Briggs weighting, a robustness parameter of R=-0.5 and the SSD deconvolution algorithm. No self-calibration was implemented. Flux measurements and source-fitting were executed using PyBDSF \footnote{\url{https://www.astron.nl/citt/pybdsf/}} and noise levels were measured within the local area surrounding the source. The restoring beam had a size of 5.87 $\times$ 4.63\,arcsecond$^2$. 

Some CVs are strong radio emitters \citep[see][for a review]{CoppejansKnigge2020arXiv200305953C}. Unfortunately, {\em V341~Ara} was not detected in these observations, unlike some other southern NLs \citep{2020MNRAS.496.2542H}. The corresponding 3$\sigma$ flux limit is 30\,$\upmu$Jy\,beam$^{-1}$.

% \begin{figure}
%        \centering
%        \includegraphics[width=0.45\textwidth]{fig/V341_colcont_inferno_3579.png}
%        \caption{Radio map at 1.284\,GHz in the vicinity of {\em V341~Ara} gathered by {\em MeerKAT} with a bandwidth of %856\,MHz. the {\em Gaia DR2} position of the optical counterpart is marked with a $+$ symbol.}
%        \label{fig:V341Ara radio}
%    \end{figure}

\begin{figure*}
\begin{minipage}{\textwidth}

    \includegraphics[width=.99\textwidth]{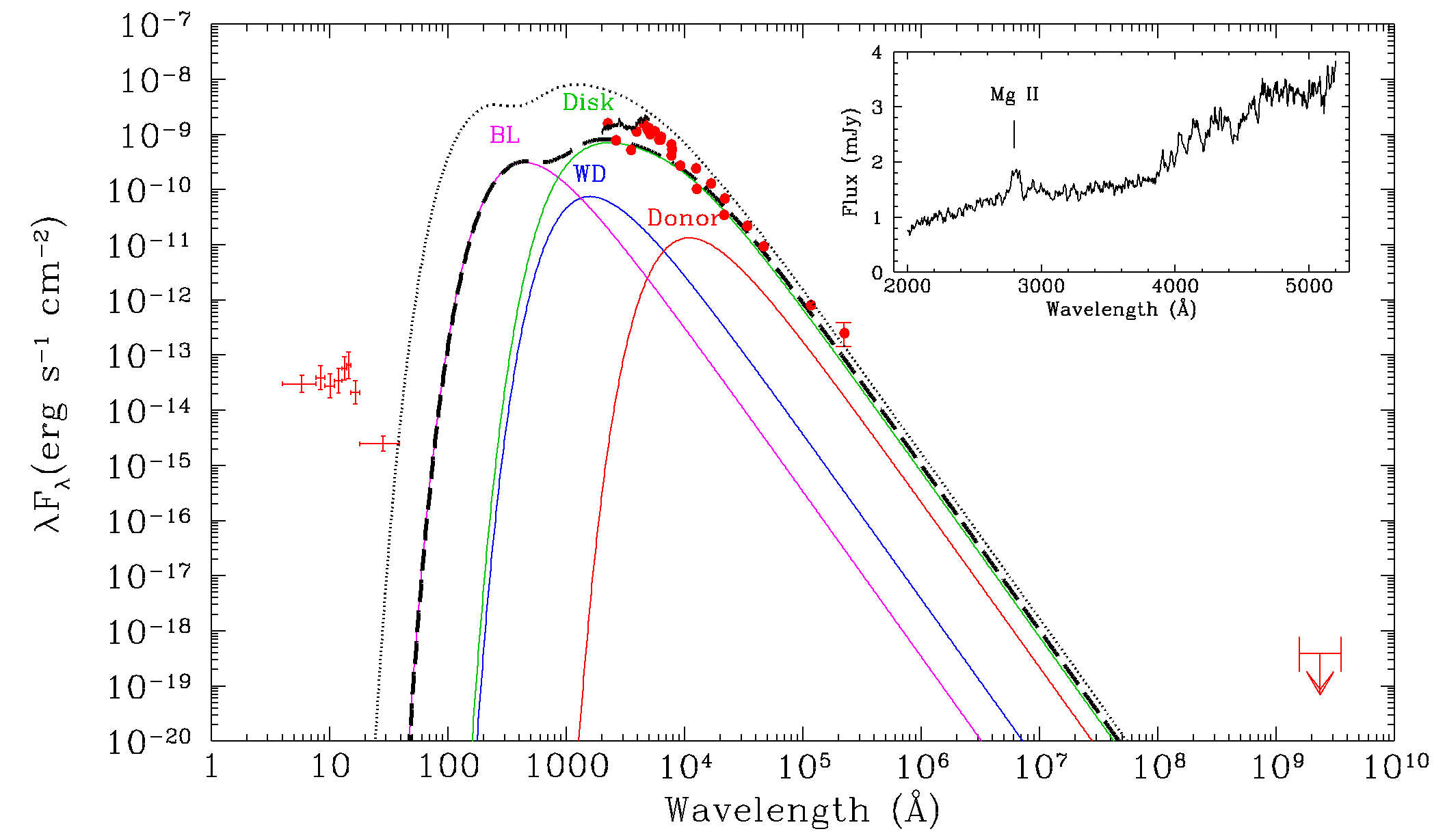}
    \caption{\small  Spectral energy distribution of {\em V341~Ara} based on archival broad band photometry (Sec. ~\ref{sec:other}), {\em Swift} and {\em MeerKAT} data sets (Sections~\ref{sec:swift} and ~\ref{sec:meerkat} respectively). The inset is a zoom into the NUV spectroscopy obtained with {\em Siwft/UVOT}. These data sets were not obtained simultaneously, so the photometric scatter is due to the intrinsic variability of the source. The different component of the system with the estimated parameters from \citet*{Knigge+2011ApJS..194...28K} are overlaid on top of the data for reference. Dashed black curve is the summed of the system components accreting at $\mathrm{10^{-9} ~M_\odot yr^{-1}}$ (green line), while dotted curve is the same but at higher accretion rate ($\mathrm{10^{-8} ~M_\odot yr^{-1}}$). 
}
    \label{fig:full SED}
\end{minipage}
\end{figure*}

%\subsection{{\em Gaia} Astrometry}
%\label{sec:gaia}
%
%{\em V341~Ara} is included in {\em Gaia} DR2 %\citep[][]{GAIADR22018AA...616A...1G}, with a measured parallax %of $\pi = 6.40 \pm 0.08$~mas, and a proper motion vector of %$\mu_{\alpha} \cos{\delta} = -48.32 \pm 0.07$~mas~yr$^{-1}$ and 
%$\mu_{\delta} = -84.91 \pm 0.08$~mas~yr$^{-1}$. For its geometric %distance, we adopt the Bayesian estimate provided by %\citet{Bailer-Jones2018AJ....156...58B}, $d = 155.5 \pm 2.0$~pc, %which is based on (but not sensitive to) an exponential space %density prior with a characteristic length scale of 1480~pc. It %also includes a correction for {\em Gaia}'s global parallax %zeropoint, $\pi_{ZP} = -0.029$~mas %\citep{Lindegren+2018AA...616A...2L}.
%
\subsection{Supplementary photometric data}
\label{sec:other}

In order to characterise the overall spectral energy distribution of {\em V341~Ara}, we have collated  additional photometric measurements from the following sources: SkyMapper DR1 \citep[optical: uvgriz;][]{SkyMapperDR12018PASA...35...10W}, APASS \citep[optical: BV+gri;][]{APASS_Cat2016yCat.2336....0H}, Gaia DR2 \citep[optical: G+B$_p$+R$_p$;][]{GAIADR22018AA...616A...1G}, 2MASS \citep[near-infrared: JHK$_s$;][]{2MASS2006AJ....131.1163S}, VISTA VHS DR4 \citep[JK$_s$;][]{VHSDR42019yCat.2359....0M}, WISE/ALLWISE \citep[mid-infrared: W1+W2+W3+W4; ][]{WISE2014yCat.2328....0C}.

\section{Analysis and Results} \label{sec: Results}

\subsection{Spectral energy distribution: luminosity and accretion rate}\label{sec: SED}

%{\em Figure: Broad-band SED}
%\begin{itemize}[leftmargin=0.4cm, label=$\bullet$]
%    \item ROSAT/XRT --> GALEX/UVOT --> WISE
%    \item include Swift grating
%    \item probably don't include optical spectroscopy (?)
%    \item do show multiple observations where we have them (X-ray, UV, opt, IR) --> illustrates variability
%    \item overlay stellar atmospher fit: could adopt that from 2017MNRAS.471..770M  -- Teff = 8169; L/Lsun = 1.148; R/Rsun = 0.536; log(g) = 5.205
%    \item perhaps also overlay BB fit (might actually be better than atmosphere; weaker Balmer jump)
%    \item perhaps also show IX Vel / RW Sex for comparison?
%    \item might want to overlay a typical (donor sequence based) donor SED (to show why it doesn't matter)
%\end{itemize}
%{\em Things to touch on:}
%\begin{itemize}[leftmargin=0.4cm, label=$\bullet$]
%    \item Weak Balmer jump (?) -- common among NLs; could be sign of wind (ref to Knigge, Matthews)
%    \item Mg II line -- emission (not actually all that common, I think)
%    \item no IR excess -- disc dominates everywhere (except X-rays)
%    \item X-ray to optical ratio --> typical of NLs? (check paper with Verbunt -- Pfefferman? Van Teeseling?)
%    \item accretion luminosity?
%    \item corresponding accretion rate for M1 = 0.8 +/- 0.2 (typical for CVs)
%\end{itemize}

%\hline

Figure~\ref{fig:full SED} shows the panchromatic SED of {\em V341~Ara}, based on the data sets described in Sections~\ref{sec:swift}, ~\ref{sec:meerkat} and ~\ref{sec:other}. These data sets were not obtained simultaneously, so the photometric scatter in Figure~\ref{fig:full SED} is consistent with the observed variability of the source. As is already clear from Figures~\ref{fig:LCs} and~\ref{fig:dynamic_power}, {\em V341~Ara} exhibits significant super-orbital variability in the optical (see Section~\ref{sec:LCs} for details). 

To provide context and guide the eye, we also show in Figure~\ref{fig:full SED} a simple model SED that is the sum of four components: (i) a geometrically thin, optically thick accretion disc; (ii) an optically thick boundary layer (BL); (iii) an accretion-heated WD; (iv) a donor star. The inclination towards {\em V341~Ara} is not known, but its absorption line dominated optical spectrum suggest a relatively face-on viewing angle \citep[c.f.][]{Beuermann+1990AA...230..326B,Beuermann+1992AA...256..433B,Matthews+2015MNRAS.450.3331M}; we therefore adopt $i \simeq 30^\circ$ in our SED model. 

Most of the remaining system parameters were taken directly from the best-fit CV evolution sequence provided by \citet*[][hereafter KBP11]{Knigge+2011ApJS..194...28K}. Specifically, we adopted a mass-transfer rate of $\mathrm{\dot{M}\simeq 10^{-9} ~M_\odot yr^{-1}}$, a donor mass of $M_2 \simeq 0.26~M_{\odot}$, a donor radius of $R_2 \simeq 0.34~R_{\odot}$ and a donor temperature of $T_{eff,2} \simeq 3380$~K. Based on the results shown in Section~\ref{sec:rv} below, the WD mass was taken to be $M_{WD} = 0.5$~M$_{\odot}$. The WD temperature, $T_{WD} = 23,2000~{\mathrm K}$ and radius, $R_{WD} = 1.2\times10^9~{\mathrm cm}$ were estimated from accretion heating models of \citet{Townsley2004ApJ...600..390T} and the WD cooling models of \citet{2006AJ....132.1221H}, respectively. 

Each of the radiating components in the model (the WD, the BL, the donor and each disc annulus) is assumed to emit as a blackbody. The BL is modelled as an cylindrical strip wrapped around the equator of the WD that generates half of the accretion luminosity. The total vertical extent of the BL is taken to be $H_{BL} = 0.1~R_{WD}$, though only the top half of the BL is visible to the observer. With these assumptions, the effective temperature of the BL in the model is $T_{eff,BL} \simeq 83,300$~K. For comparison, Figure~\ref{fig:full SED} also includes a second SED model that corresponds to a higher mass-transfer rate, $\mathrm{\dot{M} = 10^{-8} ~M_\odot yr^{-1}}$. The WD and BL temperatures are then $\mathrm{T_{eff,WD} \simeq 43,300}$~K and $T_{eff,BL} \simeq 150,700$~K, respectively.

Figure~\ref{fig:full SED} shows that -- with the exception of the X-ray band -- {\em V341~Ara}'s SED is broadly consistent with these simple SED models. The excess X-ray emission compared to the model is not unusual for a NL variable \citep[e.g.][]{Pratt2004MNRAS.348L..49P} and will be discussed further in Section~\ref{sec:BL}. Thus, based on its SED, {\em V341~Ara} appears to be a normal NL variable, with system parameters typical for its orbital period. In particular, its accretion rate is likely between $\mathrm{10^{-9} ~M_\odot yr^{-1}}$ and $\mathrm{10^{-8} ~M_\odot yr^{-1}}$.

\subsection{Optical light curves: large-amplitude, super-orbital, quasi-periodic variations} \label{sec:LCs}

\begin{figure*}
    \includegraphics[width=0.8\textwidth]{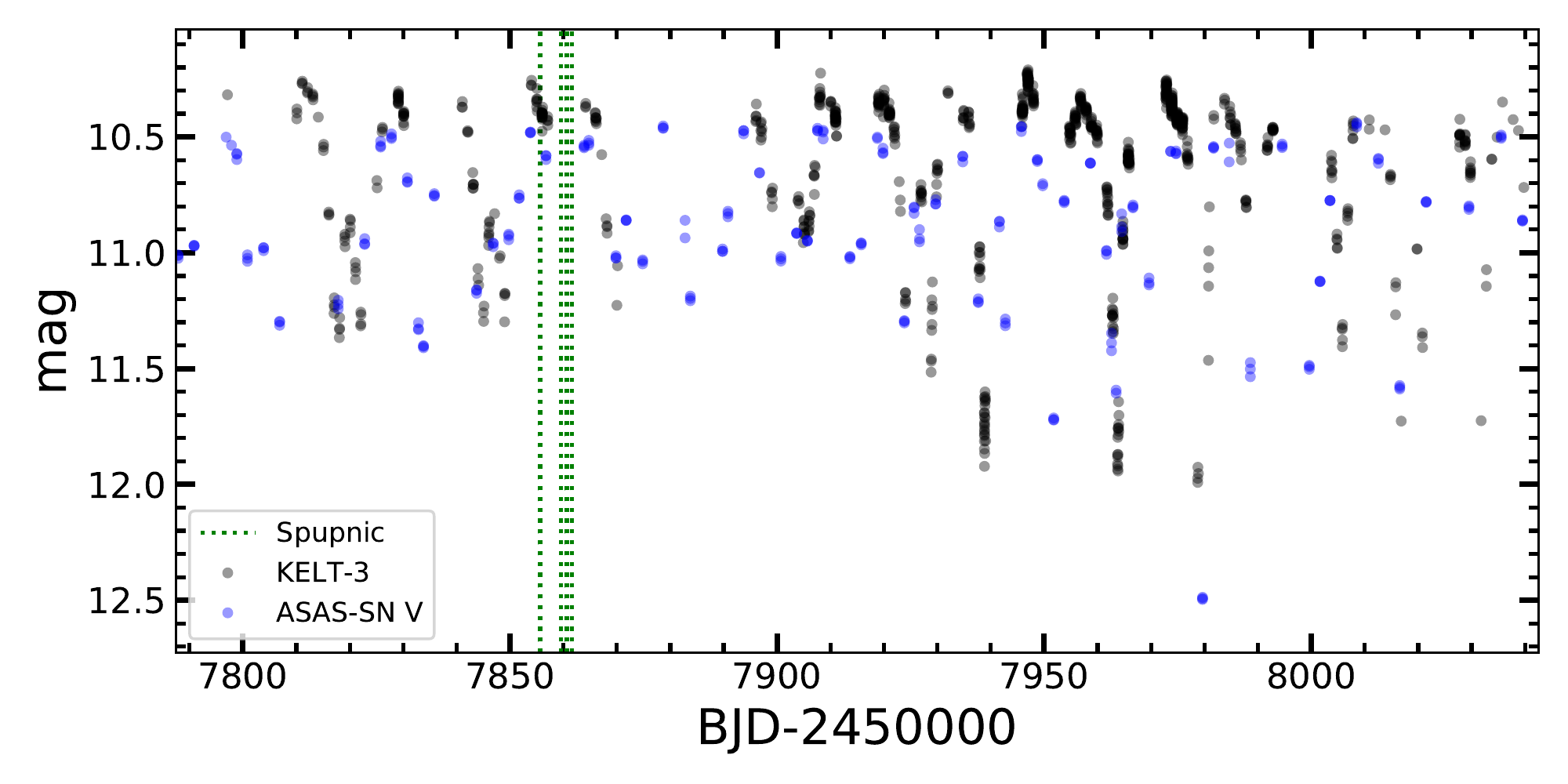}
    \includegraphics[width=0.8\textwidth]{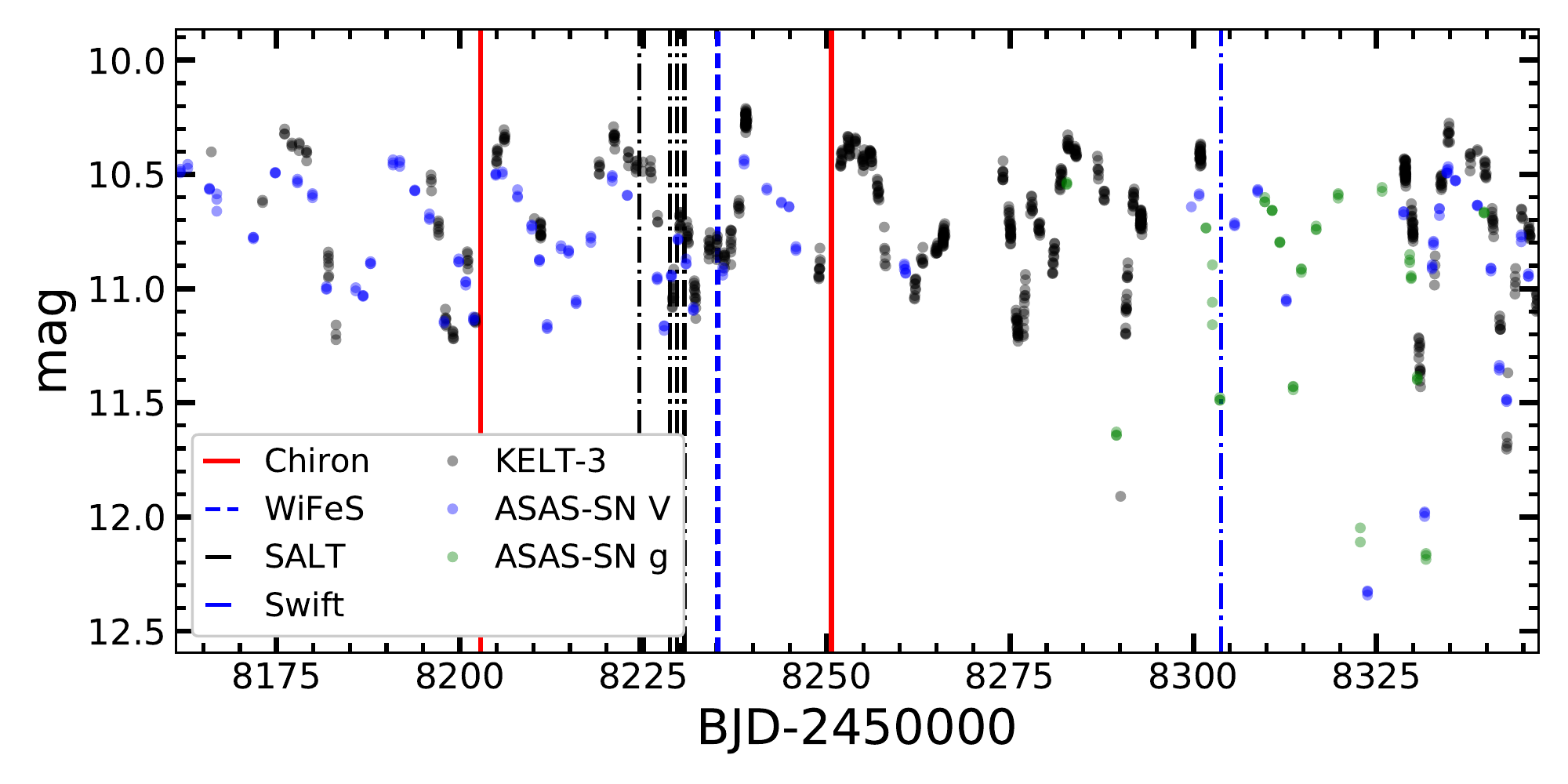}
    \includegraphics[width=0.8\textwidth]{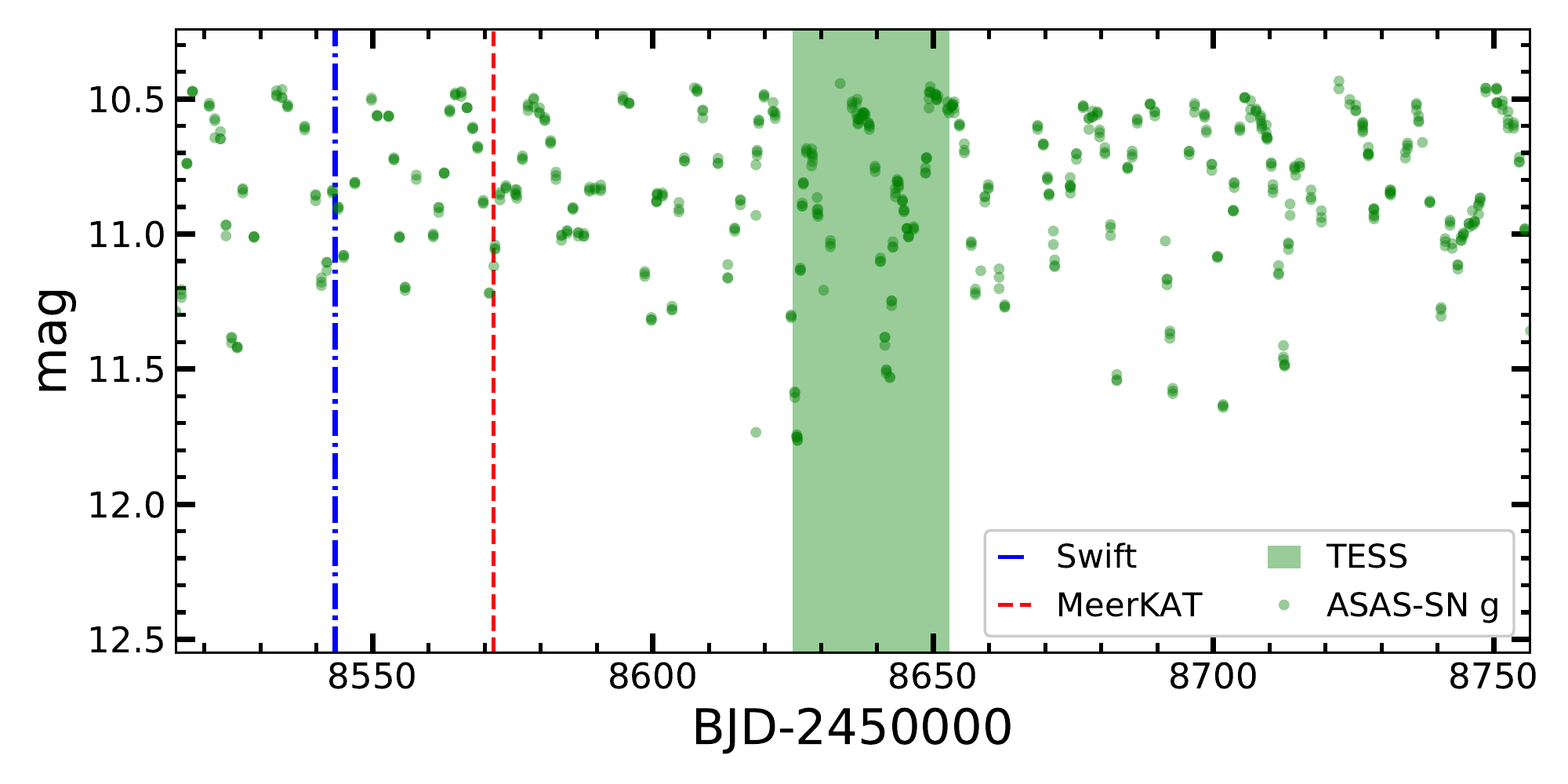}

    \caption{ \small Long term optical light curve of the seasons when the different datasets where gathered, these are marked with vertical lines. Note we only shift the baseline of the {\em KELT} data, by subtracting 3 mags, to match the median magnitude of the {\em ASAS-SN} data. We do not make any additional attempt to place the different time series photometry on the same absolute scale.}
    \label{fig:LCs}
\end{figure*}

\begin{figure*}
    \includegraphics[width=.99\textwidth]{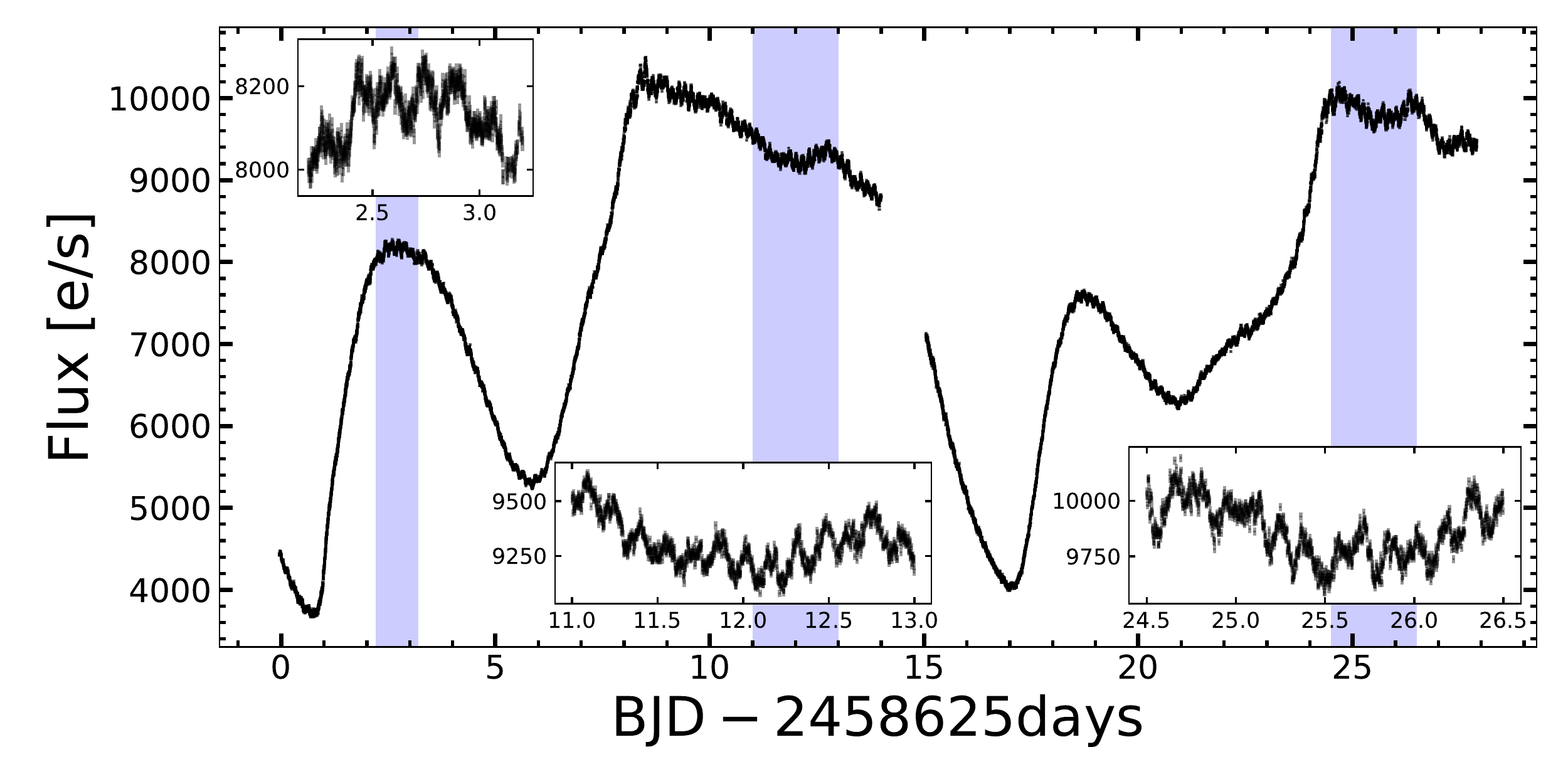}
    \caption{\small  The {\em TESS} lightcurve of {\em V341~Ara} during sector 12 illustrate the complexity of the super-orbital modulation; insets are zooms of the overlapping shaded region. Here, the orbital and superhump modulations are superimposed to the main signal with an amplitude roughly 3\% of the dominant one can be appreciated. The $\sim 1$ day data gap is due the perigee passage of the spacecraft while downloading the data.}
    \label{fig:tess}
\end{figure*}

Figure~\ref{fig:LCs}, provides an overview on each of the last 3 observing seasons, during which all of our other data sets were taken. In this figure, the {\em TESS} observing window is indicated by a vertical green band, and the dates on which spectroscopic, X-ray/UV ({\em Swift}) and radio ({\em MeerKAT}) observations were carried out, are marked by vertical lines. The 2-minute cadence, 27-day {\em TESS} light curve is shown in Figure~\ref{fig:tess}. %Finally, Figure~\ref{fig:dynamic_power} provides an overview of the entire 6-year photometric data obtained by {\em KELT} along with a dynamic power spectra. 

The first obvious and important point to note is that {\em V341~Ara}'s variability is dominated by large-amplitude -- $\delta_m \simeq 0.5 - 2.0$~mag -- super-orbital modulations (see also Figure~\ref{fig:dynamic_power}). The characteristic time-scale of these oscillations -- $\simeq$10-16~days -- is far longer than the orbital period of $P_{orb} \simeq 3.7$~hrs (B18; also see Section~\ref{sec:P_orb}). However, the amplitude and time-scale of these variations are comparable to those associated with the pulsations of Type II Cepheids, and specifically the W~Vir subclass 
\citep[e.g.][]{Berdnikov1998AcA....48..763B}%,soszinsky}
. This, in addition to the lack of statistically significant power around the orbital frequency explains the long mis-classification of {\em V341~Ara} as a Cepheid variable. 

\begin{figure*}%[H]
\begin{minipage}{\textwidth}

    %\center \includegraphics[width=.49\textwidth]{fig/KELT_power_all_yearly.pdf}\\
    
    \includegraphics[width=.49\textwidth]{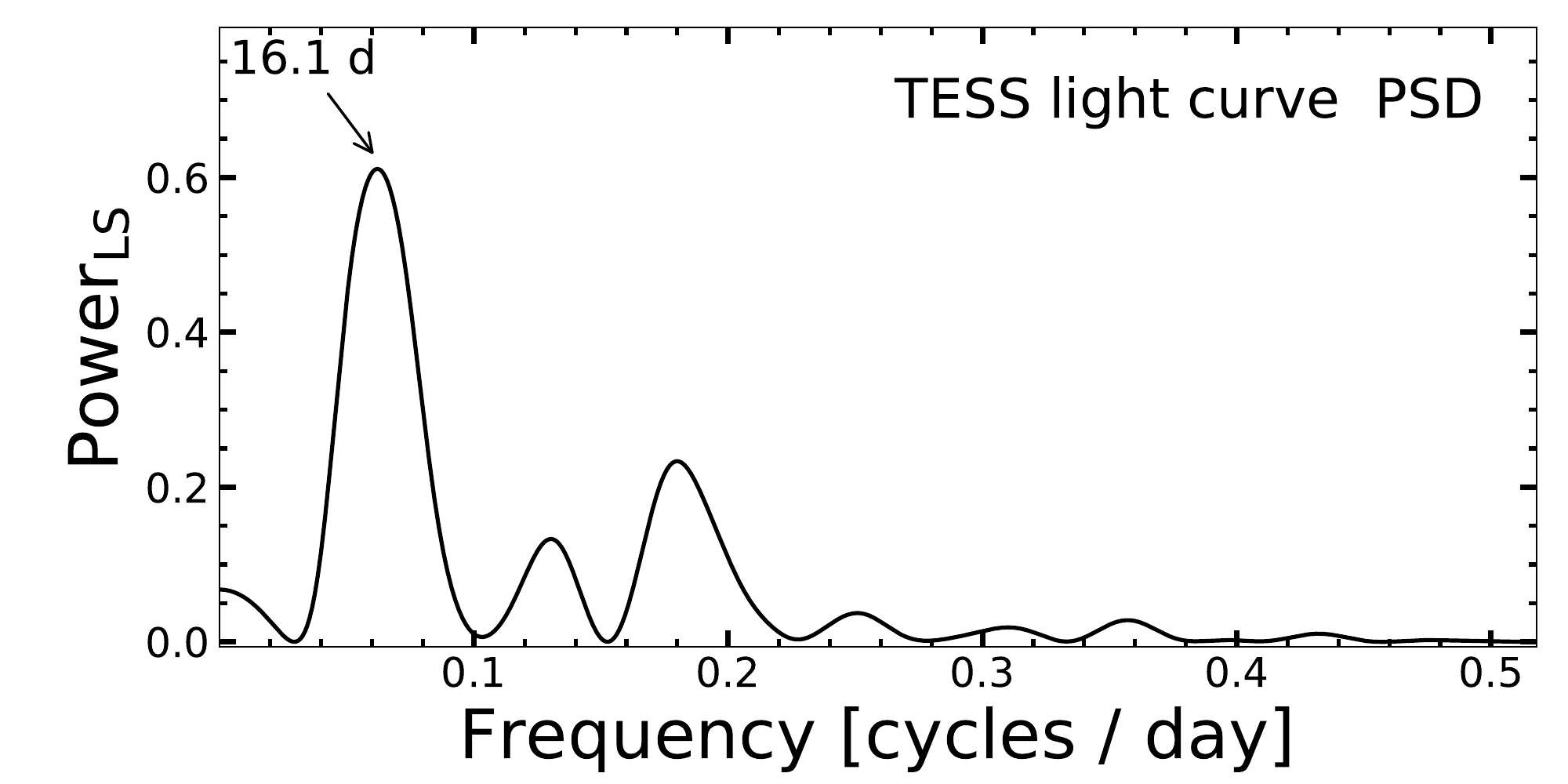}
    \includegraphics[width=.49\textwidth]{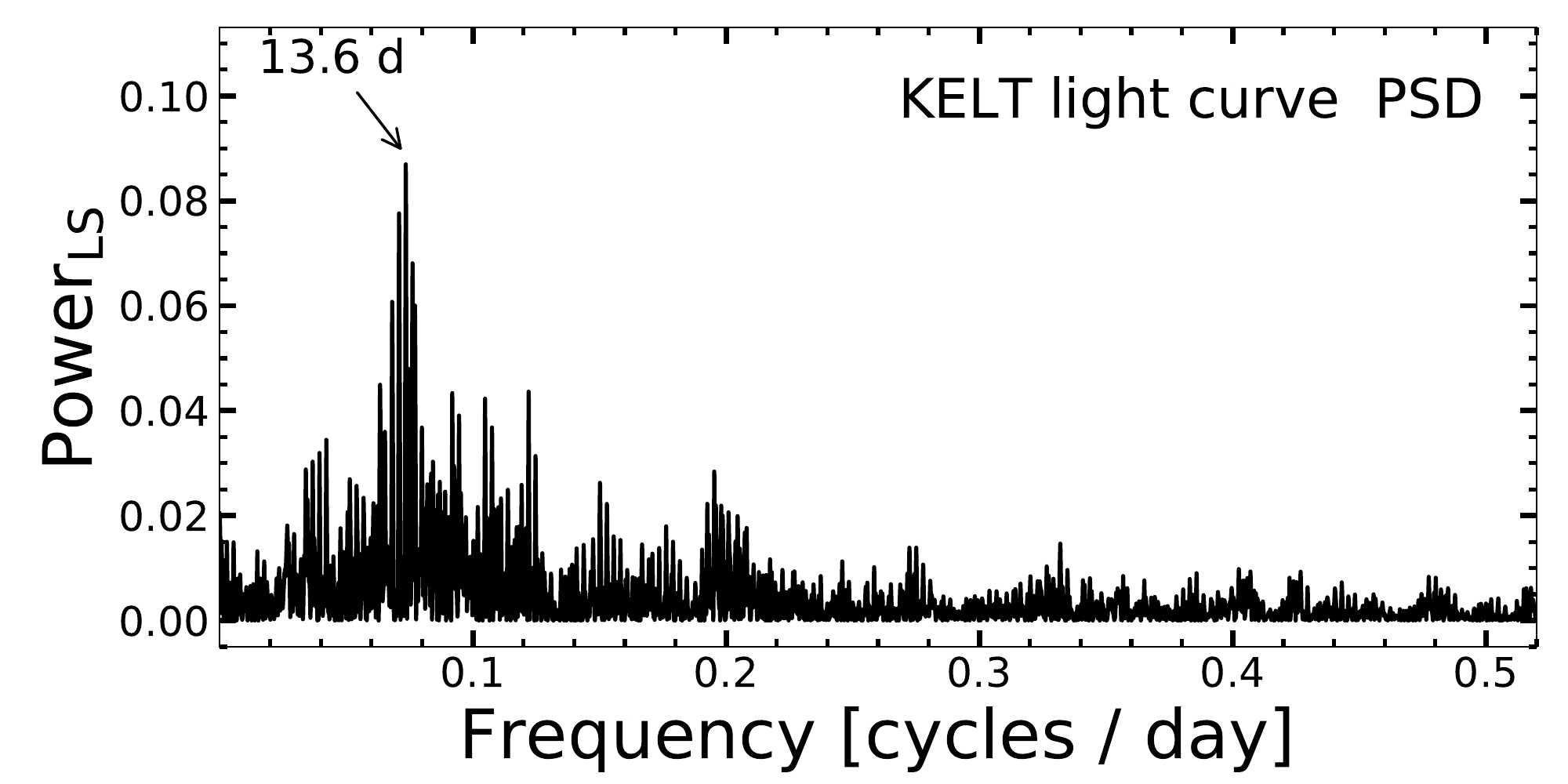}\\
    
    \includegraphics[width=.49\textwidth]{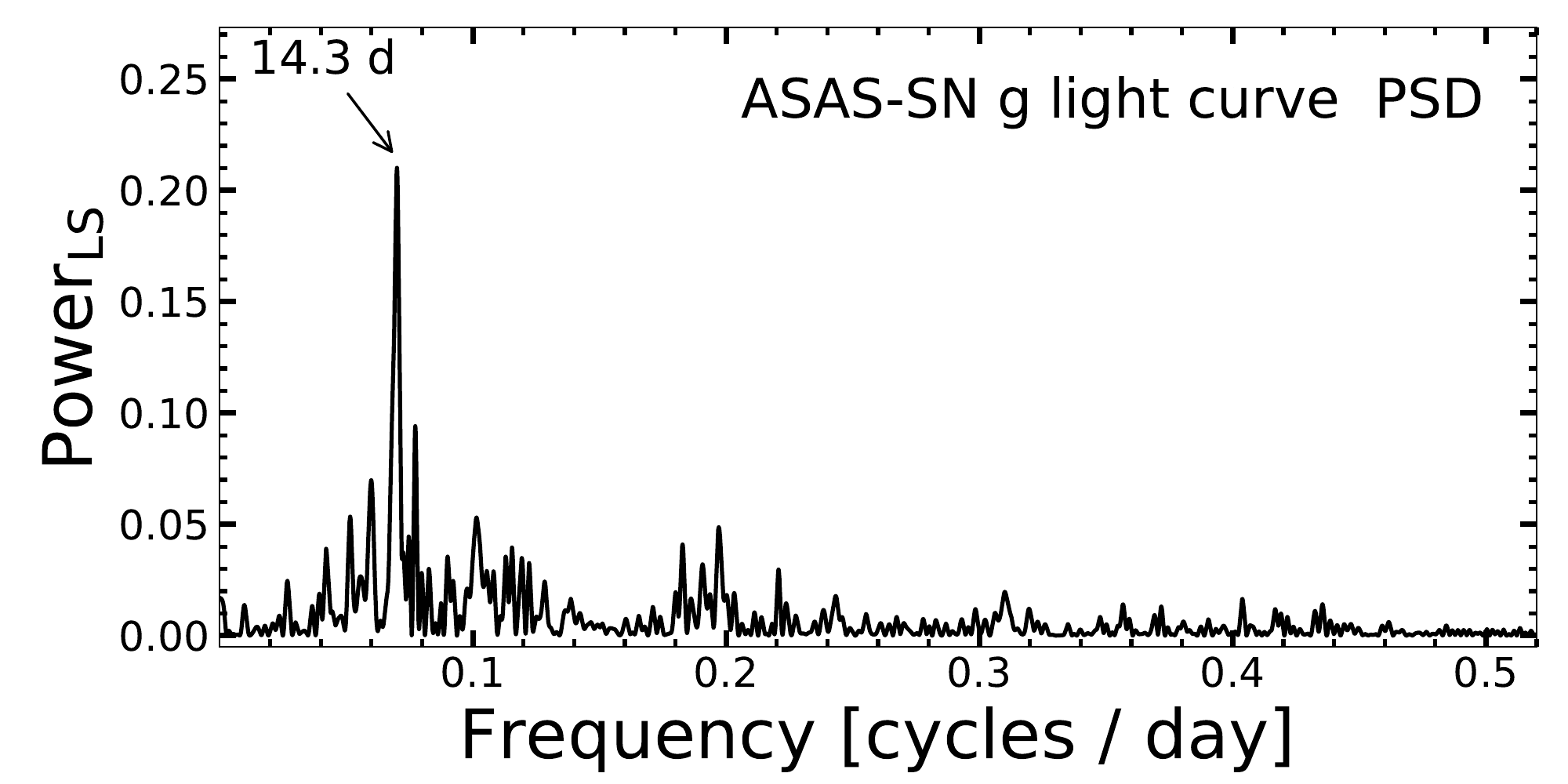}
    \includegraphics[width=.49\textwidth]{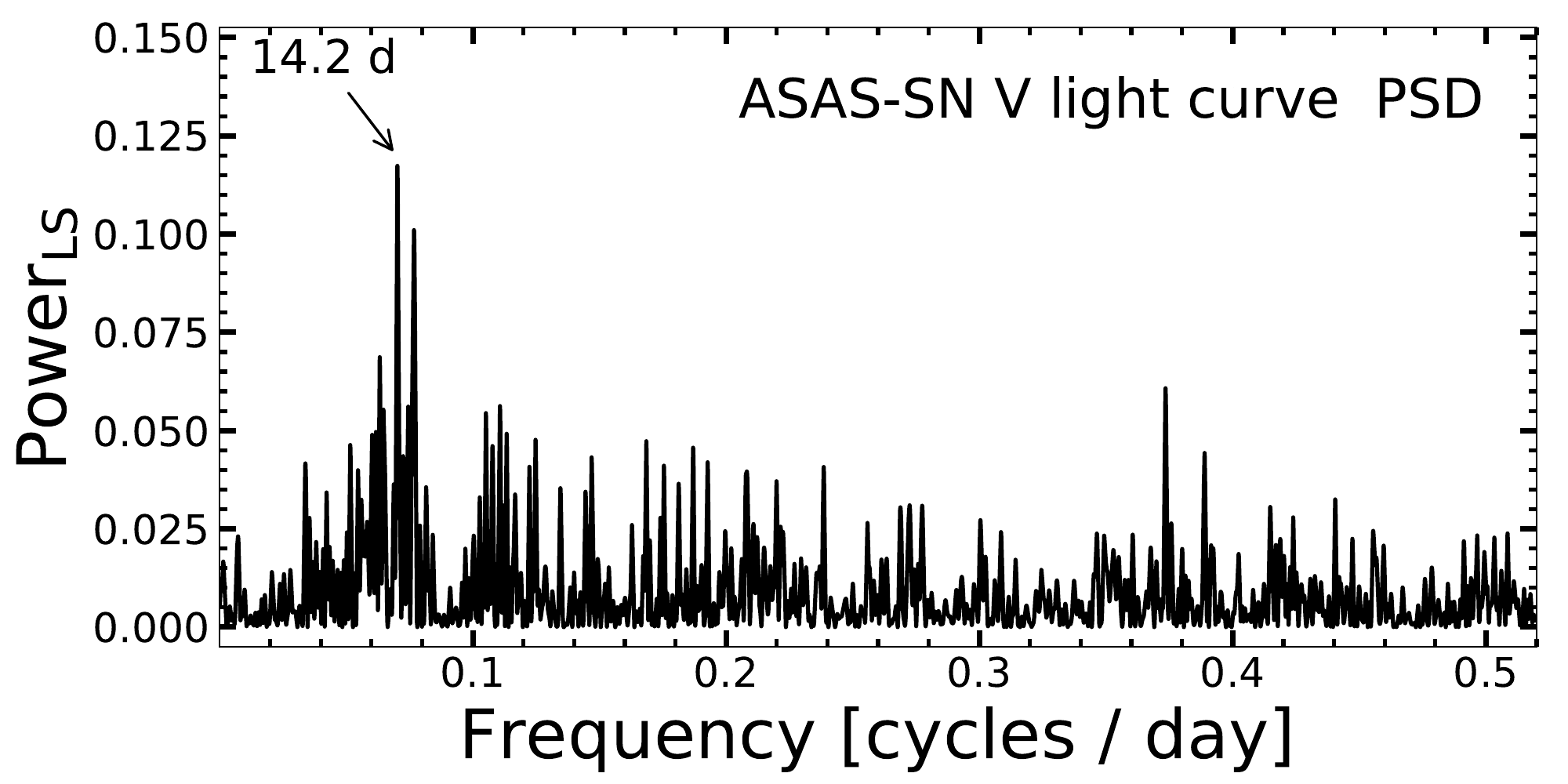}
    %\caption{\small Low frequency generalized Lomb-Scargle power spectrum of the different light curves illustrating the super-orbital period and its variability across sessions and data sets. The top panel shows seasonal power spectra from the {\em KELT} light curves, the peak frequency for each season is indicated with a vertical arrow. Remaining panels are the same, but for the whole of each data set. {\bf Note the figures are shown up to 0.55 cycles/day due to lack of statistically significant power around the orbital frequency being these frequencies dominated by strong aliases of the super-orbital modulation in the ground based observatories.}}
    \caption{\small Low frequency generalized Lomb-Scargle power spectrum of the different light curves illustrating the super-orbital period and its variability across sessions and data sets. Note the figures are shown up to 0.55 cycles/day due to lack of statistically significant power around the orbital frequency since these frequencies are dominated by strong aliases of the super-orbital modulation in the ground-based observatories.}
    \label{fig:power}
\end{minipage}
\end{figure*}

\begin{figure}
\centering
\hspace*{-0.5cm} \includegraphics[width=1.05\textwidth]{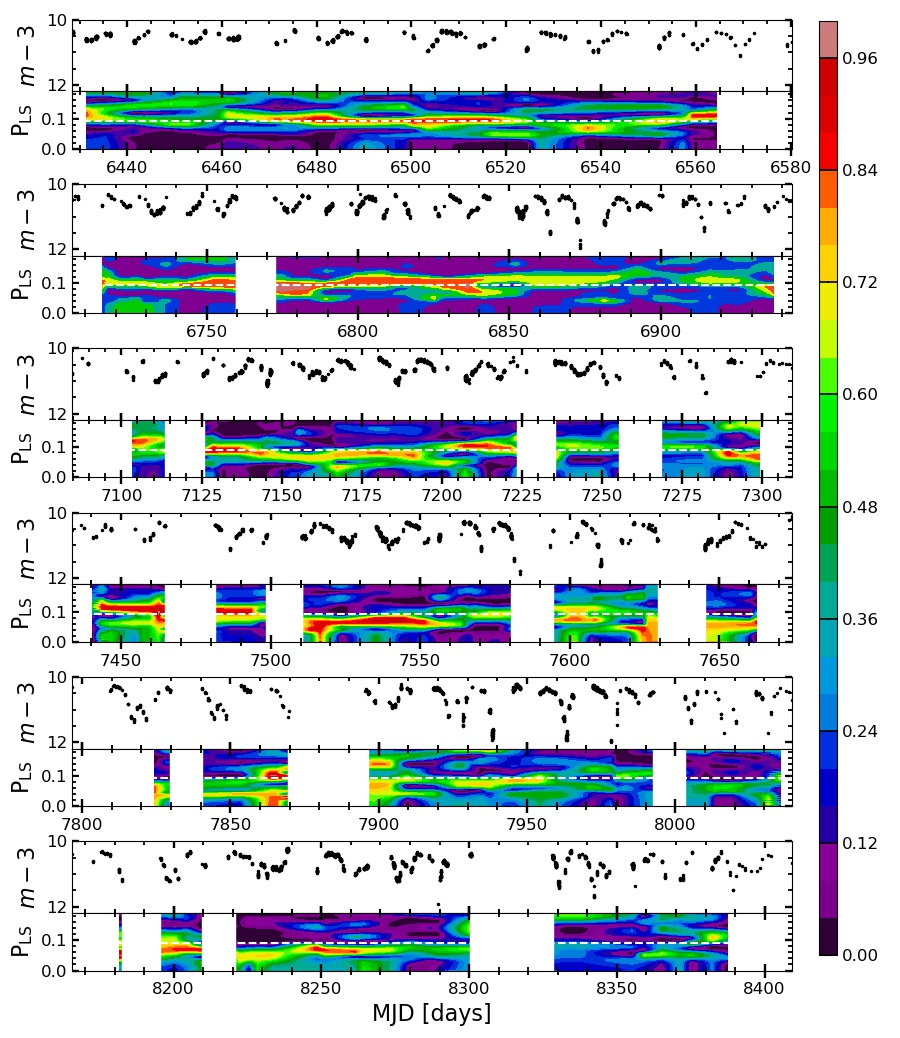}
%\hspace*{-1cm} \includegraphics[width=1.1\textwidth]{fig/Kelt_lc_power_2.png}
\caption[]{\footnotesize
{\em KELT} light curve and dynamic power spectrum. Top panels: {\em KELT} instrumental magnitude - 3 vs time. Bottom panels: dynamic Lomb-Scargle power spectrum, abscissas axis represent the time in days and the frequency in cycles per day in the ordinates, the Lomb-Scargle power is represented in the colour scale. Each bin in the periodogram includes the contiguous $\pm$15 days with a minimum of 20 data points. The horizontal white dotted line represent the frequency with highest power in the periodogram of the whole set.}
\label{fig:dynamic_power}
\end{figure}

The prevalence and quasi-periodic nature of these slow variations are apparent in Figures~\ref{fig:power} and \ref{fig:dynamic_power}, which shows generalized Lomb-Scargle \citep{Lomb1976,Scargle1982,Zechmeister_LS2009,VanderPlas2018ApJS..236...16V} periodograms for each of the light curves. Every data segment produces a strong power excess in this frequency range, but the exact location of the peak frequency changes across data sets and observing seasons. The amplitude of these variations is also highly variable. For example, Figure~\ref{fig:dynamic_power} shows that the peak-to-peak variability level in the {\em KELT} Year 1 data is $\lesssim$1~mag, but in a similar-sized segment of the {\em KELT} Year 5 data, it reaches $>1.5$~mag.

To illustrate these features more clearly, Figure~\ref{fig:dynamic_power} shows a dynamic power spectrum we have constructed for the {\em KELT} data set, along with the season-by-season light curve. In calculating this power spectrum, we require each time bin to span at least $\pm 15$ days and include a minimum of 20 data points; note that this will tend to smooth any sudden change in the periodic signal. 

\begin{figure}
        \centering
        \includegraphics[width=0.45\textwidth]{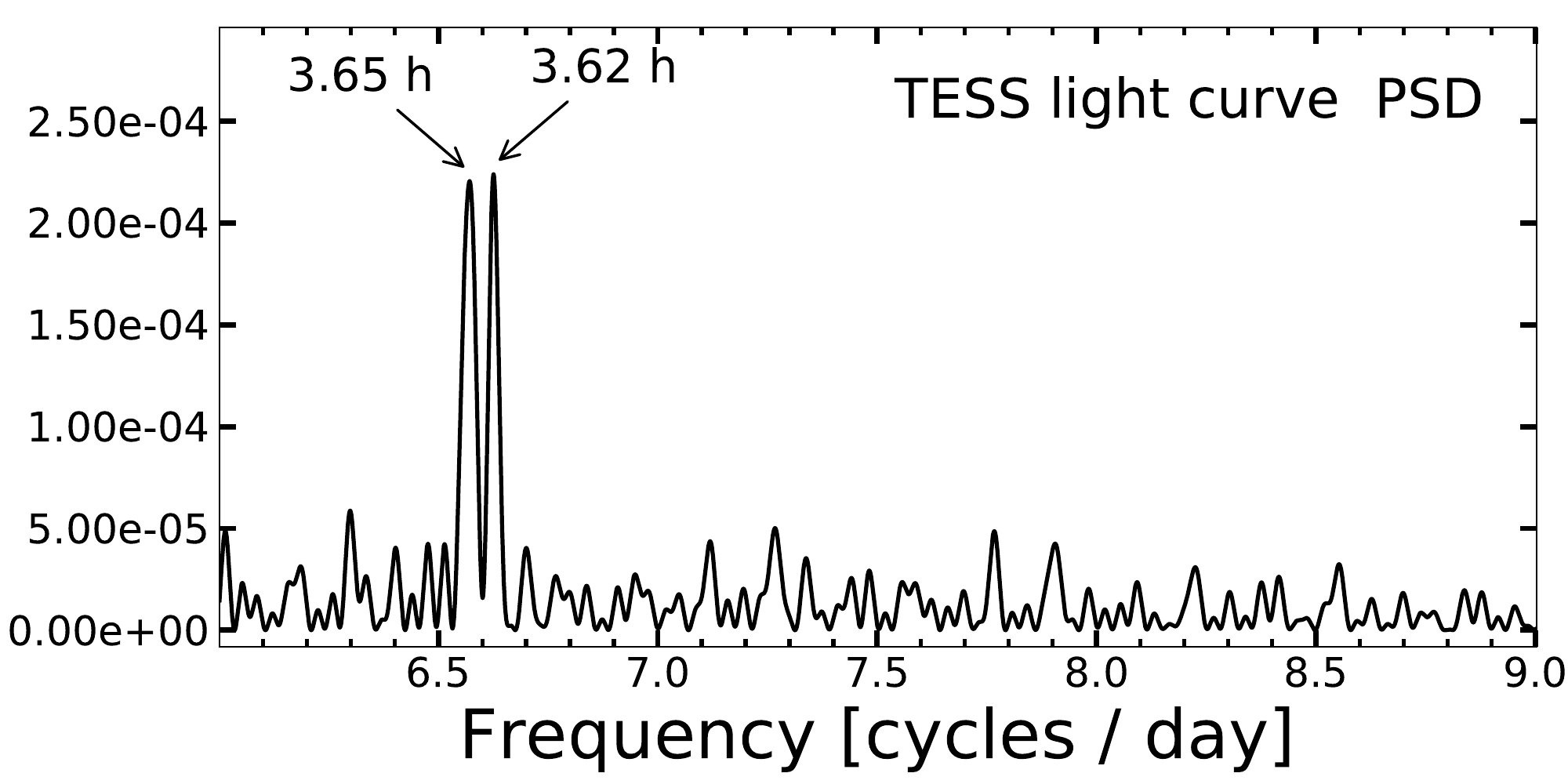}
        \caption{High frequency generalized Lomb-Scargle power spectrum of {\em {\em TESS}} light curve. The $3.65$ peak is consistent with the orbital period, while the $3.62$ h is consistent with the beat frequency of the  super-orbital modulation with the orbital period.}
        \label{fig:TESS power high}
    \end{figure}

By contrast, only the high-cadence, high-S/N {\em TESS} data set reveals a (weak) orbital signal (see insets in Figure~\ref{fig:tess} and  Figure~\ref{fig:TESS power high}). The characteristic amplitude of this signal is just $\simeq 1$\%, which suggests a relatively low (face-on) orbital inclination for {\em V341~Ara}, in agreement with the assumption in previous section. 

Perhaps more importantly, however, the power spectrum of the {\em TESS} data actually exhibits {\em two} peaks near the orbital frequency. One of these --- $f_{orb,TESS} = 6.569 \pm 0.007\, \mathrm{c/d}$), corresponding to $P_{orb} = 3.654 \pm 0.004$~hrs --- is consistent with the orbital period obtained from our multi-year spectroscopic data (see Section~\ref{sec:P_orb}). The other --- $f_{SH-} = 6.628 \pm 0.006\, \mathrm{c/d}$, corresponding to $P_{SH-} = 3.621 \pm 0.003$~hrs --- is roughly 1\% {\em faster} than the orbital signal. This signal is similar to the so-called ``negative superhumps'' seen in the light curves of several other nova-like variables \cite[e.g.][]{Patterson_SHs_2005PASP..117.1204P}. In these systems, the negative superhump signal is thought to represent a beat between the binary orbit and the much slower, retrograde, nodal precession of a tilted accretion disc. 

If the same interpretation holds for {\em V341~Ara}, it is tempting to associate the dominant large-amplitude $10 - 16$~day signals in the photometric data with the disc precession period itself. Is this viable? The frequency difference between $f_{orb}$ and $f_{SH-}$ in the {\em TESS} data is $\Delta f = 0.06 \pm 0.006\,\mathrm{c/d}$, which corresponds to an implied precession period of $P_{prec} = 16.4 \pm 1.6$~d. This is completely in line with the measured period of the slow photometric variations during the {\em TESS} observations, which was $P_{slow} \simeq 16.1$~d (see Figure~\ref{fig:power}). 

%%%%%%%%%%%%%%%%%%%%%%%%%%%%%%%%%%%%%%

 %%%%%%%%%%%%%%%%%%%%%%%%%%%%%%%%%%%%%%%%%%%%%%%%%%%%%%%%%%%%%%%%%%%%% 
%%%%%%%%%%%%%%%%%%%%%%%%%%%%%%%%%%%%%%%%%%%%%%%%%%%%%%%%%%%%%%%%%%%%%

\subsection{Astrometry: distance and membership of the thin disc population}\label{sec:gaia}

   {\em V341~Ara} is included in Gaia DR2 \citep[][]{GAIADR22018AA...616A...1G}. We therefore estimate its distance from the Gaia parallax, $\pi = 6.40\ \pm 0.08\, {\mathrm mas}$, using the Bayesian approach described in  \citet{Luri+2018} to account for the covariance of the distance and proper motion. This yields a distance of $156 \pm 2$ parsecs (pc), for a prior with a length scale of $1.5$ kpc. However, given the precision of the parallax measurement, the distance estimate is highly insensitive to the prior. Based on this estimate, {\em V341~Ara} is the 3$\mathrm{^{rd}}$ closest non-magnetic NL.
    
    The proper motion of {\em V341~Ara} provided by Gaia is $\mu_{\alpha} \cos{\delta} = -48.32 \pm 0.07\, \mathrm{mas~yr^{-1}}$ and $\mu_{\delta} = -84.91 \pm 0.08\, \mathrm{mas~yr^{-1}}$. Given its distance, this corresponds to a transverse velocity of $72 \pm 1~\mathrm{km~s^{-1}}$ 
    
    We also checked for possible distant companions to the {\em V341~Ara} binary system by searching for Gaia sources within 1$^\circ$ of its position that also have similar proper motions and distances. No such sources were found. 

    In order to identify the Galactic population to which {\em V341~Ara} belongs (thin disc, thick disc or halo), we carried out Galactic orbit integrations with {\sc galpy}, assuming a standard Milky Way potential \citep[MWPotential2014,][]{Galpy2015ApJS..216...29B}. The starting point for these integrations is the system's current position in the full six-dimensional phase space, which is defined by proper motion (2 components), position (2 components), distance (1 component) and systemic radial velocity (1 component). All but the last of these are provided by Gaia. For the systemic radial velocity, we adopt the value obtain by our spectroscopic analysis in Section~\ref{sec:rv}, $v_{sys} = 42 \pm 2\, \mathrm{km~s^{-1}}$. 
    
    The Galactic orbit implied by these parameters is characterized by a pericenter distance of 5.32~kpc and an apocenter radius of 8.46 kpc, which corresponds to an eccentricity of $e = 0.23$. The maximum height of the orbit above the mid-plane of the Galactic disc, $h_{max} = 57.3\, {\mathrm pc}$, clearly suggests membership of the thin disc population. This is in line with expectations for long-period CVs from \citet{Pretorius+2007MNRAS.382.1279P}. 
    
%\hspace{-6mm}
\subsection{Optical spectroscopy: orbital ephemeris and mass ratio} \label{sec:spectroscopy}

\subsubsection{The mean spectrum}

\begin{figure*}
    \includegraphics[width=0.9\textwidth]{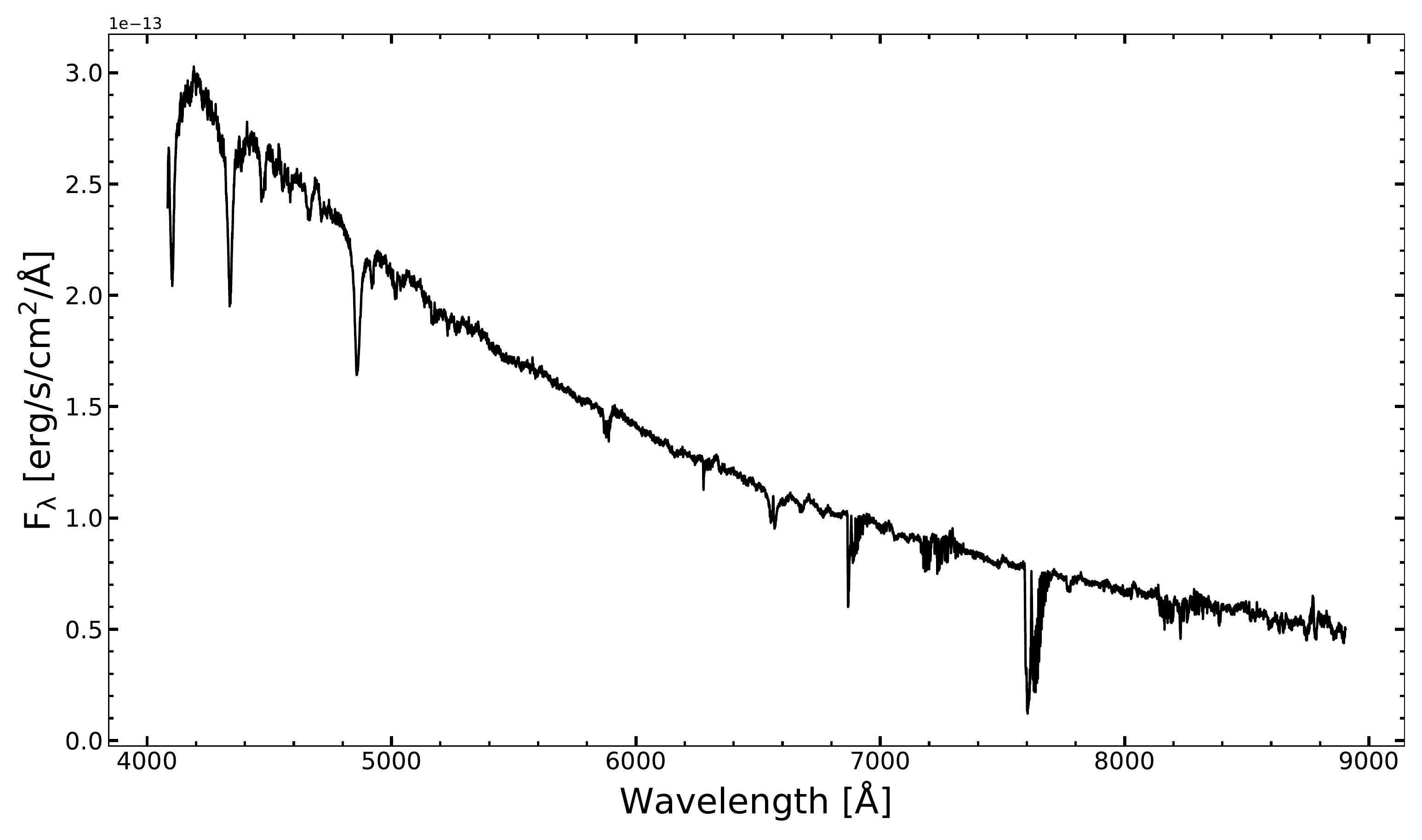}
    \caption{\small Median spectrum of all the observations taken with {\em Chiron spectrograph}. In contrast to the {\em WiFeS} campaign (see fig. \ref{fig:V341Ara line fit}), the combined spectrum is dominated by the high flux state in where the absorption lines from the accretion disc's atmosphere strongly dominates the observed spectrum. Note strong features at wavelengths above $\lambda\simeq 6800 \mathrm{\AA}$ are due to telluric absorption.
    %CK: TYPO IN XLABEL! ALSO, CAN WE MAKE THE LINE THINNER? 
}
    \label{fig:meanspec}
\end{figure*}

Figure~\ref{fig:meanspec} shows the mean optical spectrum of {\em V341~Ara}, as obtained by averaging all of the observations we obtained with {\em Chiron} at the CTIO. Prior to combining the individual spectra, each was shifted to the rest frame of the secondary, based on the radial velocity curve derived in Section~\ref{sec:rv} below.

As expected from the overall SED (Figure~\ref{fig:full SED}), the optical spectrum is very blue. The continuum shape is reasonably approximated by $F_{\lambda} \simeq \lambda^{-7/3}$, the spectrum expected for the central part (between the Wien and Rayleigh-Jeans tails) of a Shakura-Sunyaev multi-temperature black-body disc.
 
The other key aspect of Figure~\ref{fig:meanspec} is that it shows an {\em absorption} line spectrum. This, again, suggests a face-on for {\em V341~Ara}. Overall, the optical spectrum is reminiscent of that produced by RW~Sex, another bright, nearby, long-period, low-inclination non-magnetic NL \citep{Beuermann+1992AA...256..433B}.

%%%%%%%%%%%%%%%%%%%%%%%%%%%%%%%%%%%%%%%%%%%%%%%%%%%%%%%%%%%%%%%%%%%

\subsubsection{Radial velocity measurements} %\label{sec:rv}

\begin{figure*}
    \includegraphics[width=.9\textwidth]{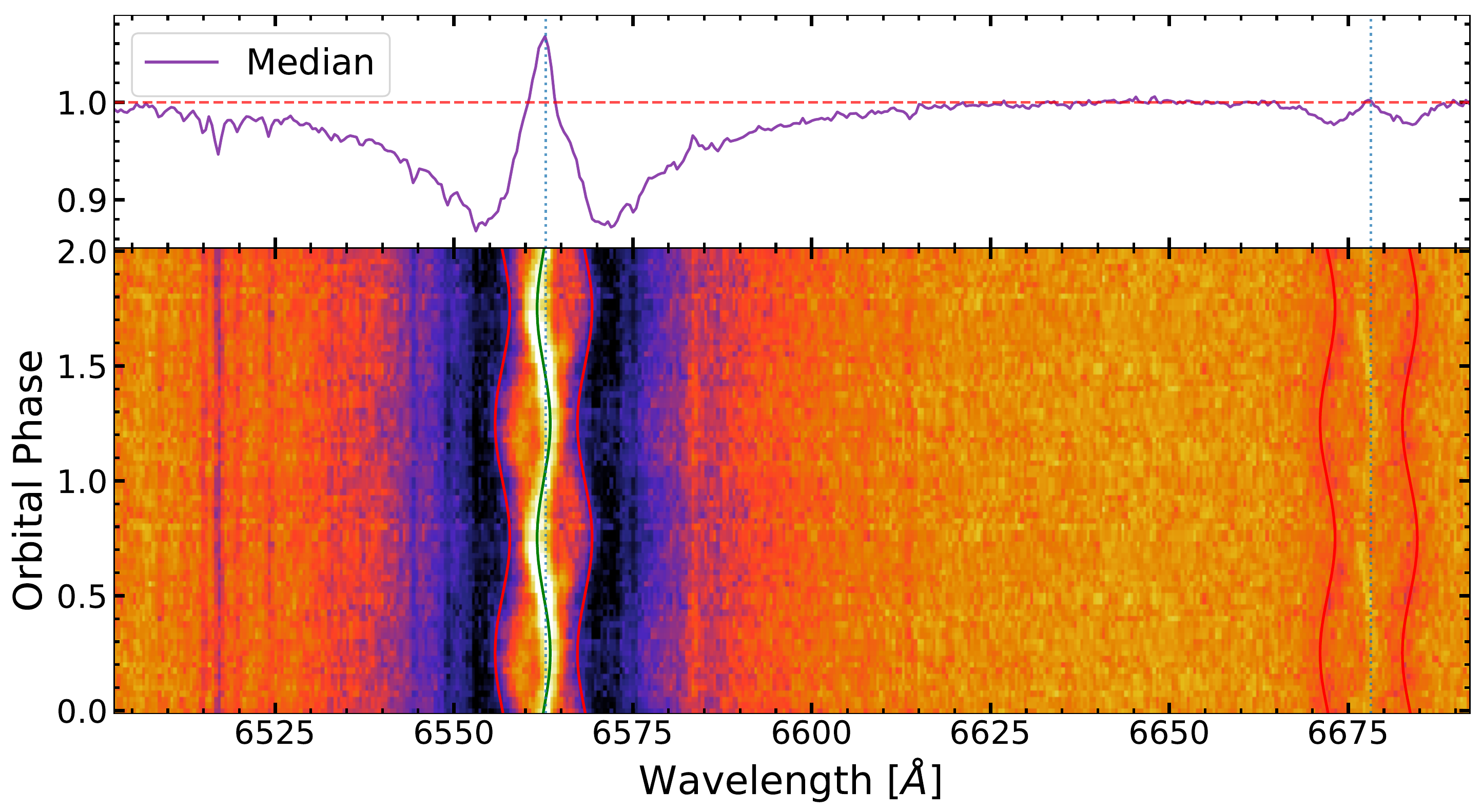}
    \caption{\small Phase folded trailed spectrum of the {\em WiFeS} time-resolved spectroscopy (bottom), normalized median in arbitrary units (top). Rest position of H$\alpha$ $\lambda 6562 $\AA\ and He {\sc i} $\lambda 6678 $\AA\ are marked with vertical dotted lines, sine-waves of the primary (red) and secondary (green) components from parameters quoted in Table \ref{tab:system param} overlaid to guide the eye. The He {\sc i} component illustrate better the ``S-wave'' of the absorption component. For clarity, the primary (absorption) component is plotted twice with a shift in order to lay on top of the absorption wings. 
}
    \label{fig:trailed}
\end{figure*}

Figure~\ref{fig:trailed} shows a trail of the H$\alpha$ and He {\sc i} $\lambda 6678 $\AA\ region in our 4.5~hrs of continuous observations with {\em WiFeS}. For clarity, the data are already shown here folded on the orbital period of $P_{orb} \simeq 3.65$~hrs. 

It is immediately obvious from Figure~\ref{fig:trailed} that {\em V341~Ara} is a double-lined spectroscopic binary. In both transitions, the central emission peaks and broad absorption wings display clear, anti-phased S-waves. The obvious interpretation is that the broad absorption lines are produced in the atmosphere of the optically thick accretion disc, whereas the narrow emission line cores are formed in the irradiated face of the secondary star. The same phenomenology is seen in other accreting compact binaries, including bright NLs, such as IX~Vel \citep{Beuermann+1990AA...230..326B} and RW~Sex \citep{Beuermann+1992AA...256..433B}.

\begin{figure}
    \hspace*{-2mm}\includegraphics[width=.49\textwidth]{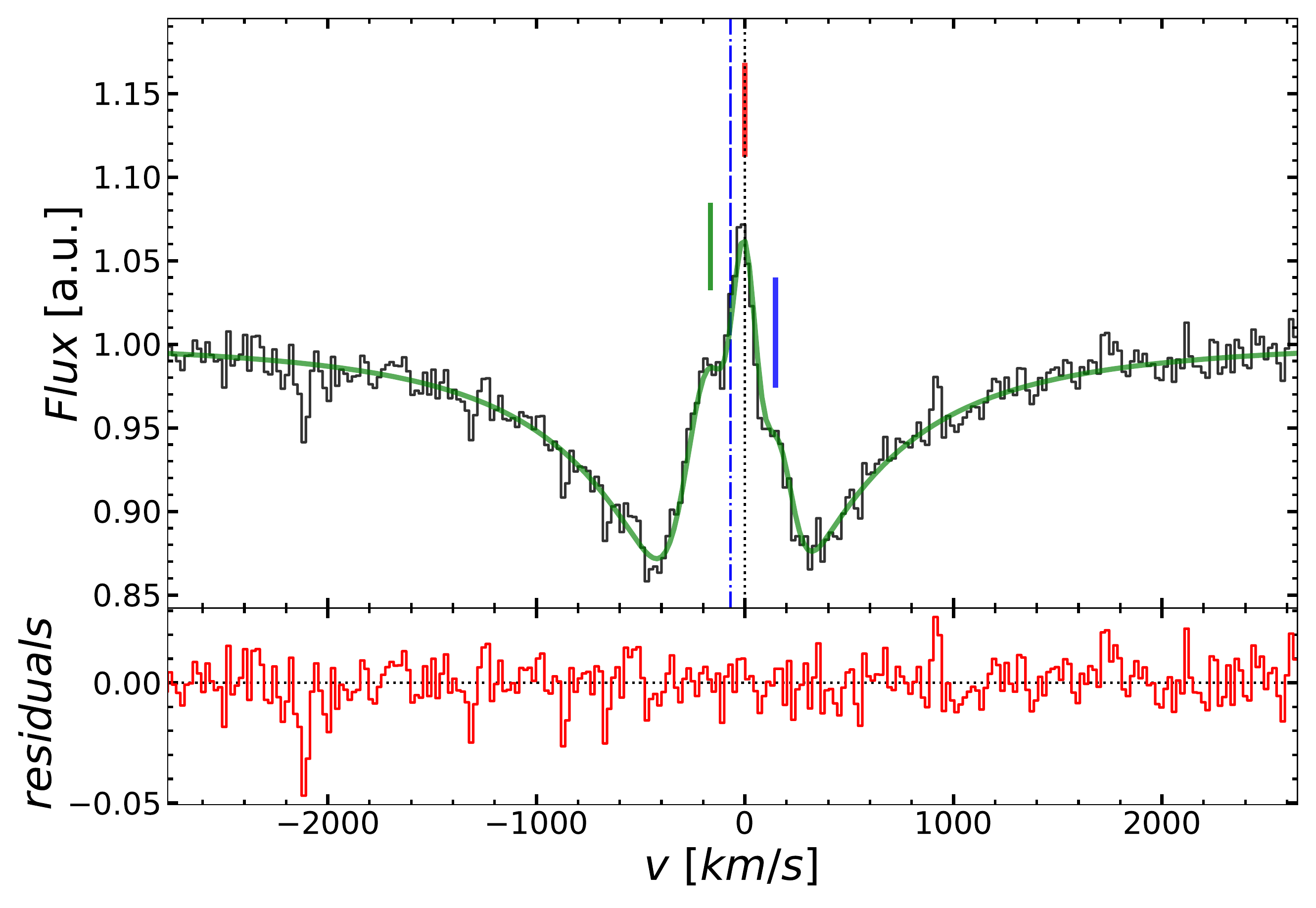}
    \caption{\small  Zoom in to H$\alpha$ emission component illustrating the evidence of simultaneous three emission emission components filling the broad absorption. Black histogram is the observed spectrum while green line is the best fit to a composite model with one absorption plus three emission components. Rest velocity of the transition is marked with a vertical black dotted line, the blue dash dotted line indicate the estimated position for the absorption component, while the colour vertical bars are centered in the position of each component in the model. Residuals of the fits are shown in the bottom panel. %The spectrum correspond to an orbital phase of 0.07.  
}
    \label{fig:V341Ara line fit}
\end{figure}

Closer inspection of Figure~\ref{fig:trailed} suggests the presence of additional emission line components. For example, a faint S-wave that is roughly anti-phased to that associated with the main emission core, but shares a similar velocity amplitude, can be seen near the center of the H$\alpha$ line. As illustrated in Figure~\ref{fig:V341Ara line fit}, this and at least one other component can be more clearly identified in some of our high resolution, high signal-to-noise spectra. These components are probably associated with asymmetric line-emitting structures in the accretion disc, such as the bright spot at the point where the stream impacts on the outer disc.   

If not accounted for, the presence of these additional components could bias radial velocity measurements of the two main components, which we intend to use as tracers of the primary and the secondary. This concern applies particularly to the disc-formed, ``double-dipped'', broad absorption line, which is harder to centroid than the narrow central emission line core. In order to avoid such biases, we therefore fit each spectrum in each of our data sets ({\em Chiron}, {\em SALT}, {\em SpUpNIC}, {\em WiFeS}) with a combination of a single Lorentzian absorption component and three Gaussian emission components. An initial fit in which all parameters were allowed to vary freely revealed that the components could be split into three distinct groups with different widths: ${\rm FWHM}<150~{\rm km
~s^{-1}}$, $150~{\rm km ~s^{-1}} < {\rm FWHM} < 220~{\rm km
~s^{-1}}$ and ${\rm FWHM}>250~{\rm km
~s^{-1}}$. Visual inspection of the data also provided limits on the centroid of the main emission component. This information was then used to set the initial parameter guesses and constrain a refined fit in which the posterior probability distributions for all model parameters were obtained via Markov Chain Monte Carlo with {\sc emcee}\footnote{http://dfm.io/emcee/current/} \citep{emcee2013PASP..125..306F}. Simple non-informative uniform priors were adopted for all fit parameters. The uncertainties on the fit parameters were then estimated from the 16\% and 84\% quantiles of the sampled posterior. These posteriors exhibit roughly normal distribution for the parameters of interest.
%\enlargethispage{\baselineskip}
%\pagebreak
%..............................

%%%%%%%%%%%%%%%%%%%%%%%%%%%%%%%%%%%%%%%%%%%%%%%%%%%%%%%%%%%%%%%%%%%

\onecolumn

%%%%%%%%%%%%%%%%%%%%%%%%%%%%%%%%%%%%%%

\twocolumn
%%%%%%%%%%%%%%%%%%%%%%%%%%%%%%%%%%%%%%

%%%%%%%%%%%%%%%%%%%%%%%%%%%%%%%%%%%%%%
%%%%%%%%%%%%%%%%%%%%%%%%%%%%%%%%%%%%%%

\subsubsection{Orbital period and ephemeris} \label{sec:P_orb}

In order to determine an accurate orbital period and ephemeris for {\em V341~Ara}, we focused on the main emission component of the H$\alpha$ line. This allowed us to combine our radial velocity estimates for this component with those obtained by B18, which greatly extends the time base of the analysis, and hence increases the precision of the result. All observation times were transformed to a barycentric reference frame using the algorithm described by \citet{Eastman_time_acccuracy2010PASP..122..935E}.
    
Before combining all of these measurements, we first estimated periods separately for the 2006, 2017 and 2018 data sets. We found that all of these agreed to within their respective errors, and also with the period estimate for the 2006 data previously reported by B18. 

The combined data from 2017 and 2018 provide sufficient precision to phase these observations without cycle count ambiguities across the $\simeq$27000 orbit gap to the 2006 CTIO/RC data set. Applying the same Lomb-Scargle periodogram and bootstrapping techniques described in Section~\ref{sec:LCs} to the combined radial velocity data allowed us to determine the orbital period to millisecond accuracy, $P_{orb}=3.6529454^{+0.0000016}_{-0.0000013}$ hrs. The corresponding full orbital ephemeris is
\begin{equation}
T_0(BJD) = 2457098.01142(39) + 0.15220606(6) E%(^{+6.6}_{-5.4}) E,
%NOTE: I strongly suggest you pick a time-constant in the middle of the interval here, so that %we don't have nasty correlations between the errors on the constant and the error on the %period.
\end{equation}
%where $T_0$ refers to the red-to-blue crossing of the broad absorption line. This should correspond to inferior conjunction of the secondary, provided the line faithfully traces the orbital motion of the accretion disc, and hence of the primary.
where $T_0$ refers to the blue-to-red crossing of the narrow emission line. This should correspond to inferior conjunction of the secondary, provided the line faithfully traces the orbital motion of the companion star.

%%%%%%%%%%%%%%%%%%%%%%%%%%%%%%%%%%%%%%%%%%%%%%%%%%%%%%%%%%%%%%%%%%%%%% 
%\begin{figure} 
%    \includegraphics[width=.5\textwidth]{fig/galactic_orbit.pdf}
%    %\includegraphics[width=.48\textwidth]{rv_folded.png}
%    \caption{\small  Vertical vs radial projection of {\em V341~Ara} orbit around The Galaxy integrated for 1 Gyr in a Milky Way potential produced with ``\emph{galpy}''. Horizontal dash-dotted line correspond to the scale height for long period CVs from \citet{Pretorius+2007MNRAS.382.1279P}, while the dotted line represent galactic thin disc population scale height from \citet{GilmoreReid1983MNRAS.202.1025G}.
%%CK: Could you show this with fewer orbits, maybe just something like 1 Gyr? Hard to see what's happening...
%}
%    \label{fig:galpy}%\label{fig:V341Ara galpy}
%\end{figure}
%%%%%%%%%%%%%%%%%%%%%%%%%%%%%%%%%%%%%%%%%%%%%%%%%%%%%%%%%%%%%%%%%%%%%%

\subsubsection{K-velocities and system parameters} \label{sec:rv}

%%%%%%%%%%%%%%%%%%%%%%%%%%%%%%%%%%%%%%%%%%%%%%%%%%%%%%%%%%%%%%%%%%%%% 

%\onecolumn
%\twocolumn
%%%%%%%%%%%%%%%%%%%%%%%%%%%%%%%%%%%%%%%%%%%%%%%%%%%%%%%%%%%%%%%%%%%%%

%%%%%%%%%%%%%%%%%%%%%%%%%%%%%%%%%%%%%%%%%%%%%%%%%%%%%%%%%%%%%%%%%%%%% 

%\twocolumn
%%%%%%%%%%%%%%%%%%%%%%%%%%%%%%%%%%%%%%%%%%%%%%%%%%%%%%%%%%%%%%%%%%%%%

In order to determine the K-velocities of the primary and secondary, we fit the radial velocity measurements of the absorption and main emission components with sinusoids, 
\begin{equation}
    v_{abs} = \gamma_1 + K_{1} \sin{(2\pi / P_{orb} + \phi_1)}\ , 
\end{equation}
and 
\begin{equation}
    v_{em} = \gamma_2 + K_{2,obs} \sin{(2\pi / P_{orb} + \phi_2)}.
\end{equation}
All parameters were allowed to vary in the fit, except the orbital period, which was sampled from the probability distribution obtained above. The resulting parameters and uncertainties are listed in Table~\ref{tab:system param}.

The measured phase offset between the absorption and emission line radial velocity curves is $180.0^\circ \pm 0.3^\circ$. This gives us confidence that these components are tracing the motions of the primary and secondary, respectively \cite[c.f.][and references therein]{2007ApJ...670..727C}.

If the absorption lines are formed in the atmosphere of an optically thick, symmetric accretion disc, the measured $K_1 = 42 \pm 2\, {\mathrm{km~s^{-1}}}$, is an unbiased estimate of the projected orbital velocity of the primary. However, if the main emission line component arises only on the irradiated front face of the secondary star, $K_{2,obs}= 47.4 \pm 0.6\, {\mathrm{km~s^{-1}}}$, is {\em not} an unbiased estimate of the secondary's projected orbital velocity. In order to determine the mass ratio of the binary system from our data, we therefore first need to correct $K_{2,obs}$ to the center-of-mass frame of the secondary. This ``{\em K-correction}'' \citep{K-Correction_MD2005ApJ...635..502M} depends on the opening angle of the shadow cast by the accretion disc on the secondary and also (weakly) on the inclination of the system.

For the low-inclination case considered by \citet{K-Correction_MD2005ApJ...635..502M}, $i=40^\circ$, a moderate opening angle of $\alpha = 10^\circ$ would correspond to a corrected estimate of $K_2 = 82.5~\mathrm{km~s^{-1}}$. Changing $\alpha$ to $6^\circ$ or $14^\circ$ would change this estimate to $K_2 = 85.8~\mathrm{km~s^{-1}}$ or $K_2 = 76.7~\mathrm{km~s^{-1}}$, respectively. We therefore adopt $\sigma_{K_2,sys} = ^{+3.3}_{-5.8}~\mathrm{km~s^{-1}}$ as a characteristic systematic uncertainty associated with this correction. For no shadowing at all ($\alpha = 0^\circ$), 
$K_2 = 88.7~\mathrm{km~s^{-1}}$.

%14 - 76.7 
%12 - 79.8 
%10 - 82.5 
%8  - 84.5 
%6  - 85.8 
% 
%2  - 87.8
%0  - 88.7
The ratio ratio implied by these K-velocities is $q\,=\,M_2/M_1\,=\,K_1/K_2 = 0.51 \pm 0.03 \mathrm{(stat)} ^{+0.4}_{-0.2} \mathrm{(sys)}$. This is consistent with the upper limit on the mass ratio imposed by the requirement for stability against dynamical or thermal time-scale mass transfer, $q < q_{crit} \simeq 2/3$ \citep{Politano1996,Hjellming1989PhDT.........7H}. However, it is somewhat higher than expected for a typical CV at this period. For example, as already noted in Section~\ref{sec: SED}, a typical donor mass at this {\em V341~Ara}'s orbital period would be $M_2 \simeq 0.26 \, M_{\odot}$ \citep{Knigge2006,Knigge+2011ApJS..194...28K}. Combining this with the inferred mass ratio would suggest a low WD mass of $M_1 \simeq 0.5\, M_{\odot}$. Even allowing for the maximum K-correction (no disc shadowing, i.e. $\alpha = 0^\circ$) would only increase this estimate to $M_1 \simeq 0.55\, M_{\odot}$. 

Both of these numbers are considerably lower than the $\simeq 0.8\,M_{\odot}$ that is typical for WDs in CVs \citep{zorotovic2011}. Moreover, the presence of a nova shell around {\em V341~Ara} would tend to suggest a relatively {\em high} WD mass for the system. At any given accretion rate, nova eruptions on massive WDs repeat with shorter recurrence times, so {\em observed} samples of classical novae will tend to be dominated by CVs with high-mass primaries \citep{Ritter1991}. Even though the absorption and emission lines are perfectly anti-phased, this result suggests their radial velocity amplitudes are only proxies for the true $K_1$ and $K_2$. Given all this, it is clearly important to confirm the apparently anomalous mass of {\em V341~Ara}'s WD.
%Given all this, it is clearly important to confirm the apparently anomalous mass of {\em V341~Ara}'s WD.

%Despite being the lines perfectly antiphased, this result may suggest an observational bias in the estimation of $K_1$. % If this is the case, the accretion disc's atmosphere would be a proxy for $K_1$.
%Given all this, it is clearly important to confirm the apparently anomalous mass of {\em V341~Ara}'s WD.

\begin{table}
 \centering
  \caption{ System parameters derived from spectroscopy in Section \ref{sec:spectroscopy}.}
  \label{tab:system param}
  \begin{tabular}{lllll}
  \hline
  Parameter & units & value & error\\
  
  \hline
  \hline
             $\delta\phi$& degrees & $180.0^\circ$ &    $0.3^\circ$\\ 
   %Primary   &     &         &   &\\ 
             $\gamma_1$  &   km/s     &  $-12.6$ & $1.4$ \\
             $K_1$  &   km/s     &  $42$ & $2$ \\
   %Secondary   &     &         &   &\\ 
             $\gamma_2$  &   km/s     &  $-24.43$ & $+0.52/-0.57$ \\
             $K_{em}$  &   km/s     &  $47.4$ & $0.6$ \\
             $K_{2}$  &   km/s     &  $84.5$ & $1.6$  \\
             $q=K_1/K_2$  &        &  $0.5$ & $0.02$  \\
   
 \hline
 \end{tabular}
 \begin{tabular}{l l l l}
 $\mathrm{P_{orb}}$&$=$&$3.6529454^{+0.0000016}_{-0.0000013}$ & hours\\
 $\mathrm{T_0^{BJD}} $&$=$&$ 2457098.01142 \pm 0.00039$& days\\
 
 \hline
 \end{tabular}
\end{table}

  %\subsubsection{Analysis}    
 
%%%%%%%%%%%%%%%%%%%%%%%%%%%%%%%%%%%%%%%%%%%%%%%%%%%%%%%%%%%%%%%%%%%%% 
%%%%%%%%%%%%%%%%%%%%%%%%%%%%%%%%%%%%%%%%%%%%%%%%%%%%%%%%%%%%%%%%%%%%%

\subsection{X-ray properties\label{sec:BL}}

Figure~\ref{fig:xrt_all} shows the {\em Swift/XRT} spectra of {\em V341~Ara} obtained during our two observing epochs. There was no statistically significant variability in the X-ray data within or between these epochs, so we analysed both spectra simultaneously. The average XRT count rate was only $\simeq 2\times 10^{-2}\, \mathrm{c~s^{-1}}$, which corresponds to a rather low X-ray luminosity of $\mathrm{L_X,0.3-10~keV} \simeq 10^{30}\, {\rm erg~s^{-1}}$. As a result, the signal-to-noise of the spectra is relatively poor. Nevertheless, it is 
clear that there are at least two components contributing to the X-ray emission, with the softer one being dominant. Using 1~keV as boundary between soft and hard photons gives a hardness ratio of $\simeq 0.6$, which is typical for accreting white dwarfs (AWDs) in a high state \citep[e.g.][]{Patterson1985}. 

%%%Outline of suggested revision:

Could the dominant soft component be the signature of the optically thick BL? We begin by noting that a relatively cool BL -- $T_{BL} \simeq (1-2) \times 10^{5}$~K, as adopted in our SED models in Figure~\ref{fig:full SED} -- would be undetectable in our {\em Swift/XRT} data. Its emission would lie almost exclusively in the extreme ultraviolet band, where even modest interstellar Hydrogen columns are sufficient to extinguish its flux. 

How hot would an optically thick emitter need to be in order to account for the soft component in our XRT spectra? We address this by modelling the soft X-ray component as a simple BB and the harder component with a single-temperature Bremsstrahlung model.
Both components were assumed to be affected by a combination of intrinsic, partially ionised absorption ({\tt tbabs} in {\sc xspec}), plus extinction by the neutral interstellar medium ({\tt phabs} in {\em XSPEC},  with $N_H = 2.9 \times 10^{20}\, \mathrm{cm^{-2}}$ fixed at a fraction of the Galactic line-of-sight value; \citealt{SF_extinction2011ApJ...737..103S,Ext_to_nH2017MNRAS.471.3494Z}). Intrinsic absorption of this type is fairly common in disc-accreting CVs \citep[e.g.][and references therein]{Pratt2004MNRAS.348L..49P,Mukai2017PASP..129f2001M}. 

This model achieves a reasonable fit, $\chi^2_\nu =1.13$, and the resulting model parameters are summarized in Table \ref{tab:xrt fit}. The $2-\sigma$ lower limit on the temperature of the soft component is $T_{BB} > 600,000$~K  This is much hotter than theoretically expected for an optically thick BL \citep{Balsara2009ApJ...702.1536B,Hertfelder2013AA...560A..56H}.

%%%%%%%%%%%%%%%%%%%%%%%%%%%%%%%%%%%%%%%%%%%%%%%%%%%%%%%%%%%%%%%%%%%%% 
\begin{figure}
    \hspace*{-2mm}\includegraphics[width=.5\textwidth]{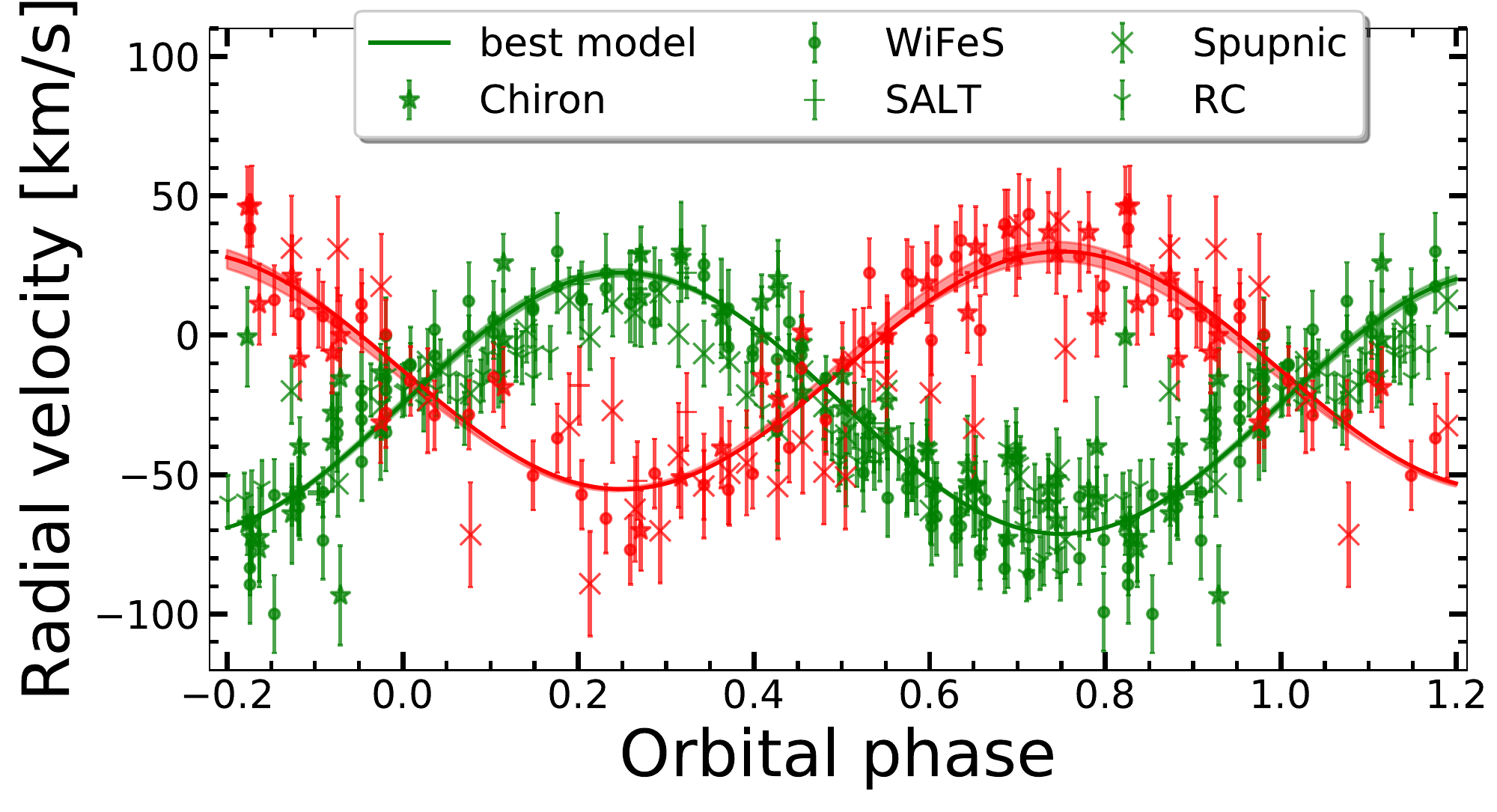}
    \caption{\small  Phase folded radial velocity curve of the main emission component (green, including H$\alpha$ and H$\beta$) and broad absorption (red, only H$\beta$). Best fit sinusoidal model is represented with a solid line with $1\sigma$ uncertainties as shaded region. 
}
    \label{fig:V341Ara rv}
\end{figure}
%%%%%%%%%%%%%%%%%%%%%%%%%%%%%%%%%%%%%%%%%%%%%%%%%%%%%%%%%%%%%%%%%%%%%

We conclude that the {\em Swift/XRT} spectra of {\em V341~Ara} are {\em not} dominated by optically thick BL. Instead, we suspect that the observed spectrum is produced by non-thermal processes (e.g. shocks in the accretion disc wind of the system; e.g. \citealt{Mauche2004RMxAC..20..174M}, also see Section~\ref{sec: SED}) and/or by optically thin thermal emission (e.g. due to an optically thin ``chromosphere'' or ``corona'' on top of the optically thick disc and/or boundary layer; e.g. \citealt{Aranzana2018MNRAS.481.2140A}). As a qualitative test of these ideas, we also fit the {\em Swift/XRT} spectra with an optically thin, multi-temperature plasma emission line model ({\tt cemekl}), which has been successfully applied to X-ray observations of other NL CVs in the past \citep{Pratt2004MNRAS.348L..49P}. Such a model can produce an acceptable fit, but the signal-to-noise of our data is too poor to provide physically interesting constraints on the model parameters.

%%%%%%%%%%%%%%%%%%%%%%%%%%%%%%%%%%%%%%%%%%%%

\begin{figure}
\centering
\hspace*{-4mm}\includegraphics[width=.5\textwidth]{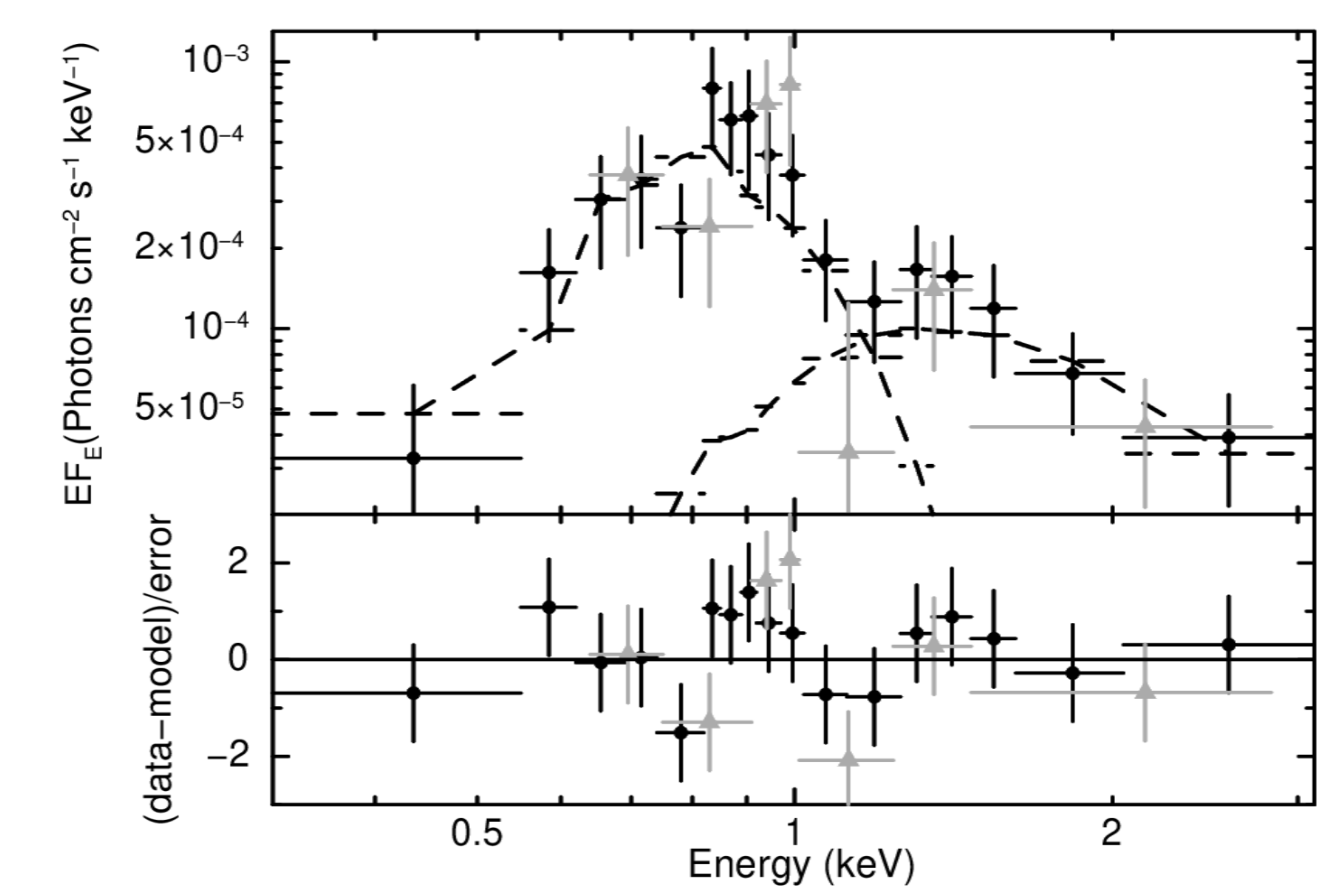}

\caption[]{\footnotesize
Combined {\em Swift/XRT}. Top panel: measured X-ray spectrum. 2019 and 2018 data sets are represented with black dot and grey triangles respectively, the dashed lines are the different components of the model. We use the XSpec model which consists of the Galactic absorption, a blackbody and thermal bremsstrahlung being absorbed with partially ionized gas. In the bottom panel the residuals of the modelling are shown.}
%CK: NO Y-AXIS LABEL; TOP LABEL SHOULD BE REMOVED. ALSO, COULD WE ADD TWO BIG DOTS OR CROSSES OR SOMETHING HERE, ONE FOR EACH OF THE MODELS OF THE BL WE USE IN OUR SED MODELLING? The numbers were Tbb = 209,000 (18eV) Lbl = 4e34 and Tbb = 115,000 (9.9 eV), Lbl = 4e33}} ### WE HAVE THIS INFO IN A EMAIL FROM MAYUKH
\label{fig:xrt_all}
\end{figure}  
%%%%%%%%%%%%%%%%%%%%%%%%%%%%%%%%%%%%%%%%%%%%%

\begin{figure}
\centering
\hspace*{-4mm}\includegraphics[scale=0.32,angle=-90]{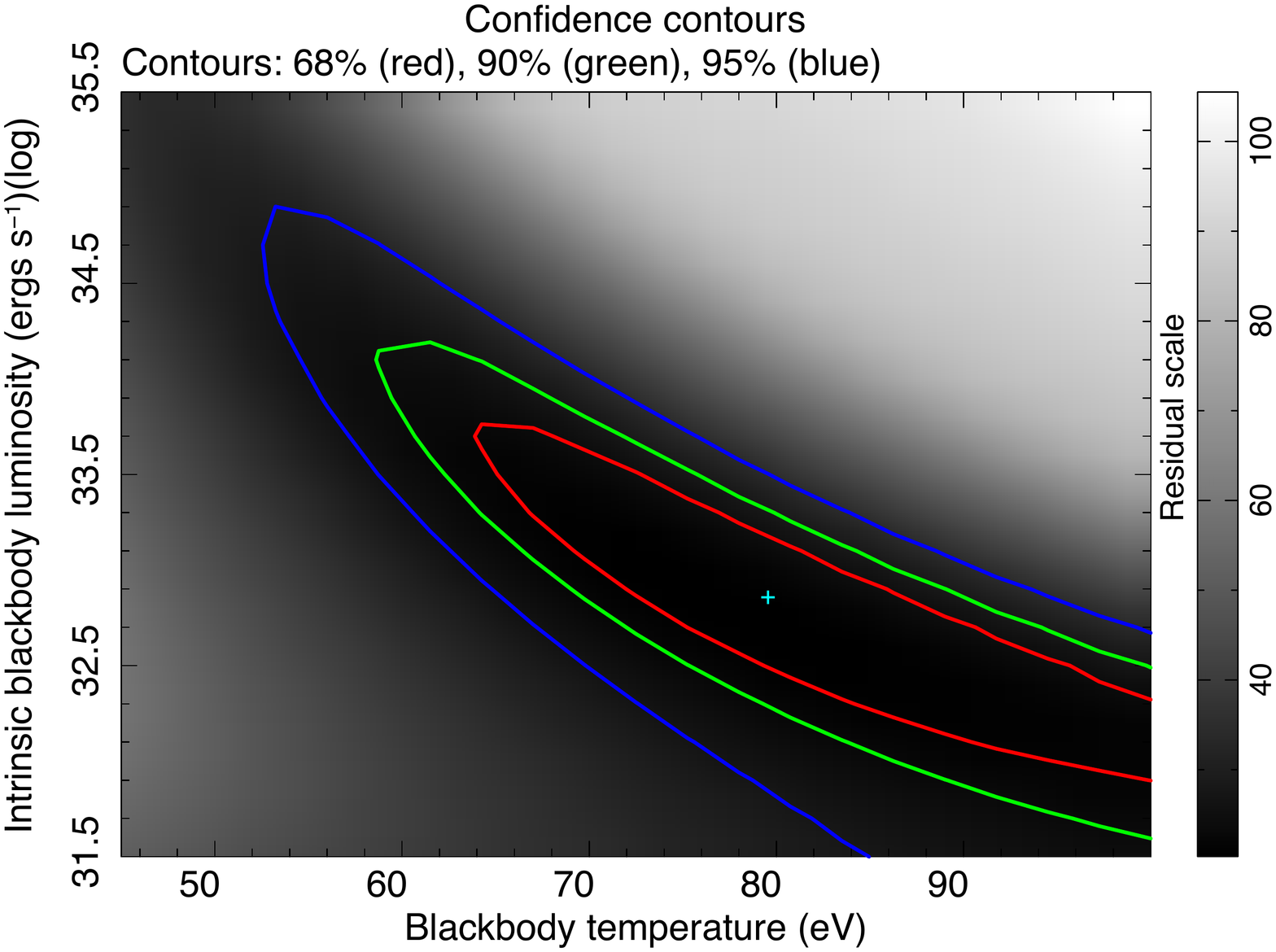}

\caption[]{\footnotesize
Intrinsic blackbody luminosity versus effective temperature for an optically thick boundary layer, contours represent the confidence intervals, while the colour scale the residuals for the tested model. }
\label{xrt}

\end{figure}

%%%%%%%%%%%%%%%%%%%%%%%%%%%%%%%%%%%%%%%%%%%%%

\begin{table}
 \centering
  \caption{ Best parameters estimated from the fit of the two {\em Swift/XRT} epochs simultaneously (fig. \ref{fig:xrt_all}). $^*$ The column density in the line of sight was fixed at a fraction of the Galactic line-of-sight value in order to accommodate the intrinsic ${\rm n_H}$.
  %CK: HOW WAS THE VALUE OF NH FOR THE ISM EXTINCTION ESTIMATED? IF IT'S THE TOTAL ALONG THE LOS, ISN'T THAT AN OVERESTIMATE SINCE THE SYSTEM IS SO CLOSE?
  }
  \label{tab:xrt fit}
  \begin{tabular}{lllll}
  \hline
  Model & parameter & units & value &\\
  
  \hline
  \hline
   phabs$^*$& nH    & cm$^{-2}$  &  $2.9$ &$\times 10^{20}$ \\
   tbabs    & nH    & cm$^{-2} $ &  $1.08_{-0.5}^{+0.6}$ & $\times 10^{22}$\\
   bremss   & kT    & keV        &  $0.73_{-0.26}^{+1.25}$ &\\ 
            & norm  &            &  $1.2_{-1.0}^{+4.1}$ & $\times 10^{-3}$\\
   bbody    & kT    & eV     &  $80_{-20}^{+40}$ &\\%\times $10^{-2}$\\
            & norm  &          &   $2.3_{-2.0}^{+6.4}$ & $\times 10^{-3}$\\
 \hline
 \end{tabular}
 \begin{tabular}{c c c c c}
 $\chi^2$ & $\chi^2_\nu$ &  C-stats &  PHA &  DOF \\
 21.45 & 1.129 & 20.25 & 24 & 19\\
 \hline
 \end{tabular}
\end{table}

%%%%%%%%%%%%%%%%%%%%%%%%%%%%%%%%%%%%%%%%%%%%%
    
\subsection{Optical imaging: nova shell and bow-shock} 
%{\em Figure: IFF Fig 1 from the intro is good enough, no need for another one here}
%{\em Things to touch on:}
%\begin{itemize}[leftmargin=0.4cm, label=$\bullet$]
%\item New independent estimate of overall Halpha nebula mass? -- nova shell vs PN: probably not worth redoing, but %worth mentioning again
%\item Asymmetry of bow-shock relative to source
%\item New independent estimate of shock stand-off distance
%\item Estimate of ISM/shell density where wind runs into it?: probably not worth redoing, but worth mentioning again
%\item Proper motion is consistent with nova shell idea -- time to reach current position ~1000 years
%\end{itemize}
%\hline
\subsubsection{The extended nebula (Fr 2-11)} \label{sec:NEBULA}

As discussed by F08 and B18, there are two viable scenarios for the origin of the large-scale H$\alpha$ nebula around {\em V341~Ara}: (i) a small H{\sc ii} region composed of interstellar gas that is photoionised by the radiation field of the CV; (ii) material ejected in a relatively recent nova eruption of the system. A third scenario, in which the emitting material is a planetary nebula (PN), can be ruled out, since the mass estimate for this material (based on its H$\alpha$ luminosity) is only $M_{shell} \simeq 0.005\,M_{\odot}$, about 40$\times$ less than what is typical for a PN (F08). In both scenarios, the smaller-scale bow-shock is produced where the accretion disc wind associated with the fast-moving CV interacts with the (now) more-or-less stationary large-scale nebula \citep[c.f.][]{Shara+2017MNRAS.465..739S,Wareing+2007MNRAS.382.1233W}.

F08 favours the first scenario, on two grounds. First, the emission lines in the nebular spectra are relatively narrow, contrary to what may be expected for a fast, differentially expanding nova shell. Second, there is no {\em distinct} forward shock at the south-eastern edge of the nebula, nor any other sign of decelerative interaction between a nova shell and the surrounding ISM (B18). Such features might be expected if the large-scale nebula is a nova shell expanding into the ISM. B18 are somewhat more circumspect, noting that the proper motion vector {\em V341~Ara} places it near the center of the nebula approximately 800~years ago. This positional coincidence, as well as the associated time scale, not unreasonable numbers in the context of the second, nova-shell scenario.

Since identifying the center of the nebula is non-trivial, we obtained an independent estimate of this displacement time scale by overlaying ellipses on the H$\alpha$ image, centered on the position of {\em V341~Ara} at times 800, 900, 1000 and 1200~yrs in the past. The middle two of these ellipses provide the best match to the overall shape of the large-scale nebula, so our new estimate of the displacement time scale is 900-1000~years. This is broadly consistent with B18 and hence in line with the possibility that the nebula is an old nova shell.

We can further test this scenario by considering the luminosity and mass of the nebula. The H$\alpha$ flux measured by F08 for Fr~2-11 from the Southern H$\alpha$ Sky Survey Atlas  \citep[SHASSA;][]{Gaustad+2001PASP..113.1326G} plates is $\log{F_{H\alpha}} = 10.99 \pm 0.06$ (in cgs units). This corresponds to an H$\alpha$ luminosity of $L_{H\alpha} \simeq 3\times10^{31}~{\mathrm{erg~s^{-1}}}$ and, as noted above, a nebular mass of $M_{shell} \simeq 5 \times 10^{-3} M_\odot$. At first sight, this mass is somewhat high compared to other known nova shells, which tend to lie in the range  $10^{-6} \lesssim M_{shell} \lesssim 10^{-4}$ \citep{Cohen1983ApJ...268..689C,Williams1994ApJ...426..279W}. However, an eruption $\simeq 1000$~yrs ago would tend to make {\em V341~Ara} one of the oldest known classical novae, with plenty of time for the nebula to grow in mass by sweeping up material from the ISM. A useful point of comparison here is the recently discovered nova shell around the NL {\em V1315~Aql} \citep{Sahman2015MNRAS.451.2863S}, which likely has a similar age and an inferred mass of $M_{shell} \simeq 2 \times 10^{-4}~\mathrm{M_{\odot} yr^{-1}}$.

On balance, we agree with B18 that the nova-shell scenario for the large-scale nebulosity should not be discounted. As already noted by them, it would therefore  be useful to estimate (limits on) the expansion rate of the nebula from archival and/or future images. To provide some context for this, we consider a $\simeq10^{-5}\, \mathrm{M_\odot}$ nova shell that initially expands freely with a velocity of $\simeq 1000~\mathrm{km~s^{^-1}}$. This free expansion phase would then last for about 50-60 years, during which the material travels an angular distance of $2'$. After this time, the mass of swept-up material exceeds that of the ejected material, and the expanding shock enters a deceleration phase \citep[the ``Taylor-Sedov'' phase;][]{Taylor1950RSPSA.201..159T,Sedov1959sdmm.book.....S}%{Taylor1950,Sedov1959}
, during which the expansion velocty scales as $v\propto t^{-0.6}$. Assuming that the ISM is at rest, and adopting a relatively high ISM density $n_e\simeq10~\mathrm{cm^{-3}}$ (F08), we then find that in $\simeq$1000 years the edge of the nebula would travel a radial distance of $\simeq3'-4'$, in line with the observations.

\subsubsection{The bow-shock} \label{sec: bow-shock}
An outflow propagating into a medium produces a shock if its speed exceeds the sound speed of that medium; if the source expelling the outflow is also moving supersonically with respect to the medium, a bow-shock forms \citep[e.g.][and references therein]{delPalacio+2018AA...617A..13D}. Such is the case of {\em V341~Ara} (see Fig. \ref{fig:V341Ara image} and \ref{fig:oiii WiFeS}), in which the supersonic wind produced by the accretion disc is interacting with a dense environment. The high proper motion of the system ($v \simeq 70$~km~s$^{-1}$) shapes the resulting structure into a bow-shock. 
  
  The collision of the wind with the surrounding medium forms an interaction region, 
  known as a forward shock, that propagates through the medium while compressing it. After traveling a distance $R'_1$ from its launching position, the high speed wind encounters the shocked medium. In the resulting interaction the kinetic energy of the wind is converted into thermal energy. This heats the wind material and produces a reverse shock \citep{Weaver+1977ApJ...218..377W}. The forward shock is radiative and slow, whereas the reverse shock is adiabatic and fast. Between these two shocks lies the so-called contact discontinuity, a surface that separates the shocked wind from the shocked medium. At the contact discontinuity, the net mass flux is zero. Following \citet{Weaver+1977ApJ...218..377W}, we denote the distance from the star to the contact discontinuity as $R'_c$. 
  
  The structure of the bow-shock is determined by the ram pressure associated with the movement of the system through the medium, which produces a quasi-parabolic shell of swept-up material \citep{vanBuren1988ApJ...329L..93V,Hollis+1992ApJ...393..217H}. %{VanBuren1988,Hollis1992THEJ.}. 
    However, when the star is not moving in the plane of the sky, projection effects play an important role in determining the observed shape of the shock \citep{Meyer2016MNRAS.459.1146M}. %\citet{Meyer2017} showed how the projection changes the observed shape of a bow wave. 
  
  In Fig.~\ref{fig:oiii WiFeS} it can be seen that there is an extra, fainter component of the shock South from the source. This suggests that the source is moving mainly in the radial direction, in agreement with the results obtained in our radial velocity study. Our narrow band imaging also shows that, even though the orientation of the bow-shock matches with the proper motion of the star, there is a slight misalignment with the apex of the bow-shock. Specifically, the peak of the enhanced [O {\sc iii}] emission lies slightly West of the source. This could be the manifestation of the biconical shape of the accretion disc wind \citep{Matthews+2015MNRAS.450.3331M}, combined with the misalignment of the accretion disc relative to the plane of the sky.

  The presence of the bow-shock allows us to estimate the energetics of the wind. 
  The wind deposits energy into the reverse shock at a rate
  \begin{equation}\label{eq:wind kinetic}
      L_w = \frac{1}{2}\dot{M}_w V_w^2 \,,
  \end{equation}
  where $L_w$ is the kinetic wind luminosity, and $\dot{M}_w$ and $V_w$ are the mass-loss rate and velocity of the wind, respectively. Under the assumption of an intrinsically parabolic bow-shock, \citet{Weaver+1977ApJ...218..377W} derived the minimum distance from the stellar system to the apex of the nebula or the stagnation point ($R'_2$) in terms of the other relevant physical parameters:
  \begin{equation} \label{eq:l2}
    R'_2 = \frac{\sqrt{20L_w}}{3\sqrt{33\pi\rho_{ISM}V_{ISM}^3}} \,.
  \end{equation}
  Here, $\rho_{ISM}$ and $V_{ISM}$ are the density of the ISM and its velocity relative to the star, respectively. \citet{Dopita1977ApJS...33..437D} showed that [O{\sc iii}] originates closer to the front shock compared to other shock-excited emission lines. Therefore, we can estimate $R_2 \simeq 7.5\pm 0.5 ''\simeq 1.7 \pm 0.1\times 10^{16}\,\mathrm{cm}$ from the [O{\sc iii}] image (Fig. \ref{fig:oiii WiFeS}). F08 obtained optical spectroscopy of the nebular region around {\em V341~Ara}. Given that the heliocentric radial velocities of the extended nebular lines are negligible, he concluded that the ISM around the source is at rest with respect to the Sun. He also estimated the electron density of the ISM to be $n_e\simeq10\, cm^{-3}$ from the extended H$\alpha$ emission imaging. In Sections~\ref{sec:gaia} and \ref{sec: bow-shock}, we measured a relative velocity $V_{ISM} \simeq 76\pm 1\, km\,s^{-1}$ with an inclination angle of $\theta = 18.7 \pm 0.4 \degr$ from the plane of the sky. Rearranging equation \ref{eq:l2}, we can therefore estimate the kinetic power of the wind as
  \begin{equation} \label{eq:Lw}
    L_w = 14.85 \pi (R_2 \sec \theta)^2 \rho_{ISM} V_{ISM}^3  = 1.2\pm 0.2 \times 10^{32}\,\mathrm{erg\,s^{-1}} \,,
  \end{equation}
  where the $\sec \theta$ term accounts for the symmetry axis of the parabola not being aligned with the plane of the sky, and we have approximated the ISM density as $\rho_{ISM} \simeq n_e m_H$, where $m_H$ is the mass of the hydrogen atom.
  
  To estimate the wind mass-loss rate ($\dot{M}_w$), we need to %have an estimation of the velocity of the wind, 
  estimate the terminal wind velocity. Ideally, this would be estimated from wind-formed P-Cygni lines, which are normally seen in the far-UV region \citep[e.g.][]{Prinja2000MNRAS.312..316P}. Unfortunately, no FUV spectra of this source have been obtained to date.

  In the absence of such data, we adopt a different approach to set an approximate upper limit to the mass loss rate of the source. 
  We consider the region in the flow where the momentum flux of the wind is exactly balanced by the ISM ram pressure: $\rho_w V_w^2 = \rho_{ISM} V_{ISM}^2$. Here, $\rho_w$ is the density of the wind at this point. Approximating the wind as spherical, we can use the continuity equation to express the wind density as $\rho_w \simeq \dot{M}_w/4\pi R_1^2V_w$. Combining the two expressions allows us to express the mass-loss rate as 
  \begin{equation} \label{eq:force balance}
    \dot{M}_w \simeq \frac{4\pi R_1^2V_{ISM}^2\rho_{ISM}}{V_w} \,.
  \end{equation}

  We can now combine Equations~\ref{eq:wind kinetic} and \ref{eq:l2} to obtain an expression for $V_w$. Substituting this into Equation~\ref{eq:force balance} we obtain
  \begin{equation}
      \dot{M}_w = \frac{0.54\pi\rho_{ISM} V_{ISM} R_1^4}{R_2^2 \cos^2 \theta} \,.
  \end{equation}
  This include a factor of $\sec^2\theta$ to account for the inclination of the system. Unfortunately, $R_1$ is not observable, since the high-temperature shocked wind does not produce any optical emission \citep{Weaver+1977ApJ...218..377W,Hollis+1992ApJ...393..217H}. However, since $R_1<R_c<R_2$, we can set an upper limit on the mass loss rate by estimating $R_c= R_2 - \Delta R$, where $\Delta R = 3\pm0.5''$ is the thickness of the [O {\sc iii}] bow-shock. With $R_c = 4.5\pm0.7''$, the wind mass loss rate has to be $\dot{M}_w < 1.5 \pm 0.9\times 10^{-10}\, M_\odot\,yr^{-1}$. %$\dot{M}_w \lesssim 1.5 \pm 0.9  \times 10^{-10}\, M_\odot\,yr^{-1}$. 
This, in turn, implies a lower limit on the terminal wind velocity of $V_w > 1600 \pm 500 \,\mathrm{km\,s}^{-1}$.

 We can extract physical information from the radio upper-limit by introducing a radiative model for the bow-shock. For this purpose, we adapt the one-zone model developed by \citet{delPalacio+2018AA...617A..13D} for bow-shocks produced by runaway massive stars. In a nutshell, relativistic electrons are expected to be accelerated at the strong shock and produce synchrotron emission that dominates the low-frequency radio spectrum. 
 The flux upper limit $S_{1.28\mathrm{GHz}} < 30~\mu$Jy allow us to constrain parameters related to the synchrotron luminosity: the magnetic field strength, $B$, and the amount of power injected in non-thermal particles, $L_\mathrm{NT}$. To break the degeneracy between these two parameters, we assume a minimum energy condition for the relativistic particles and the magnetic field \citep[$U_\mathrm{NT} = 0.75\, U_\mathrm{mag}$;][]{Longair2011hea..book.....L}. Adopting $R_0 \sim 10^{16}$~cm, $V_w \approx 1600$~km\,s$^{-1}$ and $\dot{M}_w \approx 10^{-10}\, M_\odot$\,yr$^{-1}$, we constrain the non-thermal luminosity to $L_\mathrm{NT} < 0.7\% \, L_\mathrm{w}$ and $B < 0.5$~mG. This result suggests that disc winds in CVs are not efficient cosmic-ray accelerators.
  
\section{Discussion} \label{sec:discuss}
\subsection{{\em V341~Ara} as an old nova}

Almost all CVs that are classified as novae were {\em discovered} on the basis of an observed eruption. This introduces a strong selection bias, since it restricts the time-frame over which we can estimate typical nova rates and recurrence times. In particular, large-area, high-cadence photometric data has only started to become available over the last few decades, a much shorter period than the typical inter-outburst recurrence times expected from nova models ($\tau_{rec} \sim 10^4$~yrs). The population of known novae is thus heavily biased towards systems with unusually short $\tau_{rec}$, i.e. systems with high $M_{WD}$ and/or high $\dot{M}_{acc}$ \citep[e.g.][]{Townsley2004ApJ...600..390T,Yaron2005ApJ...623..398Y,Wolf+2013ApJ...777..136W,Kato+2014ApJ...793..136K}.%(REF nova models -- Townsley, also Yaron).

We can mitigate this selection effect by identifying older novae among the ``normal'' CV population, i.e. among systems where no  eruption has so far been noted. One way to achieve this is to match known CVs to bright ``guest stars'' in old historical records \citep[e.g.][]{Patterson2013MNRAS.434.1902P,Vogt+2019AN....340..752V}. However, probably the best and most systematic method is to search for nova shells around known CVs \citep[e.g.][]{Cohen1983ApJ...268..689C}. These shells are composed of material ejected in the nova eruption (plus any swept-up ISM). They are usually observed as emission line nebulae and can remain detectable for centuries \citep[][]{Downes2001JAD.....7....6D,Duerbeck2003ASPC..292..323D}.

However, despite several systematic searches \citep{Shara2012ApJ...758..121S,Sahman2015MNRAS.451.2863S,Schmidtobreick2015MNRAS.449.2215S}, nova shells have so far been identified around only 6 CVs without previously known outbursts: Z Cam \citep{Shara2007Natur.446..159S}, AT Cnc \citep{Shara2012ApJ...758..121S}, V1315 Aql \citep{Sahman2015MNRAS.451.2863S,Sahman2018MNRAS.477.4483S}, Te~11 \citep{Miszalski2016MNRAS.456..633M}, IPHASX J210204.7+471015 \citep{Guerrero+2018ApJ...857...80G,Santamara2019MNRAS.483.3773S} and a CV in the globular M22 \citep{Gottgens2019AA...626A..69G}. The identification of {\em V341~Ara} as another member of this small group is therefore significant, particularly because it permits two valuable checks.

First, as discussed in Section~\ref{sec:NEBULA}, the offset of {\em V341~Ara} from the center of the nebula, combined with its proper motion, suggests that the nova eruption took place $\simeq 1000$~years ago. This makes the system one of the oldest ``recovered'' novae known. It is also a useful constraint for nova models. Specifically, recurrence times are predicted to decrease with increasing $M_{WD}$ and increasing $\dot{M}_{acc}$. Adopting $t_{rec} \gtrsim 1000$~yrs as a lower limit on the recurrence time therefore sets a mass-dependent upper limit on the accretion rate averaged over a nova cycle. For example, interpolating on the model grid of \citet{Yaron2005ApJ...623..398Y}, we estimate that $\overline{\dot{M}_{acc}} \lesssim 1.7 \times 10^{-8} \left[(M_{Ch} - M_{WD}] / (0.44 M_{\odot})\right]^{1.76}$,
in agreement with the SED (Sec. \ref{sec: SED}). Here, $M_{Ch} = 1.44\, M_{\odot}$ is the Chandrasekhar mass, and this specific fit is for WD models with a core temperature of $T_{\rm core} = 3\times10^{7}$~K.

Second, {\em V341~Ara} must have been exceptionally bright during its nova eruption, so it might have been recorded as a "guest star" in historical records. At optical wavelengths, nova eruptions typically have amplitudes of $\Delta m \simeq 11$~mag \citep{Vogt+2019AN....340..752V}.
This implies that {\em V341~Ara} probably reached $m_{peak} < 0$ during its nova eruption. This would have temporarily placed the system among the top-5 brightest stars visible from Earth; it may even have made it {\em the} brightest star, other than the Sun. For comparison, Kepler's and Tycho's supernovae are thought to have reached $m_{peak} \simeq -3$ and at $m_{peak} \simeq -4$, respectively \citep{Green2003LNP...598....7G}.
%(http://cds.cern.ch/record/603187/files/0301603.pdf). 
%Unfortunately, {\em V341~Ara} is located in an area of the sky where historical observations are sparse. 
We thus checked for positional coincidences in several compilations of historical guest star observations \citep{Tse-Tsung1957AZh....34..159T,Stephenson1976QJRAS..17..121S,Stephenson1971ApL.....9...81S,Clark1976QJRAS..17..290C,Yang2005AA...435..207Y,Stephenson2009JHA....40...31S,Hoffmann2020AN....341...79H}. There were no matches closer than about $18\degr$, but we did uncover one potential counterpart: \# 45 in \citet{Stephenson2009JHA....40...31S}.
%http://articles.adsabs.harvard.edu/pdf/2009JHA....40...31S

This guest star was recorded by Chinese astronomers on Aug 17 of AD 1240 -- during the Southern Song dynasty, at a right ascension in the range $16:50 \lesssim \alpha \lesssim 18:05$. Unfortunately, its peak apparent magnitude is not known, and its declination is almost unconstrained. On one hand, V341~Ara would not have been visible on this date from the observatories established in the capital Lin'an (now Hangzhou) during this dynasty (Dao 2019). On the other hand, its altitude at meridian crossing was only $\simeq 1\degr$ below the horizon as seen from Hangzhou (allowing for proper motion, precession and refraction). Thus if the guest star was originally reported by observers located just slightly south of the capital, V341~Ara would be a viable counterpart. In this case, it would be the fourth oldest recovered nova \citep{Vogt+2019AN....340..752V} and an excellent laboratory for testing nova theory.
%https://arxiv.org/pdf/1910.13464.pdf

%REFS: http://articles.adsabs.harvard.edu/pdf/1957SvA.....1..161T
%http://articles.adsabs.harvard.edu/pdf/1976QJRAS..17..121S
%https://arxiv.org/pdf/1912.03139.pdf
%http://articles.adsabs.harvard.edu/pdf/1976QJRAS..17..290C
%https://www.aanda.org/articles/aa/full/2005/19/aa2455/aa2455.html
%http://articles.adsabs.harvard.edu/pdf/1971ApL.....9...81S
%http://articles.adsabs.harvard.edu/pdf/2009JHA....40...31S

\subsection{The wind-driven bow-shock: calibrating disc wind models}

%    Accretion disc winds are ubiquitous in accreting sources on all scales. They act as sinks for mass, energy %and angular momentum for the accretion process and provide a way for these systems to interact with their %environments. 
    
    The most common method to determine the mass and energy budget of CV accretion disk winds is by modelling wind-formed UV resonance lines \citep[e.g.][]{Knigge1995MNRAS.273..225K,Knigge+1997ApJ...476..291K,LongKnigge2002ApJ...579..725L}. However, this modelling is based on parameterized kinematic models that depend on many free parameters, and it requires computationally expensive radiative transfer simulations. 

    As illustrated in Section~\ref{sec: bow-shock}, wind-driven bow-shocks can provide independent estimates of these quantities. Systems that exhibit such bow-shocks -- like {\em V341~Ara}, {\em BZ Cam} and {\em V1838~Aql} -- are therefore crucial for testing and calibrating the estimates provided by spectral models.
     
    In {\em V341~Ara}, specifically, the bow-shock suggests a kinetic wind luminosity of $L_{w}\simeq 10^{32}~\mathrm{erg~s^{-1}}$ and yields limits on the mass-loss rate ($\dot{M}_{w} \lesssim 10^{-10}~\mathrm{M_{\odot} \, yr^{-1}}$) and terminal velocity ($v_{w} \gtrsim 1800~\mathrm{km~s^{-1}}$). UV spectroscopy of the system is strongly encouraged, so that these constraints can be tested against spectral models of the wind-formed UV resonance lines. 
    
\subsection{Large-amplitude super-orbital variations from tilted precessing discs}
\label{sec: SHs}
%{\em Things to touch on:}
%\begin{itemize}[leftmargin=0.4cm, label=$\bullet$]
%\item Other systems: "stunted outbursts" (Honeycutt etc); perhaps comment that we've also found similar behaviour in %other systems now; could be quite common, actually
%\item Physical origin 1 -- "stunted" DN outburss: probably not much to say about this, other than the variations %don't really look like that (light curve shape); plus outbursts aren't usually this quasi-periodic (not sure about %that, actually). 
%\item Physical origin 2 -- precession? Discuss briefly in the context of positive/negative superhumps; implied mass %ratio; difficulty explaining the huge amplitude (and perhaps also the extreme instability of the period?)
%\item Other possibilities? Magnetic cycles on donor probably ruled out -- expected time scale much longer; 2 week %time-scale isn't really anything other than perhaps viscous at outer disc
%\item Conclusion: Origin remains a mystery
%\end{itemize}
%\hline

As noted in Section~\ref{sec:LCs}, {\em V341~Ara} exhibits large-amplitude ($\Delta m \simeq 0.5 - 2.0$~mag), super-orbital quasi-periodic oscillations on a characteristic time-scale of $P_{slow} \simeq 10 - 16$~days. Moreover, the high-quality {\em TESS} light curve additionally reveals {\em two} signals at higher frequencies. One of these corresponds to the orbital period ($P_{orb} \simeq 3.654$~hrs), while the other ($P_{SH-} \simeq 3.621$~hrs) is $\simeq$1\% faster and consistent with the beat frequency between $P_{slow}$ and $P_{orb}$. 

Photometric signals near $P_{orb}$ have been detected in many CVs. They are generally interpreted as beats between the orbital signal and an intrinsically much slower signal. More specifically, photometric signals slightly slower than $P_{orb}$ are usually called {\em ``positive superhumps''}. The underlying slow signal in this case is thought to be associated with the {\em prograde} precession of an {\em eccentric} accretion disc. By contrast,  photometric signals slightly faster than $P_{orb}$ are usually called {\em ``negative superhumps''}.  The underlying slow signal in this case is thought to be the {\em retrograde} precession of a disc that is 
{\em tilted} (or perhaps warped) relative to the orbital plane. A comprehensive discussion of superhumps can be found in \citet{Patterson1999dicb.conf...61P}.
%https://ui.adsabs.harvard.edu/abs/1999dicb.conf...61P/abstract
%here is the actual paper:
%https://cbastro.org/wp-content/uploads/2018/01/permanent-superhumps-in-cataclysmic-%variables.pdf

It is clearly tempting to interpret {\em V341~Ara}'s large-amplitude super-orbital variations on $P_{slow}$ as the signature of a tilted disc undergoing retrograde precession. The slightly faster-than-orbital variations on $P_{SH-}$ would then correspond to the usual negative superhump signal at the beat period between $P_{orb}$ and $P_{slow}$. However, there are two serious  problems with this idea. First, the so-called period deficit we infer from the {\em TESS} data, $\epsilon_{-} = (P_{orb}-P_{SH-})/P_{orb}) \simeq 0.01$ is unusually small. It is significantly smaller than theoretically expected for a tilted disc in a system with $q = M_2/M_1 \simeq 0.5$ \citep{Wood2009MNRAS.398.2110W}, and it is also smaller than the deficits typically seen in other NL systems \citep[][and references therein]{Patterson_SHs_2005PASP..117.1204P,Wood2009MNRAS.398.2110W}. Second, the amplitude of the super-orbital variations is exceptionally large. In most NLs exhibiting negative superhumps, the actual precession signal on $P_{slow}$ is either weak or not detected at all, and even the stronger signal on $P_{SH-}$ exhibits amplitudes of $\Delta m \lesssim 0.6$~mag \citep{Harvey1995PASP..107..551H}. %http://articles.adsabs.harvard.edu/pdf/1995PASP..107..551H). 

We suggest that the solution to these problems is connected to the recent discovery of slow, large-amplitude oscillations in a subset of Z~Cam stars \citep{Simonsen2011JAVSO..39...66S,Szkody2013PASP..125.1421S,Kato2019PASJ...71...20K}. Z~Cam stars are DNe that occasionally "get stuck" near the plateau phase of their outbursts. The photometric variations first noticed by \citet{Simonsen2011JAVSO..39...66S} in the proto-type {\em IW~And} (and also in {\em V513~Cas}) can be roughly described as large-amplitude, damped, quasi-periodic slow oscillations in this stand-still phase \citep[e.g.][]{Kato2019PASJ...71...20K}. We will refer to this as the {\em IW~And} phenomenon. Several additional systems exhibiting this phenomenon were identified by \citet{Kato2019PASJ...71...20K}. He proposed that these variations might be produced by the same thermal-viscous instability that is responsible for ordinary dwarf nova eruptions, if the instability could be somehow confined to the outer disc in these systems (with the inner disc remaining in a hot stable state). 

Building on this idea, and noting that several systems exhibiting the {\em IW~And} phenomenon also display negative superhumps, \citet{Kimura2020PASJ..tmp..156K} suggested that spatially confined disc instabilities may occur naturally in {\em tilted} accretion discs. The accretion stream from the donor star will impact a tilted accretion disc at a distance from the accretor that depends on both the tilt angle and the precession phase. \citet{Kimura2020PASJ..tmp..156K} therefore carried out 1-dimensional disc instability simulations in which mass is added in this time- and radius-dependent fashion. The light curves produced by their simulations depend strongly on the adopted system parameters and tilt angles, and some of them show promise as models for the {\em IW~And} phenomenon.

Most of the system parameters investigated by \citet{Kimura2020PASJ..tmp..156K} lie in a range where ordinary dwarf nova eruptions would be expected for non-tilted discs. However, the highest accretion rates they consider would produce {\em steady} accretion in the absence of any disc tilt. Systems accreting at these rates would normally be NL variables. Figure~8 in \citet{Kimura2020PASJ..tmp..156K} shows the predicted light curves for such systems, from simulations designed to mimic tilt angle between $0^\circ$ and $15^\circ$. Remarkably, the light curves for moderate to high tilt angles exhibit large-amplitude variations on the precession period, reaching $\Delta \simeq 2 $~mag peak-to-peak. These variations occur because, with mass being deposited mostly at smaller radii, the outer disc is unable to maintain a stable thermal equilibrium and instead periodically drops into a low state. 
We propose that this provides a natural explanation for the super-orbital variations we observe in {\em V341~Ara}. 

What about the anomalously small ``superhump'' period deficit we have found in {\em V341~Ara}? The super-orbital variations in the NL simulations presented in \citet{Kimura2020PASJ..tmp..156K} are driven entirely by the behaviour of the outer disc. Moreover, \citet{Kimura2020PASJ..tmp..156K} note that a gap may form in the accretion disc near the radius of the stream-disc impact point. This raises the possibility that the outer disc may precess effectively as a thin {\rm ring}. This is interesting, because \citet{Montgomery2009ApJ...705..603M} predicts period deficits of $\epsilon_- \simeq 0.01$ for this case, with only a weak dependence on mass ratio, in line with what we observe in {\em V341~Ara}. However, Montgomery's calculations are for a ring located in the inner disc, and the period deficit scales roughly as $R_{ring}^{3/2}$ (\citealt{Montgomery2009ApJ...705..603M}; Eq 30), where $R_{ring}$ is the radius of the ring. If this is the explanation for the small period deficit, the outer disc in {\em V341~Ara} must either be truncated well inside the tidal limit or its precession rate must be driven by conditions at smaller radii (perhaps the stream-disc impact radius).

%https://ui.adsabs.harvard.edu/abs/1999dicb.conf...61P/abstract
%here is the actual paper:
%https://cbastro.org/wp-content/uploads/2018/01/permanent-superhumps-in-cataclysmic-v%ariables.pdf

%https://ui.adsabs.harvard.edu/abs/1993ApJS...86..235P/abstract
%%%https://iopscience.iop.org/article/10.1088/0004-637X/705/1/603
%https://onlinelibrary.wiley.com/doi/10.1111/j.1365-2966.2009.15252.x
%https://academic.oup.com/mnras/article-pdf/394/4/1897/4034058/mnras0394-1897.pdf
%https://academic.oup.com/mnras/article-pdf/398/4/2110/3059379/mnras0398-2110.pdf
%https://iopscience.iop.org/article/10.1086/311534/fulltext/985394.text.html (refs %therein)
%https://iopscience.iop.org/article/10.1086/516723/pdf
%https://ui.adsabs.harvard.edu/abs/2002MNRAS.335..247M/abstract
%https://ui.adsabs.harvard.edu/abs/2015ApJ...803...55T/abstract

\section{Summary} \label{sec:summary}
%\begin{itemize}[leftmargin=0.4cm, label=$\bullet$]
%\item Very nearby member of the thin disc population
%\item Surrounded by Halpha nebula into which it is drivinga bow-shock
%\item Luminosity & accretion rate
%\item High-precision orbital ephemeris
%\item K1 and K2 --> Mdot
%\item Halpha nebulosity is probably a nova-shell (Mshell ~ XXX)
%\item If so, it would have been a V ~ -2 - 0 mag nova around 1000 years ago (but %only 2 marginal candidates in historical records, apparently)
%\item Wind-mass loss rate from bow-shock = XXX --> provides crucial calibration %point for UV wind modelling
%\item large-amplitude, super-orbital, quasi-periodic variations on a time-scale %of 10-14 days; hard to explain as either disc instability or precession
%\end{itemize}

We have presented the first comprehensive multi-wavelength study of the recently discovered cataclysmic variable {\em V341~Ara}. This system is remarkable because it is the third closest non-magnetic nova-like known, is surrounded by a probable nova shell, and displays a clear bow-shock where its disk wind interacts with the nova shell. The main results and conclusions of our study are as follows.

\begin{enumerate}
    
    \item Super-orbital, quasi-periodic oscillations are clearly present in all photometric data sets. {\em TESS} data additionally revealing both the orbital period and the beat between these two signals. The amplitude of the super-orbital signal is large: $\simeq 0.5$~mag variations are typical, but much larger drops in flux are seen occasionally.
    
    \item The tilted disc instability model recently proposed by \citet{Kimura2020PASJ..tmp..156K} to explain slow, large-amplitude oscillations in a subset of dwarf novae may also account for the super-orbital variations of {\em V341~Ara}.

    \item Based on its position and proper motion, {\em V341~Ara} is a member of the Galactic thin disc population. 
    
    \item We confirm the $\simeq 3.65~\mathrm{hr}$ orbital period of the system and provide a high-precision orbital ephemeris.
    
    \item We clearly detect anti-phased absorption and emission line components in our spectroscopy, allowing us to establish the K-velocity of both the accreting primary ($K_1 =42 \pm 2$) and the irradiated face of the donor star ($K_{2,obs} = 47.4 \pm 0.6$). This allows us to estimate the center-of-mass orbital velocity of the donor, and thus determine the spectroscopic mass ratio of the system, $q = K_1/K_{2,true} = M_2/M_1 0.51 \pm 0.03(stat) ^{+0.4}_{-0.2}(sys)$. For the expected companion based in the evolutionary tracks, this would imply a WD mass of $M_1 \simeq 0.5 M_\odot$
    
    \item With the exception of the X-ray band, the SED of {\em V341~Ara} is broadly consistent with expectations based on the evolutionary tracks from KPB11, implying an accretion rate in the range $\dot{M}_{acc} \simeq 10^{-9}-10^{-8} M_\odot~yr^{-1}$.
    
    \item {\em V341~Ara}'s X-ray spectrum is composed of at least two components. If the brighter soft component is modelled as a blackbody, its temperature ($kT_{BB} = 80_{-20}^{+40}$~eV, i.e. $T_{BB} \simeq 9 ^{+4}_{-2} \times 10^5$~K) is significantly higher -- and its luminosity ($\log{L_{BB} [{\rm ergs~s^{-1}}]} \simeq 32.9$) lower -- than expected for an optically thick boundary layer (BL). The fainter hard component might be associated with shocks in the disc wind.
    
    \item The large-scale nebulosity surrounding {\em V341~Ara} is thought to have been produced during a nova eruption $\simeq 1000$~yrs ago. Near maximum, it was probably one of the 5 brightest stars in the sky. We tentatively suggest that this eruption might be associated with a "guest star" recorded by Chinese astronomers in AD 1240. This would make {\em V341~Ara} the fourth oldest recovered nova.
    
    \item We have redetermined the stand-off distance of the bow-shock, $R_{stand-off} \simeq 2 \times 10^{16} {\rm cm}$. This, in turn, allows us to constrain the mass-loss rate of the system's disc wind: $\dot{M}_{wind} \lesssim 10^{-10} M_\odot~yr^{-1}$.
\end{enumerate}

{\em V341~Ara} is expected to display strong, blue-shifted absorption lines in its far-UV spectrum. Modelling these lines is the standard way to determine the mass-loss rates of CV disc winds, so {\em V341~Ara} provides us with a rare opportunity to directly test this approach. Far-UV spectroscopic observations are therefore strongly encouraged.

\section*{Acknowledgements}

We would like to thank the anonymous referee for a careful and helpful feedback. We are very grateful to Dave Green for his comments on the possible identification of V341~Ara with the guest star of AD~1240, and, in particular, for pointing out that V341~Ara would not have been visible from Hangzhou, the capital at the time. We are also thankful to Ken Freeman for useful discussions regarding the WiFeS observations. We thank Simone Scaringi and Jamie Court for useful discussions regarding the mechanism driving the long term variability.

NCS \& CK acknowledge support by the Science and Technology Facilities Council (STFC), and from STFC grant ST/M001326/1. DMH acknowledges financial assistance from the South African National  Research  Foundation (NRF) and the South African Astronomical Observatory. NPMK work was supported by the UK Space Agency. AAP acknowledges support of the STFC consolidated grant ST/S000488/1 and from Australian Research Council grant DP150104129. DA Acknowledges support from the Royal Society. JVHS acknowledges support from a STFC grant ST/R000824/1. DRAW was supported by the Oxford Centre for Astrophysical Surveys, which is funded through generous support from the Hintze Family Charitable Foundation. DJJ acknowledges support from National Science Foundation award AST-1440254. DAHB acknowledges research support from the South African National Research Foundation.

This research made use of {\sc astropy}, a community-developed core Python package for Astronomy \citep{Astropy2018AJ....156..123A} and {\sc matplotlib} \citep{matplotlib2007CSE.....9...90H}.

Some of the observations were obtained with SALT under programme 2016-2-LSP-001. This work has made use of data from the European Space Agency (ESA) mission Gaia (https://www.cosmos.esa.int/gaia), processed by the Gaia Data Processing and Analysis Consortium (DPAC, https://www.cosmos.esa.int/web/gaia/dpac/consortium). Funding for the DPAC has been provided by national institutions, in particular the institutions participating in the Gaia Multilateral Agreement.
%%%%%%%%%%%%%%%%%%%%%%%%%%%%%%%%%%%%%%%%%%%%%%%%%%

\section*{Affiliations}
\noindent
{\it
% List of institutions
%$^{1}$Royal Astronomical Society, Burlington House, Piccadilly, London W1J 0BQ, UK\\
$^{1}$Department of Physics \& Astronomy. University of Southampton, Southampton SO17 1BJ, UK.
\\		
$^2$Instituto de Astrof\'isica de Canarias (IAC), E-38205 La Laguna, Tenerife, Spain.				\\
$^3$Departamento de Astrof\'isica, Universidad de La Laguna (ULL), E-38206 La Laguna, Tenerife, Spain.				\\
$^{4}$Instituto Argentino de Radioastronom\'ia (CCT La Plata, CONICET; CICPBA; UNLP), C.C.5, (1894) Villa Elisa, Buenos Aires, Argentina.\\
$^5$SUPA School of Physics \& Astronomy, University of St Andrews, North Haugh, St Andrews KY16 9SS, UK.     \\
$^6$Department of Physics and Astronomy, University of Leicester, University Road, Leicester LE1 7RH, UK.\\
$^7$South African Astronomical Observatory, PO Box 9, Observatory 7935, Cape Town, South Africa.				\\
$^8$Department of Astronomy, University of Cape Town, Private Bag X3, Rondebosch 7701, South Africa.			\\	
$^9$ Department of Physics, University of Oxford, Denys Wilkinson Building, Keble Road, Oxford, OX1 3RH, UK. \\
$^{10}$ Center for Astrophysics \textbar \ Harvard \& Smithsonian, 60 Garden St, Cambridge, MA 02138, USA. \\
$^{11}$  Southern African Large Telescope, P.O. Box 9, Observatory, 7935, Cape Town, South Africa.\\
$^{12}$  Mullard Space Science Laboratory/University College London, Holmbury St. Mary, Dorking, Surrey, RH5 6NT, UK.\\
$^{13}$  Department of Physics, Lehigh University, 16 Memorial Drive East, Bethlehem, PA, 18015, USA.\\
$^{14}$  Department of Physics \& Astronomy, Vanderbilt University, 6301 Stevenson Center Ln., Nashville, TN 37235, USA.\\
$^{15}$  Jodrell Bank Centre for Astrophysics, School of Physics and Astronomy, The University of Manchester, Manchester, M13 9PL, UK.\\
$^{16}$Research School of Astronomy \& Astrophysics, Australian National University, Canberra, ACT 2611, Australia. 
}

\section*{Data availability}

The data underlying this article is publicly available in: {\em ASAS-SN} \url{https://asas-sn.osu.edu}; {TESS} \url{http://archive.stsci.edu/tess/}; {\em Swift} \url{https://www.swift.ac.uk/archive/index.php}; {\em Chiron} and {\em RC Spectrograph} \url{http://archive1.dm.noao.edu/}; {\em WIFI} \url{http://archive.eso.org/cms.html}. Data from {\em KELT} is available in Zenodo, at \url{https://doi.org/10.5281/zenodo.3982165}. {\em MeerKAT} data is subject to the standard data access policy of the South African Radio Astronomy Observatory. Remaining data from {\em SALT}, {\em SpUpNIC} and {\em WiFeS} will be shared on reasonable request to the corresponding author.
%\newpage
%~
%\newpage



\bibliographystyle{mnras}
\bibliography{My_papers_bio} % if your bibtex file is called example.bib

\begin{thebibliography}{}
\makeatletter
\relax
\def\mn@urlcharsother{\let\do\@makeother \do\$\do\&\do\#\do\^\do\_\do\%\do\~}
\def\mn@doi{\begingroup\mn@urlcharsother \@ifnextchar [ {\mn@doi@}
  {\mn@doi@[]}}
\def\mn@doi@[#1]#2{\def\@tempa{#1}\ifx\@tempa\@empty \href
  {http://dx.doi.org/#2} {doi:#2}\else \href {http://dx.doi.org/#2} {#1}\fi
  \endgroup}
\def\mn@eprint#1#2{\mn@eprint@#1:#2::\@nil}
\def\mn@eprint@arXiv#1{\href {http://arxiv.org/abs/#1} {{\tt arXiv:#1}}}
\def\mn@eprint@dblp#1{\href {http://dblp.uni-trier.de/rec/bibtex/#1.xml}
  {dblp:#1}}
\def\mn@eprint@#1:#2:#3:#4\@nil{\def\@tempa {#1}\def\@tempb {#2}\def\@tempc
  {#3}\ifx \@tempc \@empty \let \@tempc \@tempb \let \@tempb \@tempa \fi \ifx
  \@tempb \@empty \def\@tempb {arXiv}\fi \@ifundefined
  {mn@eprint@\@tempb}{\@tempb:\@tempc}{\expandafter \expandafter \csname
  mn@eprint@\@tempb\endcsname \expandafter{\@tempc}}}

\bibitem[\protect\citeauthoryear{{Aranzana}, {Scaringi}, {K{\"o}rding},
  {Dhillon}  \& {Coppejans}}{{Aranzana}
  et~al.}{2018}]{Aranzana2018MNRAS.481.2140A}
{Aranzana} E.,  {Scaringi} S.,  {K{\"o}rding} E.,  {Dhillon} V.~S.,
  {Coppejans} D.~L.,  2018, \mn@doi [\mnras] {10.1093/mnras/sty2367}, \href
  {https://ui.adsabs.harvard.edu/abs/2018MNRAS.481.2140A} {481, 2140}

\bibitem[\protect\citeauthoryear{{Astropy Collaboration} et~al.,}{{Astropy
  Collaboration} et~al.}{2018}]{Astropy2018AJ....156..123A}
{Astropy Collaboration} et~al., 2018, \mn@doi [\aj] {10.3847/1538-3881/aabc4f},
  \href {https://ui.adsabs.harvard.edu/abs/2018AJ....156..123A} {156, 123}

\bibitem[\protect\citeauthoryear{{Balsara}, {Fisker}, {Godon}  \&
  {Sion}}{{Balsara} et~al.}{2009}]{Balsara2009ApJ...702.1536B}
{Balsara} D.~S.,  {Fisker} J.~L.,  {Godon} P.,   {Sion} E.~M.,  2009, \mn@doi
  [\apj] {10.1088/0004-637X/702/2/1536}, \href
  {https://ui.adsabs.harvard.edu/abs/2009ApJ...702.1536B} {702, 1536}

\bibitem[\protect\citeauthoryear{{Berdnikov} \& {Szabados}}{{Berdnikov} \&
  {Szabados}}{1998}]{Berdnikov1998AcA....48..763B}
{Berdnikov} L.~N.,  {Szabados} L.,  1998, \actaa, \href
  {https://ui.adsabs.harvard.edu/abs/1998AcA....48..763B} {48, 763}

\bibitem[\protect\citeauthoryear{{Beuermann} \& {Thomas}}{{Beuermann} \&
  {Thomas}}{1990}]{Beuermann+1990AA...230..326B}
{Beuermann} K.,  {Thomas} H.~C.,  1990, \aap, \href
  {https://ui.adsabs.harvard.edu/abs/1990A&A...230..326B} {230, 326}

\bibitem[\protect\citeauthoryear{{Beuermann}, {Stasiewski}  \&
  {Schwope}}{{Beuermann} et~al.}{1992}]{Beuermann+1992AA...256..433B}
{Beuermann} K.,  {Stasiewski} U.,   {Schwope} A.~D.,  1992, \aap, \href
  {https://ui.adsabs.harvard.edu/abs/1992A&A...256..433B} {256, 433}

\bibitem[\protect\citeauthoryear{{Bond} \& {Miszalski}}{{Bond} \&
  {Miszalski}}{2018}]{BondMiszalski2018PASP..130i4201B}
{Bond} H.~E.,  {Miszalski} B.,  2018, \mn@doi [\pasp]
  {10.1088/1538-3873/aace3e}, \href
  {https://ui.adsabs.harvard.edu/abs/2018PASP..130i4201B} {130, 094201}

\bibitem[\protect\citeauthoryear{{Bovy}}{{Bovy}}{2015}]{Galpy2015ApJS..216...29B}
{Bovy} J.,  2015, \mn@doi [\apjs] {10.1088/0067-0049/216/2/29}, \href
  {https://ui.adsabs.harvard.edu/abs/2015ApJS..216...29B} {216, 29}

\bibitem[\protect\citeauthoryear{{Burrows} et~al.,}{{Burrows}
  et~al.}{2005}]{XRT2005SSRv..120..165B}
{Burrows} D.~N.,  et~al., 2005, \mn@doi [\ssr] {10.1007/s11214-005-5097-2},
  \href {https://ui.adsabs.harvard.edu/abs/2005SSRv..120..165B} {120, 165}

\bibitem[\protect\citeauthoryear{{Clark} \& {Stephenson}}{{Clark} \&
  {Stephenson}}{1976}]{Clark1976QJRAS..17..290C}
{Clark} D.~H.,  {Stephenson} F.~R.,  1976, \qjras, \href
  {https://ui.adsabs.harvard.edu/abs/1976QJRAS..17..290C} {17, 290}

\bibitem[\protect\citeauthoryear{{Cohen} \& {Rosenthal}}{{Cohen} \&
  {Rosenthal}}{1983}]{Cohen1983ApJ...268..689C}
{Cohen} J.~G.,  {Rosenthal} A.~J.,  1983, \mn@doi [\apj] {10.1086/160990},
  \href {https://ui.adsabs.harvard.edu/abs/1983ApJ...268..689C} {268, 689}

\bibitem[\protect\citeauthoryear{{Coppejans} \& {Knigge}}{{Coppejans} \&
  {Knigge}}{2020}]{CoppejansKnigge2020arXiv200305953C}
{Coppejans} D.,  {Knigge} C.,  2020, arXiv e-prints, \href
  {https://ui.adsabs.harvard.edu/abs/2020arXiv200305953C} {p. arXiv:2003.05953}

\bibitem[\protect\citeauthoryear{{Crause} et~al.,}{{Crause}
  et~al.}{2014}]{SALT-HRS2014SPIE.9147E..6TC}
{Crause} L.~A.,  et~al., 2014, {Performance of the Southern African Large
  Telescope (SALT) High Resolution Spectrograph (HRS)}.
p. 91476T, \mn@doi{10.1117/12.2055635}

\bibitem[\protect\citeauthoryear{{Crause} et~al.,}{{Crause}
  et~al.}{2016}]{SpUpNIC2016SPIE.9908E..27C}
{Crause} L.~A.,  et~al., 2016, {SpUpNIC (Spectrograph Upgrade: Newly Improved
  Cassegrain) on the South African Astronomical Observatory's 74-inch
  telescope}.
p. 990827, \mn@doi{10.1117/12.2230818}

\bibitem[\protect\citeauthoryear{{Cutri} \& {et al.}}{{Cutri} \& {et
  al.}}{2014}]{WISE2014yCat.2328....0C}
{Cutri} R.~M.,  {et al.} 2014, VizieR Online Data Catalog, \href
  {https://ui.adsabs.harvard.edu/abs/2014yCat.2328....0C} {p. II/328}

\bibitem[\protect\citeauthoryear{{Dhillon}}{{Dhillon}}{1990}]{Dhillon_1990PhDT}
{Dhillon} V.~S.,  1990, PhD thesis, -

\bibitem[\protect\citeauthoryear{{Dhillon}, {Smith}  \& {Marsh}}{{Dhillon}
  et~al.}{2013}]{Dhillon2013MNRAS.428.3559D}
{Dhillon} V.~S.,  {Smith} D.~A.,   {Marsh} T.~R.,  2013, \mn@doi [\mnras]
  {10.1093/mnras/sts294}, \href
  {https://ui.adsabs.harvard.edu/abs/2013MNRAS.428.3559D} {428, 3559}

\bibitem[\protect\citeauthoryear{{Dopita}}{{Dopita}}{1977}]{Dopita1977ApJS...33..437D}
{Dopita} M.~A.,  1977, \mn@doi [\apjs] {10.1086/190435}, \href
  {https://ui.adsabs.harvard.edu/abs/1977ApJS...33..437D} {33, 437}

\bibitem[\protect\citeauthoryear{{Dopita}, {Hart}, {McGregor}, {Oates},
  {Bloxham}  \& {Jones}}{{Dopita} et~al.}{2007}]{WiFeS2007Ap&SS.310..255D}
{Dopita} M.,  {Hart} J.,  {McGregor} P.,  {Oates} P.,  {Bloxham} G.,   {Jones}
  D.,  2007, \mn@doi [\apss] {10.1007/s10509-007-9510-z}, \href
  {https://ui.adsabs.harvard.edu/abs/2007Ap&SS.310..255D} {310, 255}

\bibitem[\protect\citeauthoryear{{Downes}, {Duerbeck}  \& {Delahodde}}{{Downes}
  et~al.}{2001}]{Downes2001JAD.....7....6D}
{Downes} R.~A.,  {Duerbeck} H.~W.,   {Delahodde} C.~E.,  2001, Journal of
  Astronomical Data, \href
  {https://ui.adsabs.harvard.edu/abs/2001JAD.....7....6D} {7, 6}

\bibitem[\protect\citeauthoryear{{Duerbeck}}{{Duerbeck}}{2003}]{Duerbeck2003ASPC..292..323D}
{Duerbeck} H.~W.,  2003, {Long-term evolution of nova shell and postnova
  luminosities, and the problem of nova recurrence times}.
p.~323

\bibitem[\protect\citeauthoryear{ESA}{ESA}{1997}]{Hipparcos1997ESASP1200.....E}
ESA ed. 1997, {The HIPPARCOS and TYCHO catalogues. Astrometric and photometric
  star catalogues derived from the ESA HIPPARCOS Space Astrometry Mission}  ESA
  Special Publication Vol. 1200

\bibitem[\protect\citeauthoryear{{Eastman}, {Siverd}  \& {Gaudi}}{{Eastman}
  et~al.}{2010}]{Eastman_time_acccuracy2010PASP..122..935E}
{Eastman} J.,  {Siverd} R.,   {Gaudi} B.~S.,  2010, \mn@doi [\pasp]
  {10.1086/655938}, \href
  {https://ui.adsabs.harvard.edu/abs/2010PASP..122..935E} {122, 935}

\bibitem[\protect\citeauthoryear{{Erben} et~al.,}{{Erben}
  et~al.}{2005}]{THELI2005AN....326..432E}
{Erben} T.,  et~al., 2005, \mn@doi [Astronomische Nachrichten]
  {10.1002/asna.200510396}, \href
  {https://ui.adsabs.harvard.edu/abs/2005AN....326..432E} {326, 432}

\bibitem[\protect\citeauthoryear{{Fender} et~al.,}{{Fender}
  et~al.}{2017}]{ThunderKAT2017arXiv171104132F}
{Fender} R.,  et~al., 2017, arXiv e-prints, \href
  {https://ui.adsabs.harvard.edu/abs/2017arXiv171104132F} {p. arXiv:1711.04132}

\bibitem[\protect\citeauthoryear{{Foreman-Mackey}, {Hogg}, {Lang}  \&
  {Goodman}}{{Foreman-Mackey} et~al.}{2013}]{emcee2013PASP..125..306F}
{Foreman-Mackey} D.,  {Hogg} D.~W.,  {Lang} D.,   {Goodman} J.,  2013, \mn@doi
  [\pasp] {10.1086/670067}, \href
  {https://ui.adsabs.harvard.edu/abs/2013PASP..125..306F} {125, 306}

\bibitem[\protect\citeauthoryear{{Frew}}{{Frew}}{2008}]{Frew2008PhDT.......109F}
{Frew} D.~J.,  2008, PhD thesis, Department of Physics, Macquarie University,
  NSW 2109, Australia

\bibitem[\protect\citeauthoryear{{Gaia Collaboration} et~al.,}{{Gaia
  Collaboration} et~al.}{2018}]{GAIADR22018AA...616A...1G}
{Gaia Collaboration} et~al., 2018, \mn@doi [\aap]
  {10.1051/0004-6361/201833051}, \href
  {https://ui.adsabs.harvard.edu/abs/2018A&A...616A...1G} {616, A1}

\bibitem[\protect\citeauthoryear{{Gaustad}, {McCullough}, {Rosing}  \& {Van
  Buren}}{{Gaustad} et~al.}{2001}]{Gaustad+2001PASP..113.1326G}
{Gaustad} J.~E.,  {McCullough} P.~R.,  {Rosing} W.,   {Van Buren} D.,  2001,
  \mn@doi [\pasp] {10.1086/323969}, \href
  {https://ui.adsabs.harvard.edu/abs/2001PASP..113.1326G} {113, 1326}

\bibitem[\protect\citeauthoryear{{G{\"o}ttgens} et~al.,}{{G{\"o}ttgens}
  et~al.}{2019}]{Gottgens2019AA...626A..69G}
{G{\"o}ttgens} F.,  et~al., 2019, \mn@doi [\aap] {10.1051/0004-6361/201935221},
  \href {https://ui.adsabs.harvard.edu/abs/2019A&A...626A..69G} {626, A69}

\bibitem[\protect\citeauthoryear{{Green} \& {Stephenson}}{{Green} \&
  {Stephenson}}{2003}]{Green2003LNP...598....7G}
{Green} D.~A.,  {Stephenson} F.~R.,  2003, {Historical Supernovae}.
pp 7--19, \mn@doi{10.1007/3-540-45863-8_2}

\bibitem[\protect\citeauthoryear{{Guerrero} et~al.,}{{Guerrero}
  et~al.}{2018}]{Guerrero+2018ApJ...857...80G}
{Guerrero} M.~A.,  et~al., 2018, \mn@doi [\apj] {10.3847/1538-4357/aab669},
  \href {https://ui.adsabs.harvard.edu/abs/2018ApJ...857...80G} {857, 80}

\bibitem[\protect\citeauthoryear{{Harvey}, {Skillman}, {Patterson}  \&
  {Ringwald}}{{Harvey} et~al.}{1995}]{Harvey1995PASP..107..551H}
{Harvey} D.,  {Skillman} D.~R.,  {Patterson} J.,   {Ringwald} F.~A.,  1995,
  \mn@doi [\pasp] {10.1086/133591}, \href
  {https://ui.adsabs.harvard.edu/abs/1995PASP..107..551H} {107, 551}

\bibitem[\protect\citeauthoryear{{Henden}, {Templeton}, {Terrell}, {Smith},
  {Levine}  \& {Welch}}{{Henden} et~al.}{2016}]{APASS_Cat2016yCat.2336....0H}
{Henden} A.~A.,  {Templeton} M.,  {Terrell} D.,  {Smith} T.~C.,  {Levine} S.,
  {Welch} D.,  2016, VizieR Online Data Catalog, \href
  {https://ui.adsabs.harvard.edu/abs/2016yCat.2336....0H} {p. II/336}

\bibitem[\protect\citeauthoryear{{Hern{\'a}ndez Santisteban}
  et~al.,}{{Hern{\'a}ndez Santisteban}
  et~al.}{2019}]{HernandezSantisteban2019MNRAS.486.2631H}
{Hern{\'a}ndez Santisteban} J.~V.,  et~al., 2019, \mn@doi [\mnras]
  {10.1093/mnras/stz798}, \href
  {https://ui.adsabs.harvard.edu/abs/2019MNRAS.486.2631H} {486, 2631}

\bibitem[\protect\citeauthoryear{{Hertfelder}, {Kley}, {Suleimanov}  \&
  {Werner}}{{Hertfelder} et~al.}{2013}]{Hertfelder2013AA...560A..56H}
{Hertfelder} M.,  {Kley} W.,  {Suleimanov} V.,   {Werner} K.,  2013, \mn@doi
  [\aap] {10.1051/0004-6361/201322542}, \href
  {https://ui.adsabs.harvard.edu/abs/2013A&A...560A..56H} {560, A56}

\bibitem[\protect\citeauthoryear{{Hewitt} et~al.,}{{Hewitt}
  et~al.}{2020}]{2020MNRAS.496.2542H}
{Hewitt} D.~M.,  et~al., 2020, \mn@doi [\mnras] {10.1093/mnras/staa1747}, \href
  {https://ui.adsabs.harvard.edu/abs/2020MNRAS.496.2542H} {496, 2542}

\bibitem[\protect\citeauthoryear{{Hjellming}}{{Hjellming}}{1989}]{Hjellming1989PhDT.........7H}
{Hjellming} M.~S.,  1989, PhD thesis, Illinois Univ. at Urbana-Champaign,
  Savoy.

\bibitem[\protect\citeauthoryear{{Hoffmann}, {Vogt}  \& {Protte}}{{Hoffmann}
  et~al.}{2020}]{Hoffmann2020AN....341...79H}
{Hoffmann} S.~M.,  {Vogt} N.,   {Protte} P.,  2020, \mn@doi [Astronomische
  Nachrichten] {10.1002/asna.202013682}, \href
  {https://ui.adsabs.harvard.edu/abs/2020AN....341...79H} {341, 79}

\bibitem[\protect\citeauthoryear{{Hoffmeister}}{{Hoffmeister}}{1956}]{Hoffmeister1956}
{Hoffmeister} C.,  1956, Veroeffentlichungen der Sternwarte Sonneberg, \href
  {https://ui.adsabs.harvard.edu/abs/1956VeSon...3....1H} {3, 1}

\bibitem[\protect\citeauthoryear{{Hollis}, {Oliversen}, {Wagner}  \&
  {Feibelman}}{{Hollis} et~al.}{1992}]{Hollis+1992ApJ...393..217H}
{Hollis} J.~M.,  {Oliversen} R.~J.,  {Wagner} R.~M.,   {Feibelman} W.~A.,
  1992, \mn@doi [\apj] {10.1086/171499}, \href
  {https://ui.adsabs.harvard.edu/abs/1992ApJ...393..217H} {393, 217}

\bibitem[\protect\citeauthoryear{{Hunter}}{{Hunter}}{2007}]{matplotlib2007CSE.....9...90H}
{Hunter} J.~D.,  2007, \mn@doi [Computing in Science and Engineering]
  {10.1109/MCSE.2007.55}, \href
  {https://ui.adsabs.harvard.edu/abs/2007CSE.....9...90H} {9, 90}

\bibitem[\protect\citeauthoryear{{Jenkins} et~al.,}{{Jenkins}
  et~al.}{2016}]{TESSpipeline2016SPIE.9913E..3EJ}
{Jenkins} J.~M.,  et~al., 2016, {The TESS science processing operations
  center}.
p. 99133E, \mn@doi{10.1117/12.2233418}

\bibitem[\protect\citeauthoryear{{Jonas}}{{Jonas}}{2009}]{MeerKAT2009IEEEP..97.1522J}
{Jonas} J.~L.,  2009, \mn@doi [IEEE Proceedings] {10.1109/JPROC.2009.2020713},
  \href {https://ui.adsabs.harvard.edu/abs/2009IEEEP..97.1522J} {97, 1522}

\bibitem[\protect\citeauthoryear{{Kato}}{{Kato}}{2019}]{Kato2019PASJ...71...20K}
{Kato} T.,  2019, \mn@doi [\pasj] {10.1093/pasj/psy138}, \href
  {https://ui.adsabs.harvard.edu/abs/2019PASJ...71...20K} {71, 20}

\bibitem[\protect\citeauthoryear{{Kato}, {Saio}, {Hachisu}  \& {Nomoto}}{{Kato}
  et~al.}{2014}]{Kato+2014ApJ...793..136K}
{Kato} M.,  {Saio} H.,  {Hachisu} I.,   {Nomoto} K.,  2014, \mn@doi [\apj]
  {10.1088/0004-637X/793/2/136}, \href
  {https://ui.adsabs.harvard.edu/abs/2014ApJ...793..136K} {793, 136}

\bibitem[\protect\citeauthoryear{{Kimura}, {Osaki}, {Kato}  \&
  {Mineshige}}{{Kimura} et~al.}{2020}]{Kimura2020PASJ..tmp..156K}
{Kimura} M.,  {Osaki} Y.,  {Kato} T.,   {Mineshige} S.,  2020, \mn@doi [\pasj]
  {10.1093/pasj/psz144}, \href
  {https://ui.adsabs.harvard.edu/abs/2020PASJ..tmp..156K} {}

\bibitem[\protect\citeauthoryear{{Kiraga}}{{Kiraga}}{2012}]{kiraga2012AcA....62...67K}
{Kiraga} M.,  2012, \actaa, \href
  {https://ui.adsabs.harvard.edu/abs/2012AcA....62...67K} {62, 67}

\bibitem[\protect\citeauthoryear{{Knigge}}{{Knigge}}{2006}]{Knigge2006}
{Knigge} C.,  2006, \mn@doi [\mnras] {10.1111/j.1365-2966.2006.11096.x}, \href
  {https://ui.adsabs.harvard.edu/abs/2006MNRAS.373..484K} {373, 484}

\bibitem[\protect\citeauthoryear{{Knigge}, {Woods}  \& {Drew}}{{Knigge}
  et~al.}{1995}]{Knigge1995MNRAS.273..225K}
{Knigge} C.,  {Woods} J.~A.,   {Drew} J.~E.,  1995, \mn@doi [\mnras]
  {10.1093/mnras/273.2.225}, \href
  {https://ui.adsabs.harvard.edu/abs/1995MNRAS.273..225K} {273, 225}

\bibitem[\protect\citeauthoryear{{Knigge}, {Long}, {Blair}  \& {Wade}}{{Knigge}
  et~al.}{1997}]{Knigge+1997ApJ...476..291K}
{Knigge} C.,  {Long} K.~S.,  {Blair} W.~P.,   {Wade} R.~A.,  1997, \mn@doi
  [\apj] {10.1086/303607}, \href
  {https://ui.adsabs.harvard.edu/abs/1997ApJ...476..291K} {476, 291}

\bibitem[\protect\citeauthoryear{{Knigge}, {Baraffe}  \& {Patterson}}{{Knigge}
  et~al.}{2011}]{Knigge+2011ApJS..194...28K}
{Knigge} C.,  {Baraffe} I.,   {Patterson} J.,  2011, \mn@doi [\apjs]
  {10.1088/0067-0049/194/2/28}, \href
  {https://ui.adsabs.harvard.edu/abs/2011ApJS..194...28K} {194, 28}

\bibitem[\protect\citeauthoryear{{Kobulnicky}, {Chick}  \&
  {Povich}}{{Kobulnicky} et~al.}{2018}]{Kobulnicky2+018ApJ...856...74K}
{Kobulnicky} H.~A.,  {Chick} W.~T.,   {Povich} M.~S.,  2018, \mn@doi [\apj]
  {10.3847/1538-4357/aab3e0}, \href
  {https://ui.adsabs.harvard.edu/abs/2018ApJ...856...74K} {856, 74}

\bibitem[\protect\citeauthoryear{{Kochanek} et~al.,}{{Kochanek}
  et~al.}{2017}]{Kochanek+2017PASP..129j4502K}
{Kochanek} C.~S.,  et~al., 2017, \mn@doi [\pasp] {10.1088/1538-3873/aa80d9},
  \href {https://ui.adsabs.harvard.edu/abs/2017PASP..129j4502K} {129, 104502}

\bibitem[\protect\citeauthoryear{{Kuin}}{{Kuin}}{2014}]{UVOTPY2014ascl.soft10004K}
{Kuin} P.,  2014, {UVOTPY: Swift UVOT grism data reduction} (\mn@eprint {ascl}
  {1410.004})

\bibitem[\protect\citeauthoryear{{Kuin} et~al.,}{{Kuin}
  et~al.}{2015}]{UVOTPY2015MNRAS.449.2514K}
{Kuin} N.~P.~M.,  et~al., 2015, \mn@doi [\mnras] {10.1093/mnras/stv408}, \href
  {https://ui.adsabs.harvard.edu/abs/2015MNRAS.449.2514K} {449, 2514}

\bibitem[\protect\citeauthoryear{{Lasota}}{{Lasota}}{2016}]{Lasota2016}
{Lasota} J.-P.,  2016, {Black Hole Accretion Discs}.
p.~1, \mn@doi{10.1007/978-3-662-52859-4_1}

\bibitem[\protect\citeauthoryear{{Leavitt} \& {Pickering}}{{Leavitt} \&
  {Pickering}}{1907}]{Leavitt+1907}
{Leavitt} H.~S.,  {Pickering} E.~C.,  1907, Harvard College Observatory
  Circular, \href {https://ui.adsabs.harvard.edu/abs/1907HarCi.130....1L} {130,
  1}

\bibitem[\protect\citeauthoryear{{Lomb}}{{Lomb}}{1976}]{Lomb1976}
{Lomb} N.~R.,  1976, \mn@doi [\apss] {10.1007/BF00648343}, \href
  {https://ui.adsabs.harvard.edu/abs/1976Ap&SS..39..447L} {39, 447}

\bibitem[\protect\citeauthoryear{{Long} \& {Knigge}}{{Long} \&
  {Knigge}}{2002}]{LongKnigge2002ApJ...579..725L}
{Long} K.~S.,  {Knigge} C.,  2002, \mn@doi [\apj] {10.1086/342879}, \href
  {https://ui.adsabs.harvard.edu/abs/2002ApJ...579..725L} {579, 725}

\bibitem[\protect\citeauthoryear{{Longair}}{{Longair}}{2011}]{Longair2011hea..book.....L}
{Longair} M.~S.,  2011, {High Energy Astrophysics}

\bibitem[\protect\citeauthoryear{{Luri} et~al.,}{{Luri}
  et~al.}{2018}]{Luri+2018}
{Luri} X.,  et~al., 2018, \mn@doi [\aap] {10.1051/0004-6361/201832964}, \href
  {https://ui.adsabs.harvard.edu/abs/2018A&A...616A...9L} {616, A9}

\bibitem[\protect\citeauthoryear{{Matthews}, {Knigge}, {Long}, {Sim}  \&
  {Higginbottom}}{{Matthews} et~al.}{2015}]{Matthews+2015MNRAS.450.3331M}
{Matthews} J.~H.,  {Knigge} C.,  {Long} K.~S.,  {Sim} S.~A.,   {Higginbottom}
  N.,  2015, \mn@doi [\mnras] {10.1093/mnras/stv867}, \href
  {https://ui.adsabs.harvard.edu/abs/2015MNRAS.450.3331M} {450, 3331}

\bibitem[\protect\citeauthoryear{{Mauche}}{{Mauche}}{2004}]{Mauche2004RMxAC..20..174M}
{Mauche} C.~W.,  2004, in {Tovmassian} G.,  {Sion} E.,  eds,  Revista Mexicana
  de Astronomia y Astrofisica Conference Series Vol. 20, Revista Mexicana de
  Astronomia y Astrofisica Conference Series. pp 174--175

\bibitem[\protect\citeauthoryear{{McMahon} et~al.,}{{McMahon}
  et~al.}{2019}]{VHSDR42019yCat.2359....0M}
{McMahon} et~al., 2019, VizieR Online Data Catalog, \href
  {https://ui.adsabs.harvard.edu/abs/2019yCat.2359....0M} {p. II/359}

\bibitem[\protect\citeauthoryear{{McMullin}, {Waters}, {Schiebel}, {Young}  \&
  {Golap}}{{McMullin} et~al.}{2007}]{CASA2007ASPC..376..127M}
{McMullin} J.~P.,  {Waters} B.,  {Schiebel} D.,  {Young} W.,   {Golap} K.,
  2007, {CASA Architecture and Applications}.
p.~127

\bibitem[\protect\citeauthoryear{{Meyer}, {van Marle}, {Kuiper}  \&
  {Kley}}{{Meyer} et~al.}{2016}]{Meyer2016MNRAS.459.1146M}
{Meyer} D.~M.~A.,  {van Marle} A.~J.,  {Kuiper} R.,   {Kley} W.,  2016, \mn@doi
  [\mnras] {10.1093/mnras/stw651}, \href
  {https://ui.adsabs.harvard.edu/abs/2016MNRAS.459.1146M} {459, 1146}

\bibitem[\protect\citeauthoryear{{Miszalski} et~al.,}{{Miszalski}
  et~al.}{2016}]{Miszalski2016MNRAS.456..633M}
{Miszalski} B.,  et~al., 2016, \mn@doi [\mnras] {10.1093/mnras/stv2689}, \href
  {https://ui.adsabs.harvard.edu/abs/2016MNRAS.456..633M} {456, 633}

\bibitem[\protect\citeauthoryear{{Montgomery}}{{Montgomery}}{2009}]{Montgomery2009ApJ...705..603M}
{Montgomery} M.~M.,  2009, \mn@doi [\apj] {10.1088/0004-637X/705/1/603}, \href
  {https://ui.adsabs.harvard.edu/abs/2009ApJ...705..603M} {705, 603}

\bibitem[\protect\citeauthoryear{{Mu{\~n}oz-Darias}, {Casares}  \&
  {Mart{\'\i}nez-Pais}}{{Mu{\~n}oz-Darias}
  et~al.}{2005}]{K-Correction_MD2005ApJ...635..502M}
{Mu{\~n}oz-Darias} T.,  {Casares} J.,   {Mart{\'\i}nez-Pais} I.~G.,  2005,
  \mn@doi [\apj] {10.1086/497420}, \href
  {https://ui.adsabs.harvard.edu/abs/2005ApJ...635..502M} {635, 502}

\bibitem[\protect\citeauthoryear{{Mukai}}{{Mukai}}{2017}]{Mukai2017PASP..129f2001M}
{Mukai} K.,  2017, \mn@doi [\pasp] {10.1088/1538-3873/aa6736}, \href
  {https://ui.adsabs.harvard.edu/abs/2017PASP..129f2001M} {129, 062001}

\bibitem[\protect\citeauthoryear{{Offringa}}{{Offringa}}{2010}]{AOFlagger2010ascl.soft10017O}
{Offringa} A.~R.,  2010, {AOFlagger: RFI Software} (\mn@eprint {ascl}
  {1010.017})

\bibitem[\protect\citeauthoryear{{Page} et~al.,}{{Page}
  et~al.}{2013}]{Page2013MNRAS.436.1684P}
{Page} M.~J.,  et~al., 2013, \mn@doi [\mnras] {10.1093/mnras/stt1689}, \href
  {https://ui.adsabs.harvard.edu/abs/2013MNRAS.436.1684P} {436, 1684}

\bibitem[\protect\citeauthoryear{{Patterson}}{{Patterson}}{1999}]{Patterson1999dicb.conf...61P}
{Patterson} J.,  1999, in {Mineshige} S.,  {Wheeler} J.~C.,  eds, Disk
  Instabilities in Close Binary Systems. p.~61

\bibitem[\protect\citeauthoryear{{Patterson} \& {Raymond}}{{Patterson} \&
  {Raymond}}{1985}]{Patterson1985}
{Patterson} J.,  {Raymond} J.~C.,  1985, \mn@doi [\apj] {10.1086/163188}, \href
  {https://ui.adsabs.harvard.edu/abs/1985ApJ...292..550P} {292, 550}

\bibitem[\protect\citeauthoryear{{Patterson} et~al.,}{{Patterson}
  et~al.}{2005}]{Patterson_SHs_2005PASP..117.1204P}
{Patterson} J.,  et~al., 2005, \mn@doi [\pasp] {10.1086/447771}, \href
  {https://ui.adsabs.harvard.edu/abs/2005PASP..117.1204P} {117, 1204}

\bibitem[\protect\citeauthoryear{{Patterson} et~al.,}{{Patterson}
  et~al.}{2013}]{Patterson2013MNRAS.434.1902P}
{Patterson} J.,  et~al., 2013, \mn@doi [\mnras] {10.1093/mnras/stt1085}, \href
  {https://ui.adsabs.harvard.edu/abs/2013MNRAS.434.1902P} {434, 1902}

\bibitem[\protect\citeauthoryear{{Pepper} et~al.,}{{Pepper}
  et~al.}{2007}]{Pepper+2007PASP..119..923P}
{Pepper} J.,  et~al., 2007, \mn@doi [\pasp] {10.1086/521836}, \href
  {https://ui.adsabs.harvard.edu/abs/2007PASP..119..923P} {119, 923}

\bibitem[\protect\citeauthoryear{{Pepper}, {Kuhn}, {Siverd}, {James}  \&
  {Stassun}}{{Pepper} et~al.}{2012}]{Pepper+2012PASP..124..230P}
{Pepper} J.,  {Kuhn} R.~B.,  {Siverd} R.,  {James} D.,   {Stassun} K.,  2012,
  \mn@doi [\pasp] {10.1086/665044}, \href
  {https://ui.adsabs.harvard.edu/abs/2012PASP..124..230P} {124, 230}

\bibitem[\protect\citeauthoryear{{Pepper}, {Stassun}  \& {Gaudi}}{{Pepper}
  et~al.}{2018}]{Pepper2018haex.bookE.128P}
{Pepper} J.,  {Stassun} K.~G.,   {Gaudi} B.~S.,  2018, {KELT: The Kilodegree
  Extremely Little Telescope, a Survey for Exoplanets Transiting Bright, Hot
  Stars}.
p.~128, \mn@doi{10.1007/978-3-319-55333-7_128}

\bibitem[\protect\citeauthoryear{{Politano}}{{Politano}}{1996}]{Politano1996}
{Politano} M.,  1996, \mn@doi [\apj] {10.1086/177423}, \href
  {https://ui.adsabs.harvard.edu/abs/1996ApJ...465..338P} {465, 338}

\bibitem[\protect\citeauthoryear{{Pratt}, {Mukai}, {Hassall}, {Naylor}  \&
  {Wood}}{{Pratt} et~al.}{2004}]{Pratt2004MNRAS.348L..49P}
{Pratt} G.~W.,  {Mukai} K.,  {Hassall} B.~J.~M.,  {Naylor} T.,   {Wood} J.~H.,
  2004, \mn@doi [\mnras] {10.1111/j.1365-2966.2004.07574.x}, \href
  {https://ui.adsabs.harvard.edu/abs/2004MNRAS.348L..49P} {348, L49}

\bibitem[\protect\citeauthoryear{{Pretorius}, {Knigge}, {O'Donoghue}, {Henry},
  {Gioia}  \& {Mullis}}{{Pretorius}
  et~al.}{2007}]{Pretorius+2007MNRAS.382.1279P}
{Pretorius} M.~L.,  {Knigge} C.,  {O'Donoghue} D.,  {Henry} J.~P.,  {Gioia}
  I.~M.,   {Mullis} C.~R.,  2007, \mn@doi [\mnras]
  {10.1111/j.1365-2966.2007.12461.x}, \href
  {https://ui.adsabs.harvard.edu/abs/2007MNRAS.382.1279P} {382, 1279}

\bibitem[\protect\citeauthoryear{{Prinja}, {Ringwald}, {Wade}  \&
  {Knigge}}{{Prinja} et~al.}{2000}]{Prinja2000MNRAS.312..316P}
{Prinja} R.~K.,  {Ringwald} F.~A.,  {Wade} R.~A.,   {Knigge} C.,  2000, \mn@doi
  [\mnras] {10.1046/j.1365-8711.2000.03111.x}, \href
  {https://ui.adsabs.harvard.edu/abs/2000MNRAS.312..316P} {312, 316}

\bibitem[\protect\citeauthoryear{{Ricker} et~al.,}{{Ricker}
  et~al.}{2015}]{TESS2015JATIS...1a4003R}
{Ricker} G.~R.,  et~al., 2015, \mn@doi [Journal of Astronomical Telescopes,
  Instruments, and Systems] {10.1117/1.JATIS.1.1.014003}, \href
  {https://ui.adsabs.harvard.edu/abs/2015JATIS...1a4003R} {1, 014003}

\bibitem[\protect\citeauthoryear{{Ritter}, {Politano}, {Livio}  \&
  {Webbink}}{{Ritter} et~al.}{1991}]{Ritter1991}
{Ritter} H.,  {Politano} M.,  {Livio} M.,   {Webbink} R.~F.,  1991, \mn@doi
  [\apj] {10.1086/170265}, \href
  {https://ui.adsabs.harvard.edu/abs/1991ApJ...376..177R} {376, 177}

\bibitem[\protect\citeauthoryear{{Rodr{\'\i}guez-Gil}
  et~al.,}{{Rodr{\'\i}guez-Gil}
  et~al.}{2007}]{Rodriguez-Gil2007MNRAS.377.1747R}
{Rodr{\'\i}guez-Gil} P.,  et~al., 2007, \mn@doi [\mnras]
  {10.1111/j.1365-2966.2007.11743.x}, \href
  {https://ui.adsabs.harvard.edu/abs/2007MNRAS.377.1747R} {377, 1747}

\bibitem[\protect\citeauthoryear{{Roming} et~al.,}{{Roming}
  et~al.}{2005}]{UVOT2005SSRv..120...95R}
{Roming} P. W.~A.,  et~al., 2005, \mn@doi [\ssr] {10.1007/s11214-005-5095-4},
  \href {https://ui.adsabs.harvard.edu/abs/2005SSRv..120...95R} {120, 95}

\bibitem[\protect\citeauthoryear{{Sahman}, {Dhillon}, {Knigge}  \&
  {Marsh}}{{Sahman} et~al.}{2015}]{Sahman2015MNRAS.451.2863S}
{Sahman} D.~I.,  {Dhillon} V.~S.,  {Knigge} C.,   {Marsh} T.~R.,  2015, \mn@doi
  [\mnras] {10.1093/mnras/stv1150}, \href
  {https://ui.adsabs.harvard.edu/abs/2015MNRAS.451.2863S} {451, 2863}

\bibitem[\protect\citeauthoryear{{Sahman}, {Dhillon}, {Littlefair}  \&
  {Hallinan}}{{Sahman} et~al.}{2018}]{Sahman2018MNRAS.477.4483S}
{Sahman} D.~I.,  {Dhillon} V.~S.,  {Littlefair} S.~P.,   {Hallinan} G.,  2018,
  \mn@doi [\mnras] {10.1093/mnras/sty950}, \href
  {https://ui.adsabs.harvard.edu/abs/2018MNRAS.477.4483S} {477, 4483}

\bibitem[\protect\citeauthoryear{{Samus}, {Pastukhova}  \& {Durlevich}}{{Samus}
  et~al.}{2007}]{Samus+2007PZ.....27....6S}
{Samus} N.~N.,  {Pastukhova} E.~N.,   {Durlevich} O.~V.,  2007, Peremennye
  Zvezdy, \href {https://ui.adsabs.harvard.edu/abs/2007PZ.....27....6S} {27, 6}

\bibitem[\protect\citeauthoryear{{Santamar{\'\i}a}, {Guerrero}, {Ramos-Larios},
  {Sabin}, {V{\'a}zquez}, {G{\'o}mez-Mu{\~n}oz}  \&
  {Toal{\'a}}}{{Santamar{\'\i}a} et~al.}{2019}]{Santamara2019MNRAS.483.3773S}
{Santamar{\'\i}a} E.,  {Guerrero} M.~A.,  {Ramos-Larios} G.,  {Sabin} L.,
  {V{\'a}zquez} R.,  {G{\'o}mez-Mu{\~n}oz} M.~A.,   {Toal{\'a}} J.~A.,  2019,
  \mn@doi [\mnras] {10.1093/mnras/sty3364}, \href
  {https://ui.adsabs.harvard.edu/abs/2019MNRAS.483.3773S} {483, 3773}

\bibitem[\protect\citeauthoryear{{Scargle}}{{Scargle}}{1982}]{Scargle1982}
{Scargle} J.~D.,  1982, \mn@doi [\apj] {10.1086/160554}, \href
  {https://ui.adsabs.harvard.edu/abs/1982ApJ...263..835S} {263, 835}

\bibitem[\protect\citeauthoryear{{Schirmer}}{{Schirmer}}{2013}]{THELI2013ApJS..209...21S}
{Schirmer} M.,  2013, \mn@doi [\apjs] {10.1088/0067-0049/209/2/21}, \href
  {https://ui.adsabs.harvard.edu/abs/2013ApJS..209...21S} {209, 21}

\bibitem[\protect\citeauthoryear{{Schlafly} \& {Finkbeiner}}{{Schlafly} \&
  {Finkbeiner}}{2011}]{SF_extinction2011ApJ...737..103S}
{Schlafly} E.~F.,  {Finkbeiner} D.~P.,  2011, \mn@doi [\apj]
  {10.1088/0004-637X/737/2/103}, \href
  {https://ui.adsabs.harvard.edu/abs/2011ApJ...737..103S} {737, 103}

\bibitem[\protect\citeauthoryear{{Schmidtobreick}, {Rodr{\'\i}guez-Gil}  \&
  {G{\"a}nsicke}}{{Schmidtobreick}
  et~al.}{2012}]{Schmidtobreick2012MmSAI..83..610S}
{Schmidtobreick} L.,  {Rodr{\'\i}guez-Gil} P.,   {G{\"a}nsicke} B.~T.,  2012,
  \memsai, \href {https://ui.adsabs.harvard.edu/abs/2012MmSAI..83..610S} {83,
  610}

\bibitem[\protect\citeauthoryear{{Schmidtobreick}, {Shara}, {Tappert}, {Bayo}
  \& {Ederoclite}}{{Schmidtobreick}
  et~al.}{2015}]{Schmidtobreick2015MNRAS.449.2215S}
{Schmidtobreick} L.,  {Shara} M.,  {Tappert} C.,  {Bayo} A.,   {Ederoclite} A.,
   2015, \mn@doi [\mnras] {10.1093/mnras/stv250}, \href
  {https://ui.adsabs.harvard.edu/abs/2015MNRAS.449.2215S} {449, 2215}

\bibitem[\protect\citeauthoryear{{Sedov}}{{Sedov}}{1959}]{Sedov1959sdmm.book.....S}
{Sedov} L.~I.,  1959, {Similarity and Dimensional Methods in Mechanics}

\bibitem[\protect\citeauthoryear{{Shakura} \& {Sunyaev}}{{Shakura} \&
  {Sunyaev}}{1973}]{ShakuraSunyaev1973}
{Shakura} N.~I.,  {Sunyaev} R.~A.,  1973, \aap, \href
  {https://ui.adsabs.harvard.edu/abs/1973A&A....24..337S} {500, 33}

\bibitem[\protect\citeauthoryear{{Shara} et~al.,}{{Shara}
  et~al.}{2007}]{Shara2007Natur.446..159S}
{Shara} M.~M.,  et~al., 2007, \mn@doi [\nat] {10.1038/nature05576}, \href
  {https://ui.adsabs.harvard.edu/abs/2007Natur.446..159S} {446, 159}

\bibitem[\protect\citeauthoryear{{Shara}, {Mizusawa}, {Wehinger}, {Zurek},
  {Martin}, {Neill}, {Forster}  \& {Seibert}}{{Shara}
  et~al.}{2012}]{Shara2012ApJ...758..121S}
{Shara} M.~M.,  {Mizusawa} T.,  {Wehinger} P.,  {Zurek} D.,  {Martin} C.~D.,
  {Neill} J.~D.,  {Forster} K.,   {Seibert} M.,  2012, \mn@doi [\apj]
  {10.1088/0004-637X/758/2/121}, \href
  {https://ui.adsabs.harvard.edu/abs/2012ApJ...758..121S} {758, 121}

\bibitem[\protect\citeauthoryear{{Shara}, {Drissen}, {Martin}, {Alarie}  \&
  {Stephenson}}{{Shara} et~al.}{2017}]{Shara+2017MNRAS.465..739S}
{Shara} M.~M.,  {Drissen} L.,  {Martin} T.,  {Alarie} A.,   {Stephenson} F.~R.,
   2017, \mn@doi [\mnras] {10.1093/mnras/stw2753}, \href
  {https://ui.adsabs.harvard.edu/abs/2017MNRAS.465..739S} {465, 739}

\bibitem[\protect\citeauthoryear{{Simonsen}}{{Simonsen}}{2011}]{Simonsen2011JAVSO..39...66S}
{Simonsen} M.,  2011, Journal of the American Association of Variable Star
  Observers (JAAVSO), \href
  {https://ui.adsabs.harvard.edu/abs/2011JAVSO..39...66S} {39, 66}

\bibitem[\protect\citeauthoryear{{Skrutskie} et~al.,}{{Skrutskie}
  et~al.}{2006}]{2MASS2006AJ....131.1163S}
{Skrutskie} M.~F.,  et~al., 2006, \mn@doi [\aj] {10.1086/498708}, \href
  {https://ui.adsabs.harvard.edu/abs/2006AJ....131.1163S} {131, 1163}

\bibitem[\protect\citeauthoryear{{Stephenson}}{{Stephenson}}{1971}]{Stephenson1971ApL.....9...81S}
{Stephenson} F.~R.,  1971, \aplett, \href
  {https://ui.adsabs.harvard.edu/abs/1971ApL.....9...81S} {9, 81}

\bibitem[\protect\citeauthoryear{{Stephenson}}{{Stephenson}}{1976}]{Stephenson1976QJRAS..17..121S}
{Stephenson} F.~R.,  1976, \qjras, \href
  {https://ui.adsabs.harvard.edu/abs/1976QJRAS..17..121S} {17, 121}

\bibitem[\protect\citeauthoryear{{Stephenson} \& {Green}}{{Stephenson} \&
  {Green}}{2009}]{Stephenson2009JHA....40...31S}
{Stephenson} F.~R.,  {Green} D.~A.,  2009, \mn@doi [Journal for the History of
  Astronomy] {10.1177/002182860904000104}, \href
  {https://ui.adsabs.harvard.edu/abs/2009JHA....40...31S} {40, 31}

\bibitem[\protect\citeauthoryear{{Szkody} et~al.,}{{Szkody}
  et~al.}{2013}]{Szkody2013PASP..125.1421S}
{Szkody} P.,  et~al., 2013, \mn@doi [\pasp] {10.1086/674170}, \href
  {https://ui.adsabs.harvard.edu/abs/2013PASP..125.1421S} {125, 1421}

\bibitem[\protect\citeauthoryear{{Tasse} et~al.,}{{Tasse}
  et~al.}{2018}]{DDFacet2018A&A...611A..87T}
{Tasse} C.,  et~al., 2018, \mn@doi [\aap] {10.1051/0004-6361/201731474}, \href
  {https://ui.adsabs.harvard.edu/abs/2018A&A...611A..87T} {611, A87}

\bibitem[\protect\citeauthoryear{{Taylor}}{{Taylor}}{1950}]{Taylor1950RSPSA.201..159T}
{Taylor} G.,  1950, \mn@doi [Proceedings of the Royal Society of London Series
  A] {10.1098/rspa.1950.0049}, \href
  {https://ui.adsabs.harvard.edu/abs/1950RSPSA.201..159T} {201, 159}

\bibitem[\protect\citeauthoryear{{Thorstensen}, {Ringwald}, {Wade}, {Schmidt}
  \& {Norsworthy}}{{Thorstensen} et~al.}{1991}]{Thorstensen1991AJ....102..272T}
{Thorstensen} J.~R.,  {Ringwald} F.~A.,  {Wade} R.~A.,  {Schmidt} G.~D.,
  {Norsworthy} J.~E.,  1991, \mn@doi [\aj] {10.1086/115874}, \href
  {https://ui.adsabs.harvard.edu/abs/1991AJ....102..272T} {102, 272}

\bibitem[\protect\citeauthoryear{{Tokovinin}, {Fischer}, {Bonati}, {Giguere},
  {Moore}, {Schwab}, {Spronck}  \& {Szymkowiak}}{{Tokovinin}
  et~al.}{2013}]{Chiron2013PASP..125.1336T}
{Tokovinin} A.,  {Fischer} D.~A.,  {Bonati} M.,  {Giguere} M.~J.,  {Moore} P.,
  {Schwab} C.,  {Spronck} J. F.~P.,   {Szymkowiak} A.,  2013, \mn@doi [\pasp]
  {10.1086/674012}, \href
  {https://ui.adsabs.harvard.edu/abs/2013PASP..125.1336T} {125, 1336}

\bibitem[\protect\citeauthoryear{{Townsley} \& {Bildsten}}{{Townsley} \&
  {Bildsten}}{2004}]{Townsley2004ApJ...600..390T}
{Townsley} D.~M.,  {Bildsten} L.,  2004, \mn@doi [\apj] {10.1086/379647}, \href
  {https://ui.adsabs.harvard.edu/abs/2004ApJ...600..390T} {600, 390}

\bibitem[\protect\citeauthoryear{{Tse-Tsung}}{{Tse-Tsung}}{1957}]{Tse-Tsung1957AZh....34..159T}
{Tse-Tsung} H.,  1957, \azh, \href
  {https://ui.adsabs.harvard.edu/abs/1957AZh....34..159T} {34, 159}

\bibitem[\protect\citeauthoryear{{VanderPlas}}{{VanderPlas}}{2018}]{VanderPlas2018ApJS..236...16V}
{VanderPlas} J.~T.,  2018, \mn@doi [\apjs] {10.3847/1538-4365/aab766}, \href
  {https://ui.adsabs.harvard.edu/abs/2018ApJS..236...16V} {236, 16}

\bibitem[\protect\citeauthoryear{{Vogt}, {Hoffmann}  \& {Tappert}}{{Vogt}
  et~al.}{2019}]{Vogt+2019AN....340..752V}
{Vogt} N.,  {Hoffmann} S.~M.,   {Tappert} C.,  2019, \mn@doi [Astronomische
  Nachrichten] {10.1002/asna.201913635}, \href
  {https://ui.adsabs.harvard.edu/abs/2019AN....340..752V} {340, 752}

\bibitem[\protect\citeauthoryear{{Wareing}, {Zijlstra}  \& {O'Brien}}{{Wareing}
  et~al.}{2007}]{Wareing+2007MNRAS.382.1233W}
{Wareing} C.~J.,  {Zijlstra} A.~A.,   {O'Brien} T.~J.,  2007, \mn@doi [\mnras]
  {10.1111/j.1365-2966.2007.12459.x}, \href
  {https://ui.adsabs.harvard.edu/abs/2007MNRAS.382.1233W} {382, 1233}

\bibitem[\protect\citeauthoryear{{Weaver}, {McCray}, {Castor}, {Shapiro}  \&
  {Moore}}{{Weaver} et~al.}{1977}]{Weaver+1977ApJ...218..377W}
{Weaver} R.,  {McCray} R.,  {Castor} J.,  {Shapiro} P.,   {Moore} R.,  1977,
  \mn@doi [\apj] {10.1086/155692}, \href
  {https://ui.adsabs.harvard.edu/abs/1977ApJ...218..377W} {218, 377}

\bibitem[\protect\citeauthoryear{{Williams}}{{Williams}}{1994}]{Williams1994ApJ...426..279W}
{Williams} R.~E.,  1994, \mn@doi [\apj] {10.1086/174062}, \href
  {https://ui.adsabs.harvard.edu/abs/1994ApJ...426..279W} {426, 279}

\bibitem[\protect\citeauthoryear{{Wolf}, {Bildsten}, {Brooks}  \&
  {Paxton}}{{Wolf} et~al.}{2013}]{Wolf+2013ApJ...777..136W}
{Wolf} W.~M.,  {Bildsten} L.,  {Brooks} J.,   {Paxton} B.,  2013, \mn@doi
  [\apj] {10.1088/0004-637X/777/2/136}, \href
  {https://ui.adsabs.harvard.edu/abs/2013ApJ...777..136W} {777, 136}

\bibitem[\protect\citeauthoryear{{Wolf} et~al.,}{{Wolf}
  et~al.}{2018}]{SkyMapperDR12018PASA...35...10W}
{Wolf} C.,  et~al., 2018, \mn@doi [\pasa] {10.1017/pasa.2018.5}, \href
  {https://ui.adsabs.harvard.edu/abs/2018PASA...35...10W} {35, e010}

\bibitem[\protect\citeauthoryear{{Wood}, {Thomas}  \& {Simpson}}{{Wood}
  et~al.}{2009}]{Wood2009MNRAS.398.2110W}
{Wood} M.~A.,  {Thomas} D.~M.,   {Simpson} J.~C.,  2009, \mn@doi [\mnras]
  {10.1111/j.1365-2966.2009.15252.x}, \href
  {https://ui.adsabs.harvard.edu/abs/2009MNRAS.398.2110W} {398, 2110}

\bibitem[\protect\citeauthoryear{{Yang}, {Park}, {Cho}  \& {Park}}{{Yang}
  et~al.}{2005}]{Yang2005AA...435..207Y}
{Yang} H.-J.,  {Park} M.-G.,  {Cho} S.-H.,   {Park} C.,  2005, \mn@doi [\aap]
  {10.1051/0004-6361:20042455}, \href
  {https://ui.adsabs.harvard.edu/abs/2005A&A...435..207Y} {435, 207}

\bibitem[\protect\citeauthoryear{{Yaron}, {Prialnik}, {Shara}  \&
  {Kovetz}}{{Yaron} et~al.}{2005}]{Yaron2005ApJ...623..398Y}
{Yaron} O.,  {Prialnik} D.,  {Shara} M.~M.,   {Kovetz} A.,  2005, \mn@doi
  [\apj] {10.1086/428435}, \href
  {https://ui.adsabs.harvard.edu/abs/2005ApJ...623..398Y} {623, 398}

\bibitem[\protect\citeauthoryear{{Zechmeister} \& {K{\"u}rster}}{{Zechmeister}
  \& {K{\"u}rster}}{2009}]{Zechmeister_LS2009}
{Zechmeister} M.,  {K{\"u}rster} M.,  2009, \mn@doi [\aap]
  {10.1051/0004-6361:200811296}, \href
  {https://ui.adsabs.harvard.edu/abs/2009A&A...496..577Z} {496, 577}

\bibitem[\protect\citeauthoryear{{Zhu}, {Tian}, {Li}  \& {Zhang}}{{Zhu}
  et~al.}{2017}]{Ext_to_nH2017MNRAS.471.3494Z}
{Zhu} H.,  {Tian} W.,  {Li} A.,   {Zhang} M.,  2017, \mn@doi [\mnras]
  {10.1093/mnras/stx1580}, \href
  {https://ui.adsabs.harvard.edu/abs/2017MNRAS.471.3494Z} {471, 3494}

\bibitem[\protect\citeauthoryear{{Zorotovic}, {Schreiber}  \&
  {G{\"a}nsicke}}{{Zorotovic} et~al.}{2011}]{zorotovic2011}
{Zorotovic} M.,  {Schreiber} M.~R.,   {G{\"a}nsicke} B.~T.,  2011, \mn@doi
  [\aap] {10.1051/0004-6361/201116626}, \href
  {https://ui.adsabs.harvard.edu/abs/2011A&A...536A..42Z} {536, A42}

\bibitem[\protect\citeauthoryear{{del Palacio}, {Bosch-Ramon}, {M{\"u}ller}  \&
  {Romero}}{{del Palacio} et~al.}{2018}]{delPalacio+2018AA...617A..13D}
{del Palacio} S.,  {Bosch-Ramon} V.,  {M{\"u}ller} A.~L.,   {Romero} G.~E.,
  2018, \mn@doi [\aap] {10.1051/0004-6361/201833321}, \href
  {https://ui.adsabs.harvard.edu/abs/2018A&A...617A..13D} {617, A13}

\bibitem[\protect\citeauthoryear{{van Buren} \& {McCray}}{{van Buren} \&
  {McCray}}{1988}]{vanBuren1988ApJ...329L..93V}
{van Buren} D.,  {McCray} R.,  1988, \mn@doi [\apjl] {10.1086/185184}, \href
  {https://ui.adsabs.harvard.edu/abs/1988ApJ...329L..93V} {329, L93}

\bibitem[\protect\citeauthoryear{{van Dokkum}}{{van
  Dokkum}}{2001}]{LACOSMIC2001PASP..113.1420V}
{van Dokkum} P.~G.,  2001, \mn@doi [\pasp] {10.1086/323894}, \href
  {https://ui.adsabs.harvard.edu/abs/2001PASP..113.1420V} {113, 1420}

\makeatother
\end{thebibliography}

%\begin{thebibliography}{99}
%\bibliography{references.bib}

%\end{thebibliography}

%%%%%%%%%%%%%%%%%%%%%%%%%%%%%%%%%%%%%%%%%%%%%%%%%%

%%%%%%%%%%%%%%%%% APPENDICES %%%%%%%%%%%%%%%%%%%%%

\appendix
\newpage
\onecolumn
\section{List of observations}

\begin{table}
 \centering
  \caption{Spectroscopic observing log
  }
  \label{tab:sp log}
  \begin{tabular}{llllll}
  \hline
  Telescope & Instrument & Grating & $R=\lambda/\Delta\lambda$& $t_{exp}$ & N$_{exp}$ \\
  
  \hline
  \hline
   ANU & WiFeS & B7000/R7000 & 7000  & 300 & 41\\
   SALT & HRS & LR & 15000 & 1500 & 4\\
   SAAO & SpUpNIC & g4 & 2500  & 600 & 26 \\ 
   SMARTS 1.5 & Chiron & - & 27800 & 600 & 26 \\ 
   SMARTS 1.5 &  RC Spectrograph & 47/Ib & 3000 & 300 & 59$^*$ \\
   Swift & UVOT & UV & 150 & 275 & 8 \\
 \hline
 \footnotesize{$^*$ Taken from B18}
 \end{tabular}
\end{table}

\begin{table}
 \centering
  \caption{ Photometric observing log
  }
  \label{tab:phot log}
  \begin{tabular}{lllll}
  \hline
  Telescope  & Filter & $t_{exp}$ & N$_{exp}$ \\
  
  \hline
  \hline

   ASAS-SN  & V & 90 &616&\\
   ASAS-SN  & g & 90 &893&\\
   KELT  & White light & 150 & 5934&\\
   TESS  & White light & 120 &19104&\\
   %Swift/UVOT & W1 & 326 & 2\\
   %Swift/UVOT & M2 & 170  & 4\\
   %MeerKAT  & 1.284 GHz & 7200 & 1\\
 \hline
 \end{tabular}
\end{table}

%
%
%\onecolumn
%\begin{landscape}
%\begin{figure}
%\centering
%\includegraphics[width=1.5\textwidth]{fig/KELT_LC_POWER.png}
%
%\caption{\footnotesize
%{{\em KELT} light curve and power. Top panel: KELT instrumental magnitude - 3 vs time. Bottom panel: Time dependent %Lomb-Scargle power spectrum, abscissas axis represent the time in days and the frequency in cycles per day in the %ordinates. Each bin in the periodogram includes the contiguous $\pm$15 days with a minimum of 20 data points. The %horizontal white dotted line represent the frequency with highest power in the periodogram of the whole set.}}
%%\label{fig:dynamic_power}
%\end{figure}  
%\end{landscape}
%
%% Don't change these lines
%\bsp	% typesetting comment
%\label{lastpage}
\end{document}